\documentclass[10pt,journal,compsoc]{IEEEtran}
\newif\ifpeerreview

\peerreviewfalse

\usepackage[nocompress]{cite}
\usepackage{url}
\usepackage{amsmath,amssymb,graphicx}

\usepackage[switch]{lineno}
\usepackage{booktabs}
\usepackage{float}
\usepackage{xcolor}

\usepackage{nameref}
\usepackage{varioref}
\usepackage{hyperref}
\usepackage{cleveref}
\usepackage{multirow}

\newcommand{\paperID}{36}

\title{Deep Phase Coded Image Prior}

\author{Nimrod Shabtay,$^1$ Eli Schwartz,$^{1,2}$ and Raja~Giryes,$^1$~\IEEEmembership{Senior Member,~IEEE} \\
$^1$Tel-Aviv University, $^2$ IBM Research
}
\begin{document}

\IEEEtitleabstractindextext{%
\begin{abstract}
Phase-coded imaging is a computational imaging scheme designed to tackle tasks such as passive depth estimation and extended depth of field (EDOF) by inserting depth cues during image capture. Most of the current deep learning-based methods for depth estimation or all-in-focus imaging require a training dataset with high-quality depth maps and an optimal focus point at infinity for all-in-focus images. These datasets are difficult to create, often synthetic, and require external graphic software. Thus, limited datasets are being used, limiting the diversity of applicable domains.
We propose a new method named ``Deep Phase Coded Image Prior" (DPCIP) for jointly recovering the depth map and all-in-focus image from a coded-phase image using solely the captured image and the optical information of the imaging system. Our approach does not depend on any specific dataset and surpasses prior supervised techniques utilizing the same imaging system. This improvement is achieved through the utilization of a problem formulation based on implicit neural representation (INR) and deep image prior (DIP).
Due to our zero-shot method, we overcome the barrier of acquiring accurate ground-truth data of depth maps and all-in-focus images for each new phase-coded system introduced. This allows focusing mainly on developing the imaging system, and not on ground-truth data collection.
\end{abstract}

}

\maketitle
\thispagestyle{empty}

\IEEEraisesectionheading{
  \section{Introduction}\label{sec:introduction}
}
%
%
%
%
\IEEEPARstart{P}{hase}-coded imaging is a method in computational imaging, where a camera is equipped with an optical phase mask that provides unambiguous depth-related color characteristics for the captured image. While single-image passive depth estimation is an ill-posed problem, in phase-coded imaging, a phase mask is designed to create a different defocus blur for each color channel of the image. The depth cues, encoded by the phase mask, can later be decoded by a deep neural network for passive depth estimation or extended depth-of-field.

Prior works decode depth cues using a supervised learning approach, which involves training a neural network on a dataset with an accurate ground-truth to extract the depth map and/or all-in-focus image from the blurred (coded) image. . A major drawback of these methods is the need for a large and accurate dataset for training. 
Capturing such datasets can be quite complex and may require external sensors or graphic rendering software.
While synthetic datasets using graphic software are easier to create than real-world datasets with external depth sensors, they require expertise, and their ability to generalize to real-world applications is uncertain. Real-world datasets are limited in quantity. Thus, there are two options: rely on known datasets, which limits the problem domain, or create a dedicated dataset, which is limited and may suffer from a domain gap between training and test sets.

\begin{figure*}[]   
\centering
    \includegraphics[width=1.0\linewidth]{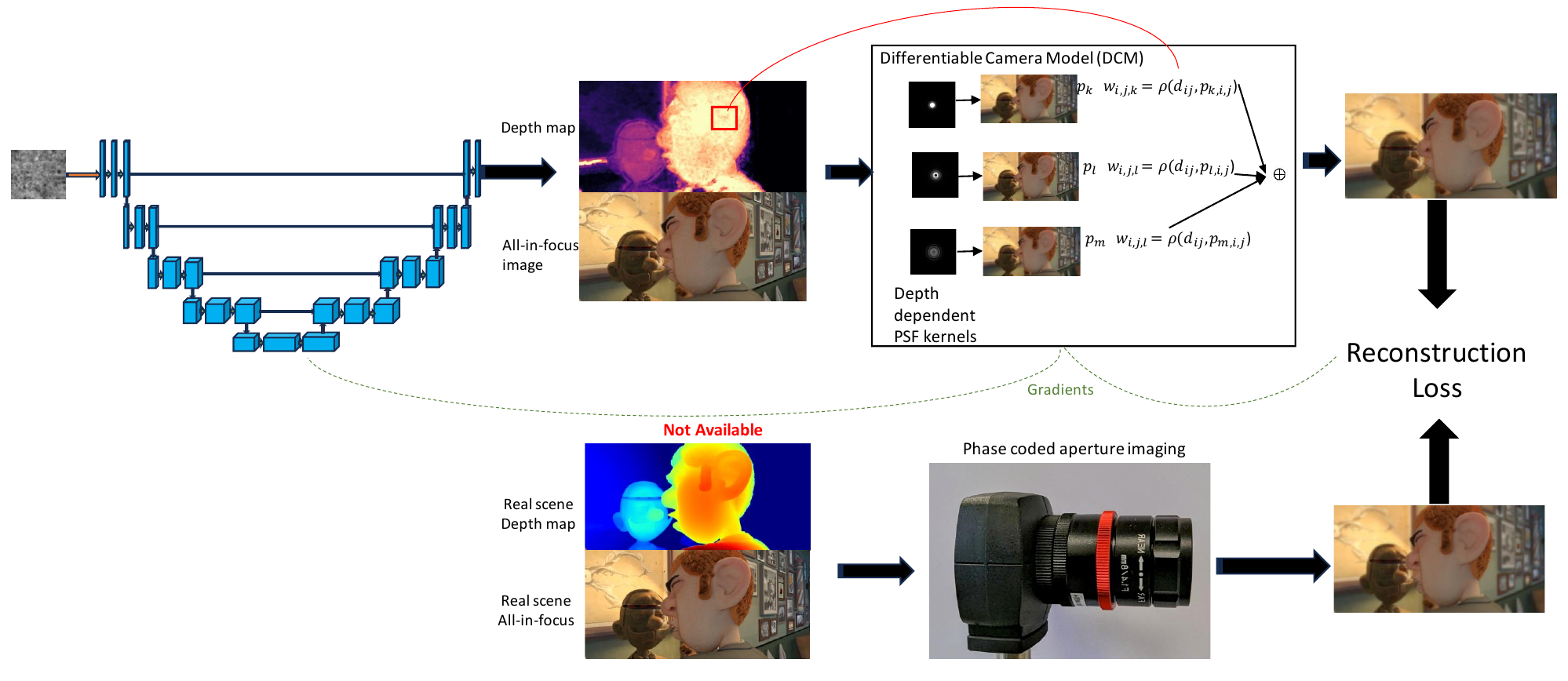}    
    \caption{\textbf{System overview.}  We follow a neural field flow, where an implicit generator first maps the encoded input image into reconstruction representations (in our case an all-in-focus image and a depth map). In the second part, A Differential Camera
Model (DCM) simulates the acquisition process of a phased-coded imaging system, provides the acquired image. The gradients from the reconstruction loss flow back to the generator forcing it to produce accurate intermediate outputs to match the acquired image after passing through the DCM. In the bottom row we illustrate the actual acquisition process compared to our neural field flow.}
    \label{fig:block_diagram}
\end{figure*}

To address the aforementioned limitations, we propose a new approach inspired by prior image restoration techniques that utilize Deep Image Prior \cite{ulyanov2018deep}. We formulate the tasks of single-image passive depth estimation and extraction of all-in-focus images as a unified inverse problem. An implicit generator is optimized to jointly create an all-in-focus image and a depth map from a given input code. The outputs from the generator are then passed through  a Differential Camera Model (DCM) that simulates phase-coded imaging from a given depth map and all-in-focus image, resulting in a blurred image with depth cues. Our reconstruction loss compares the given phase-coded image to the one generated by the DCM. Since our camera model is differential the gradients propagate all the way back to the generator to improve the mapping between the input code and the depth map and all-in-focus image such that, after passing through the DCM, the produced output matches the input. At the end of the learning process, we discard the final output and retain the intermediate results: the depth map and an all-in-focus RGB image.
Figure \ref{fig:block_diagram} illustrates this process.

\noindent \textbf{Contribution.} Our contributions can be summarized as: (i) proposing a method to extract both depth estimation and all-in-focus images in a zero-shot manner using a known DCM, reducing the need for a curated dataset for training; and (ii) outperforming prior supervised works for the same imaging system.
Our code is available at \href{https://github.com/NimrodShabtay/DPCIP}{https://github.com/NimrodShabtay/DPCIP}.

\section{Related Work}
\subsection{Coded Aperture Imaging}
Coded aperture imaging systems employ dedicated optical masks to encode light. Such imaging can decode and extract visual cues in order to reconstruct a passive depth estimation of the acquired image. Levin et al. \cite{levin2007image} use an amplitude-coded mask such that objects at different depths exhibit a distinctive spectral structure. Haim et al. \cite{haim2018depth} proposed a ring phase mask to generate depth-dependent PSF filters for encoding the depth cues. They proposed a fully-convolutional neural network to learn the decoding process to generate the depth map. A follow-up work by Gil et al. \cite{gil2019monster} leverages the encoding process in a stereo vision setup where conventional cameras fail. Chang et al. \cite{chang2019deep} proposed a depth-dependent image formation where a neural network learns to extract a depth map from a simulated acquired image produced from an all-in-focus image, a binned depth map and the lens parameters. Wu et al. \cite{wu2019phasecam3d} proposed an end-to-end architecture consisting of two parts: An optical layer that simulates depth-dependent PSFs given a learnable phase mask, and a reconstruction network, which estimates depth from the coded image.

Another line of work aims to recover an extended depth of field (EDOF) in order to reconstruct an all-in-focus image from the acquired image. Elmalem et al. \cite{elmalem2018learned} proposed to use a ring mask with a depth-dependent PSF function and a neural network to recover the depth cues and the all-in-focus image. Akpinar et al. \cite{akpinar2021wavefront} employed wavefront coding via a diffractive optical element (DOE) and a neural network for deblurring in an end-to-end learning process. Gkioulekas et al. \cite{Gkioulekas2021defocus} used a dual pixel camera and a pre-defined per depth calibration kernel set, to construct a multi-depth representation for extracting the all-in-focus image and the depth map simultaneously. Sitzmann et al. \cite{sitzmann2018end} constructed a joint optimization system composed of an optical part and a reconstruction part to simulate a real image acquisition for various image processing tasks such as super-resolution and extended depth of field.
In addition, Yosef et al. \cite{yosef2021video} proposed to use the coding in the temporal axis instead of the depth axis to reconstruct video from a coded motion blur.
An important common property of previous works is the need to create a dedicated dataset for training the reconstruction module, where in our approach we apply a self-supervised scheme consisting only of a single input image, eliminating the cumbersome preliminary step of constructing a dedicated dataset.

\subsection{Deep Image Prior}
Deep Image Prior (DIP) \cite{ulyanov2018deep} showed that a convolutional neural network (CNN) can recover a clean image from a degraded image in the optimization process of mapping random noise to a degraded image. The power of DIP is shown for several key image restoration tasks such as denoising, super-resolution, and inpainting.
DIP was extended to various applications. DoubleDIP \cite{gandelsman2019double} introduced a system composed of several DIP networks, where each learns one component of the image such that their sum is the original image. It has been used for image dehazing and foreground/background segmentation. SelfDeblur \cite{ren2020neural} performed blind image deblurring by simultaneously recovering a sharp image and the corresponding blur kernel using DIP.
In the realm of medical imaging, various studies have endeavored to harness the potential of image priors for the reconstruction of PET images \cite{Yokota_2019_ICCV, Hashimoto_2022, 8581448, 9576711}. Alternative approaches have sought to enhance the efficacy of image priors through modifications and additions to the original configuration. Mataev et al. \cite{Mataev2019DeepREDDI} and Fermanian et al. \cite{fermanian:hal-03310533} employed a combination of DIP and the plug-and-play framework to enhance DIP's performance across multiple inverse problems. In other domains, Kurniawan et al. \cite{kurniawan2022noise} introduced a demosaicing method based on DIP, while Chen et al. \cite{chen2020dip} proposed a neural architecture search guided by DIP principles.
Recently, Shabtay et al. \cite{shabtay2022pip} demonstrated the connection between DIP and implicit neural representation (INR). The connection between DIP and INR inspired the formulation of our reconstruction method as a self-supervised system consisting of a DIP generator and a DCM.

\subsection{Implicit Neural Representation}
Implicit neural representation is an area of research that studies the abilities of neural networks to map input coordinates to target values. This includes 2D images, 3D shapes, SDF maps and many more.
It was demonstrated in \cite{tancik2020fourier} that representing the input coordinates as Fourier-features with a tuneable bandwidth enables a simple MLP to generate complex target domains such as images and 3D shapes while preserving their high-level details.
In Neural Radiance Fields (NeRF) \cite{mildenhall2020nerf}, implicit functions with Fourier features are used for synthesizing novel views of a 3D scene from sparse 2D images.
SIREN \cite{sitzmann2019siren} showed that by changing the activation function of a simple MLP from the frequently used ReLU to a periodic activation (for example a sine activation), the network can represent the spatial and temporal derivatives using a simple input grid to successfully recover a wide range of target domains (images, videos, 3D surfaces, etc.).
Following NeRF, improvements were proposed \cite{Yu_2021_ICCV, barron2023zip, muller2022instant} in terms of convergence speed compared to the original NeRF by utilizing a structure for the implicit model.
Methods were also proposed \cite{Barron_2021_ICCV, Barron_2022_CVPR, Tancik_2022_CVPR} to tackle the scaling and aliasing problems that the original NeRF suffered from.
Yu et al. \cite{yu2021pixelnerf} suggested a method to reduce the need for many calibrated images from sparse views as supervision for NeRF.
Mildenhall et al. \cite{Mildenhall_2022_CVPR} suggested a method for rendering the raw images before the ISP in order to achieve much wider rendering abilities such as control on tone mapping.
It was also proposed \cite{ramasinghe2022beyond} that Gaussian activation can perform better than sine activation while having the property of being continuous everywhere and at every derivative.
SAPE \cite{hertz2021sape} and BACON \cite{lindell2021bacon} demonstrate how a combination of a coordinate network along with a limitation of its frequency spectrum can achieve a multi-scale representation of the target domain.

\section{Deep Phase Coded Image Prior}
\label{sec:method}
We consider an imaging system capable of extracting simultaneously an all-in-focus RGB image and a pixel-level depth map with a single image capture.
Our goal is to achieve single image depth estimation and extended depth of field (EDOF) simultaneously relying solely on the captured image without any training data.Eliminating the need for a large dataset is a major advantage over supervised methods, which use synthetic or real-world datasets from limited domains and are sensitive to domain shift and generalization \cite{Butler:ECCV:2012,silberman2012indoor,mayer2016large}, thus they are sensitive to domain shift and generalization. We achieve this via a joint optimization of a generation part (a U-Net shaped neural network) and a forward process part, namely, a differentiable approximation of the acquisition process of the phase-coded imaging system - DCM in short.

As shown in Figure \ref{fig:block_diagram}, we formulate our problem as an implicit neural representation problem where our system is composed of two major components: an implicit generator and a Differentiable Camera Model (DCM). 

\subsection{Phase Mask}
\begin{figure*}
    \centering
    \begin{tabular}{ccc}
        \includegraphics[width=0.3\linewidth]{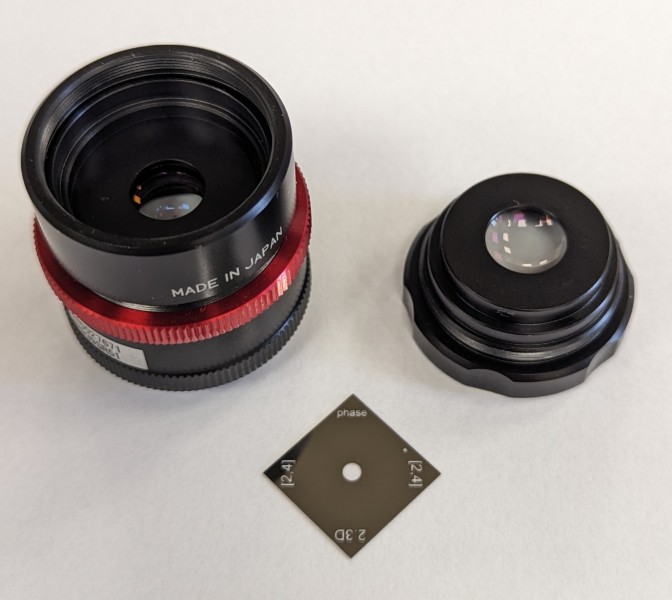} & 
        \includegraphics[width=0.3\linewidth]{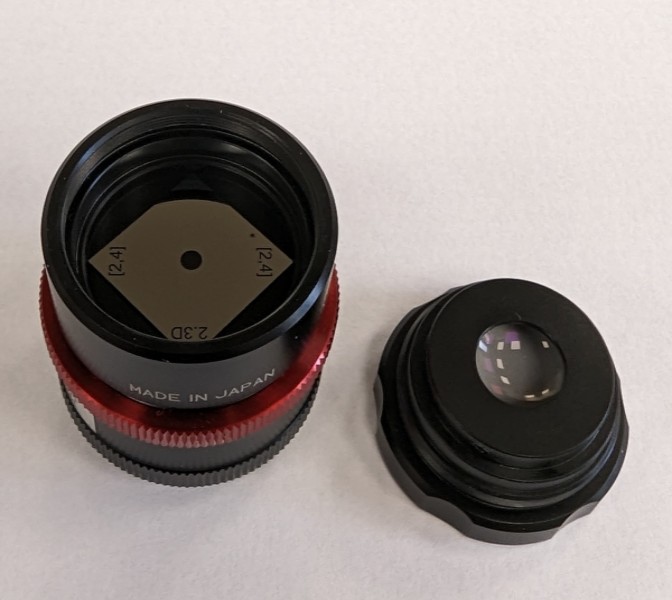} &
        \includegraphics[width=0.27\linewidth]{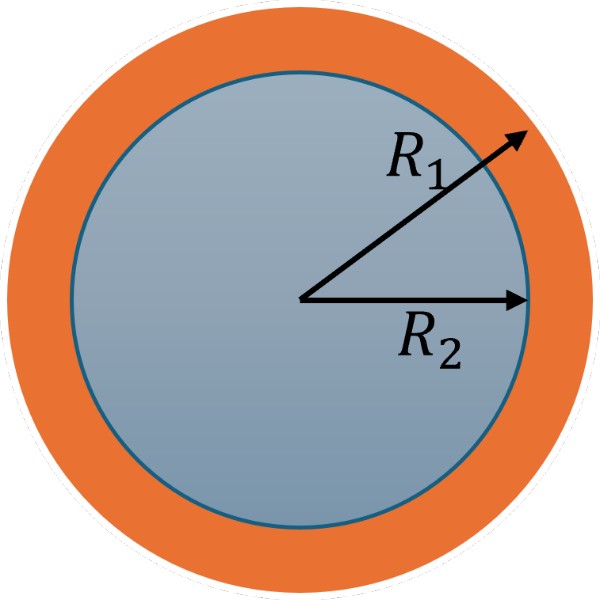} \\
    \end{tabular}
    \caption{\textbf{Hardware setup}. Our setup is composed of a phase mask embedded inside the lens on the aperture plane. On the left, the optical components are placed seperately. In the middle, the phase mask is located on the aperture plane. On the right, you can see a schematic view of the phase mask pattern.}
    \label{fig:hw_Setup}
\end{figure*}

We simulate the coded-phase acquisition process as done in \cite{haim2018depth, elmalem2018learned}. The proposed mask is a two-ring phase mask designed to manipulate the PSF of the imaging system based on color and depth. For a color RGB sensor, three separate behaviors can be generated, such that at each depth of the scene, a different channel (R/G/B) is in focus while the others are not. Figure \ref{fig:hw_Setup} describes the hardware setup of our system.

Assuming the optical parameters are known (lens parameters and the focus point), we can recover the out of focus (OOF) parameter (denoted as $\psi$) from \cite{goodman2005introduction}. For the case of a circular exit pupil with radius $R$, the following equation defines the OOF parameter:
\begin{equation}\label{eq:optic}
    \psi = \frac{\pi R^2}{\lambda}(\frac{1}{z_0} + \frac{1}{z_{img}} - \frac{1}{f})=\frac{\pi R^2}{\lambda}(\frac{1}{z_{img}}-\frac{1}{z_i})=\frac{\pi R^2}{\lambda}(\frac{1}{z_0}-\frac{1}{z_n}),
\end{equation}
where $\lambda$ is the illumination wavelength, $z_{img}$ is the sensor plane location for an object at a nominal position $z_n$, and $z_i$ is the ideal location for an image plane with respect to an object at location $z_0$.

From Eq.~\ref{eq:optic} we can derive that a focused object is located at $\psi = 0$, and an out of focus object will be located at other values of $\psi$. 
While defocus has a symmetry around the focus point, it is not possible to generate a symmetric phase mask for multiple colors ($\pi$ phase for one color is a different phase for another). Therefore, we enhance the color-dependent effects, and create an extremely nonsymmetric mask, which behaves very differently for each color. The asymmetric design can help us extend the depth of field (DOF) for each color separately (i.e. for every depth one color channel will provide good contrast, one medium contrast and one poor). These depth-color cues will serve also as the depth hints. Since closer objects will be larger in the final image, they can ‘suffer’ from more defocus so we take the longer part of the antisymmetric domain towards the closer objects. The range values [-4, 10] is a design choice, given the hardware and optics components.

\subsection{Implicit Generator}
The implicit generator is based on the model introduced in DIP \cite{ulyanov2018deep}. It is an encoder-decoder model such that the encoder has 5 levels and each level has 2 convolutional layers with a depth of 128 and 3x3 spatial kernels. The downsampling between the levels is done by stride. The decoder has also 5 levels with 2 convolutional layers with a depth of 128 and 3x3 spatial kernels followed by bilinear upsampling. 
At each level the model has skip connections as a 1x1 convolutional layer with a depth of 16. This creates a generator whose learnable parameter is the mapping between an input code and a pair consisting of an all-in-focus image and a pixel-level depth map. Like in DIP, we define the input code to be uniform random noise, $X_{input} \sim \mathcal{U}[0,1]$, where $X_{input}\in\mathbb{R}^{D \times H \times W}$.

\subsection{Differentiable Camera Model (DCM)}
The DCM is a differentiable simulation of the phase-coded optical imaging. Given a pair of an all-in-focus image and a depth map, it generates the acquired image from a phase-coded imaging system. Our method and simulations are based on \cite{haim2018depth, elmalem2018learned}, where we have a ring mask with pre-defined depth-dependent PSF kernels. Since the acquisition process has several non-differentiable operations  (i.e. `argmax'), we used a differentiable approximation, 
inspired by \cite{Gkioulekas2021defocus, chang2019deep} - The DCM takes the generated sharp RGB image and creates several blur images based on a fixed set of defocus ($/psi$) values. Each image is a single defocus blur corresponding to a particular depth. Then, with the generated depth map, the DCM linearly interpolates each pixel based on the set of blur images it created to produce the acquired RGB pixel value. 

This design allows our acquisition process to be fully differentiable, thus suitable as a forward process in a neural field flow. In this setup, the depth-dependent PSF kernels are pre-defined and we only learn the interpolation weights between the fixed set of produced images. See Figure \ref{fig:block_diagram} for visual illustration.
The differentiable variant allows us to optimize the whole system in a self-supervised manner, where the generator first maps the input encoding to a pair of intermediate outputs (an all-in-focus image and a pixel-level depth map) that are fed as inputs to the DCM in order to reproduce the acquired image taken by the camera.

The joint optimization of producing both an all-in-focus image and a pixel-level depth map forces an alignment between them, also the pass in the DCM forces the intermediate outputs not to collapse to a trivial blurry solution but to an accurate solution and yields better results compared to previous works where a model was trained to produce just a depth map or just an all-in-focus image.

To optimize our neural network weights to map a coded input to pixel values we used the non-differential acquisition simulation to produce coded-phase images to act as our supervision. 
Our objective function is to the minimize difference between a given image and the forward process output. Similarly to \cite{ren2020neural}, we first use L2 as our reconstruction loss and switch to SSIM \cite{wang2004image} after a fixed number of iterations (see Section~\ref{sec:ablation} for ablation on the benefits of the switch).
We can formulate our optimization objective as follows:     
    \begin{equation}\label{eq:objective}
    \theta^{*} = \operatornamewithlimits{argmin}_{\theta} \ell (h(f_{\theta}(z)) - y),   
\end{equation}
where $h$ is the DCM, $f_{\theta}$ is the implicit model and $z$ is the encoded input. $y$ is the blurry image acquired by our coded phase imaging system.
We evaluate our method both for PSNR and SSIM for the all-in-focus images and the mean error in meters for the depth maps.

\section{Experimental Results}
We conduct a series of experiments, both in simulation and on real-world examples using a dedicated phase-coded camera. We show only qualitative results on real-world image due to lack of ground-truth. Thus, for quantitative comparison and evaluation of our method we focused on simulated data with accurate ground-truth.
Unless stated otherwise, we used a U-net model as described in Section \ref{sec:method}. We trained with Adam optimizer \cite{kingma2014adam}, with $0.01$ learning rate. We switched the reconstruction loss from L2 to SSIM after 500 iterations.

\subsection{Simulation Results}
To evaluate the performance of our method quantitatively both in terms of depth estimation and all-in-focus image reconstruction, we first perform an evaluation on a simulated dataset, namely, `TAUAgent’ dataset\footnote{https://www.cs.toronto.edu/~harel/TAUAgent/home.html} \cite{gil2019monster}. The Agent dataset consists of a total of 530 images from five synthetic scenes created using the `Blender’ computer graphics software (named ``City", ``Headbutt", ``Sitting", ``WallSlam", ``WuManchu"). Each scene consists of a ground-truth low-noise all-in-focus image, along with its corresponding pixel-wise accurate depth map. Such data enables an exact depth-dependent imaging simulation, with the corresponding depth of focus (DOF) effects. In contrast to prior works \cite{elmalem2018learned, gil2019monster, haim2018depth}, we do not use the `ground-truth' data from the Agent dataset in our optimization. We only use the acquired coded images produced by the camera simulation. The all-in-focus GT images and the GT depth maps are only used to evaluate the performance of our method. For the all-in-focus evaluation, we used PSNR and SSIM, while for the depth estimation, we used average error in meters (as done in \cite{haim2018depth, gil2019monster}).
To make a fair comparison we simulate the imaging with the same optical parameters as done in \cite{elmalem2018learned, haim2018depth, gil2019monster}. 

Note that a crucial disadvantage of previous works is the high dependency on a large and high-quality dataset for successful learning. High-quality datasets like the `agent' dataset are usually small and cumbersome to collect, thus, making the learning much more challenging. In contrast to them, our method reduces this constraint since we explicitly incorporate the DCM into the learning scheme. Thus, we can extract both the all-in-focus RGB image and the pixel-level depth map from a single captured image.
We took a subset of 110 images (namely, ``agent subset") from the scenes mentioned above to evaluate the performance on several tasks namely passive depth estimation, image reconstruction, and image deblurring. Lastly, we also show the effect of the implicit generator on the performance of the mentioned tasks.
We ran all the experiments on RTX-2090Ti. Each sample took $\sim$ 15 minutes and the memory footprint is 7.5GB for $1024 \times 512$ RGB images.

\subsubsection{Passive Depth Estimation}
\begin{figure*}  
    \resizebox{1.0\linewidth}{!}{
    \begin{tabular}{cccccc}
    & & Supervised & & 0-shot &\\
    GT & Input & Mono from \cite{gil2019monster} & Error map (Mono-GT) & DPCIP & Error map (DPCIP-GT)\\
    \includegraphics[width=0.25\linewidth]{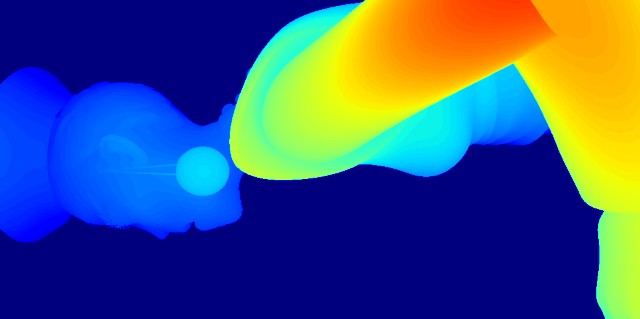} & 
    \includegraphics[width=0.25\linewidth]{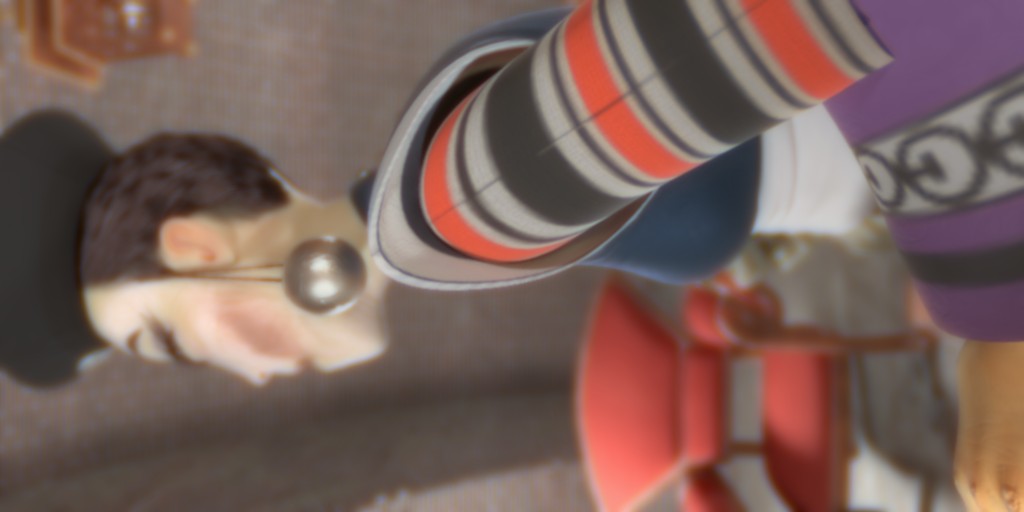} & 
    \includegraphics[width=0.25\linewidth]{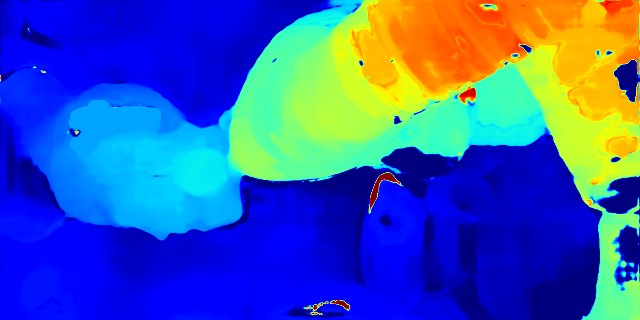} & 
    \includegraphics[width=0.285\linewidth]{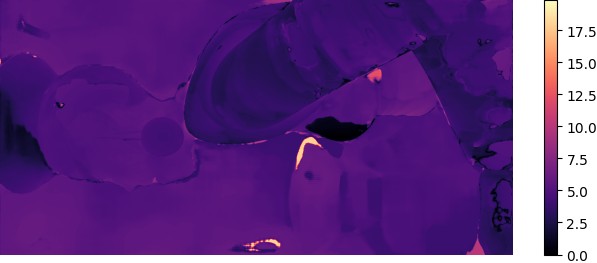} & 
    \includegraphics[width=0.25\linewidth]{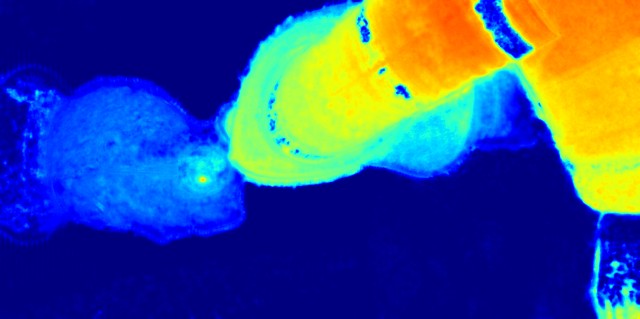} &
    \includegraphics[width=0.285\linewidth]{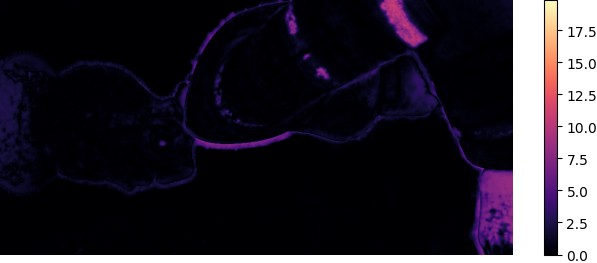} \\     
    \includegraphics[width=0.25\linewidth]{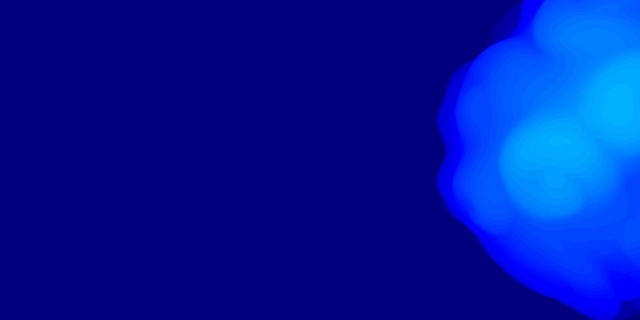} & 
    \includegraphics[width=0.25\linewidth]{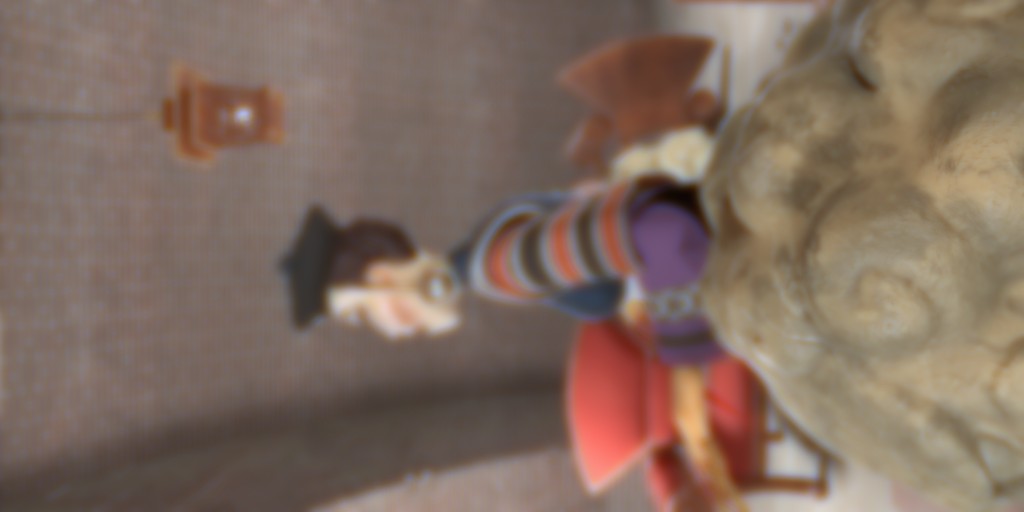} & 
    \includegraphics[width=0.25\linewidth]{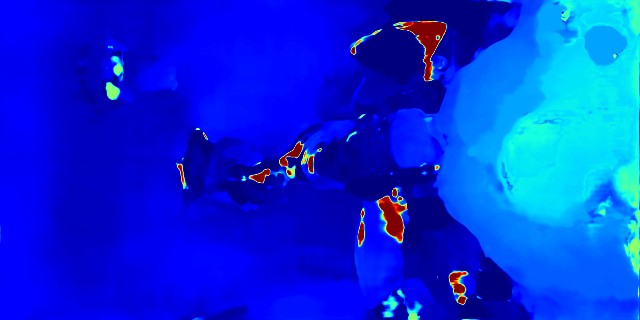} & 
    \includegraphics[width=0.285\linewidth]{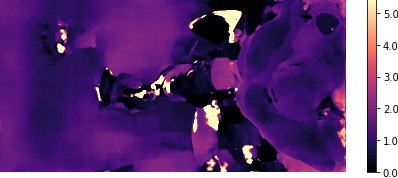} & 
    \includegraphics[width=0.25\linewidth]{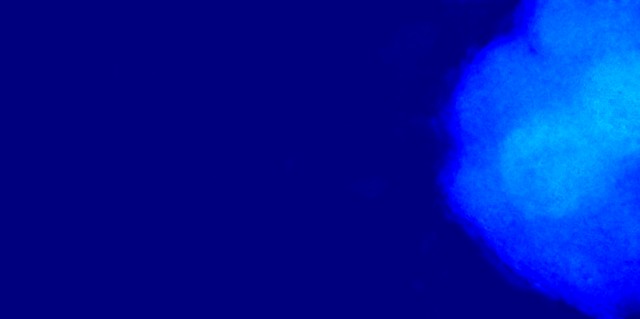} & 
    \includegraphics[width=0.285\linewidth]{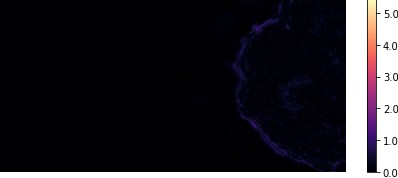} \\ 
    \end{tabular}
    }
    \caption{\textbf{A qualitative comparison of depth map estimation} reveals that our method generates more precise depth maps compared to the mono network from \cite{gil2019monster}, In order to see the performance gap between the methods, we added an absolute difference error map for each method, we can see that DPCIP tends to produce much more accurate depth maps in the background and on the main object compared to the mono network. Note that the depths are clipped to the imaging system's physical range.}
    \label{fig:depth}
\end{figure*}

To assess our method's performance in passive depth estimation we compared ourselves to the monocular depth estimation network for phase-coded imaging from \cite{gil2019monster} which built upon the same phase-coded imaging system. We also followed the same optical parameters to produce the same acquired images. The evaluation was done on the relevant part of the ``agent subset" (``WuManchu" scene) that was not used for training \cite{gil2019monster}. Table \ref{tab:depth_estimation} shows a quantitative comparison in terms of RMSE error in meters. Visual results can be found in Figure \ref{fig:depth}.
Overall, our method outperformed the mono network by two orders of magnitude and effectively managed to produce an accurate pixel-level depth map solely from a given captured image. More examples can be found in the supplementary materials.

\begin{table}[!b]
 \caption{\textbf{A quantitative comparison of depth estimation. Results are reported in terms of depth RMSE error [m] ($\downarrow$).} Our method outperforms the mono network from \cite{gil2019monster} in depth estimation on the same imaging system, even though the Monocular depth estimation network from \cite{gil2019monster} was trained in a supervised manner on the same dataset. This is likely due to the fact that the dataset is small and the model is prone to overfitting.}
 \centering
    \resizebox{0.7\linewidth}{!}{
    \begin{tabular}{l|c|c}
    \toprule
    & Supervised & 0-shot \\
         & Mono from \cite{gil2019monster} & DPCIP \\ 
    \midrule
    WuManchu & 0.1639 & \textbf{0.0003}   \\ 
    \bottomrule
\end{tabular}
}
\label{tab:depth_estimation}
\end{table}

\subsubsection{Extended Depth Of Field (EDOF)}
\begin{figure*}  []
    \resizebox{1.0\linewidth}{!}{
    \begin{tabular}{cccc}
    & & Supervised & 0-shot\\
    GT & Input & EDOF \cite{elmalem2018learned}  & DPCIP \\
    \includegraphics[width=0.25\linewidth]{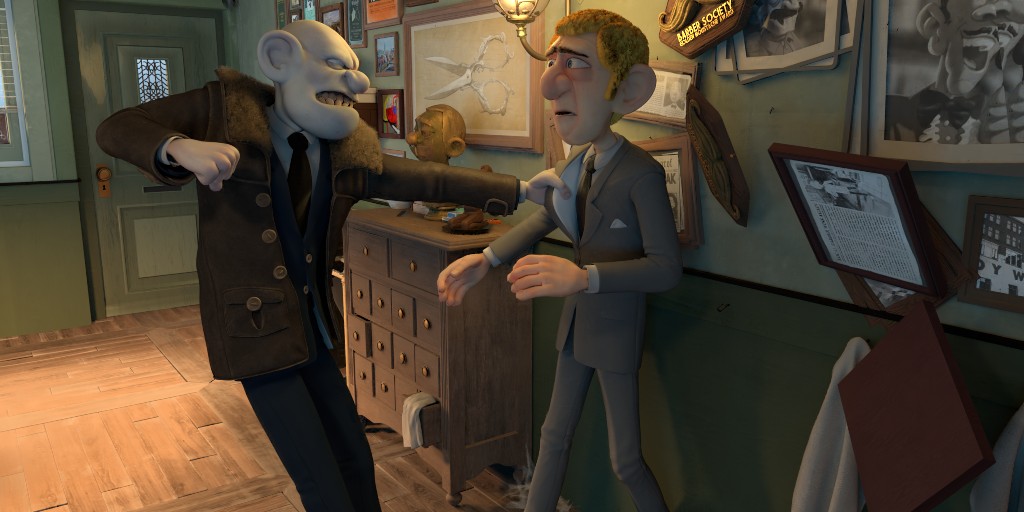} & 
    \includegraphics[width=0.25\linewidth]{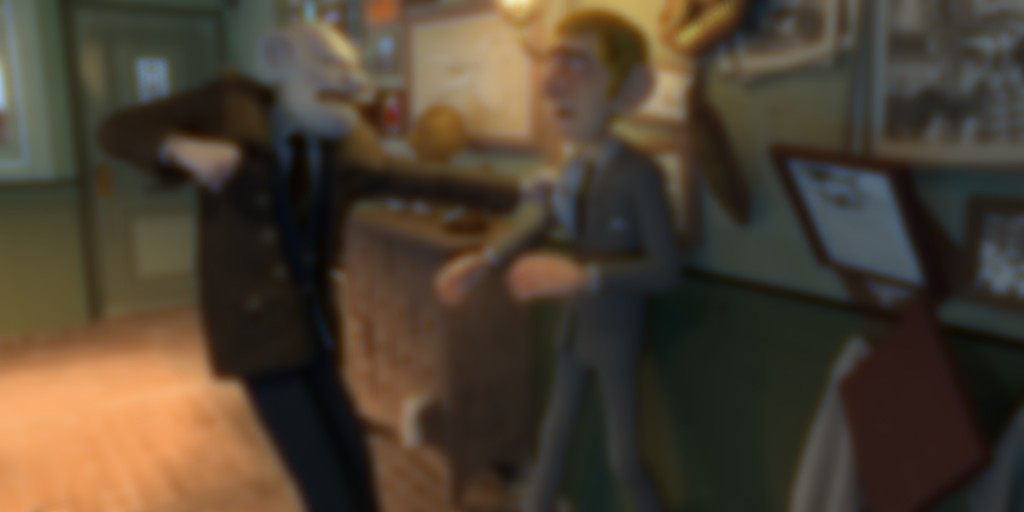} & 
    \includegraphics[width=0.25\linewidth]{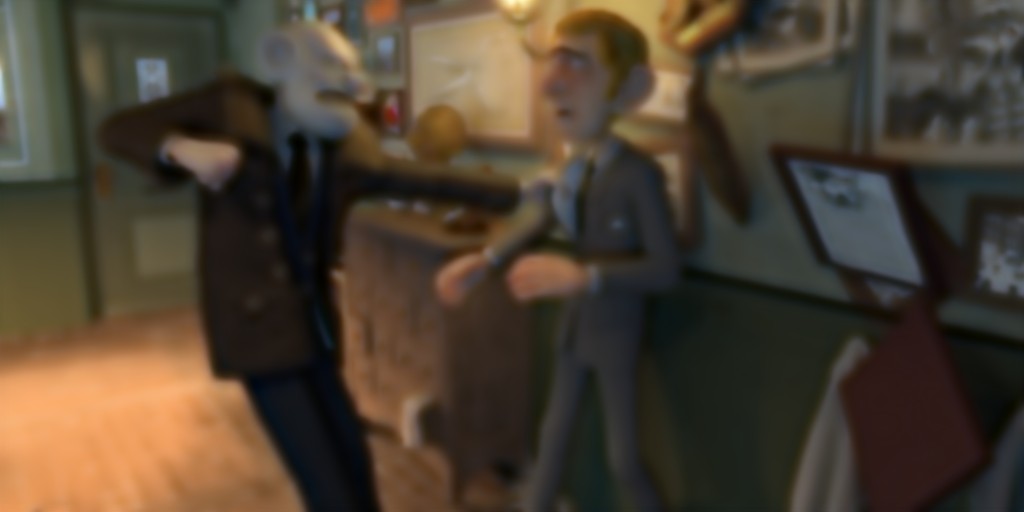} & 
    \includegraphics[width=0.25\linewidth]{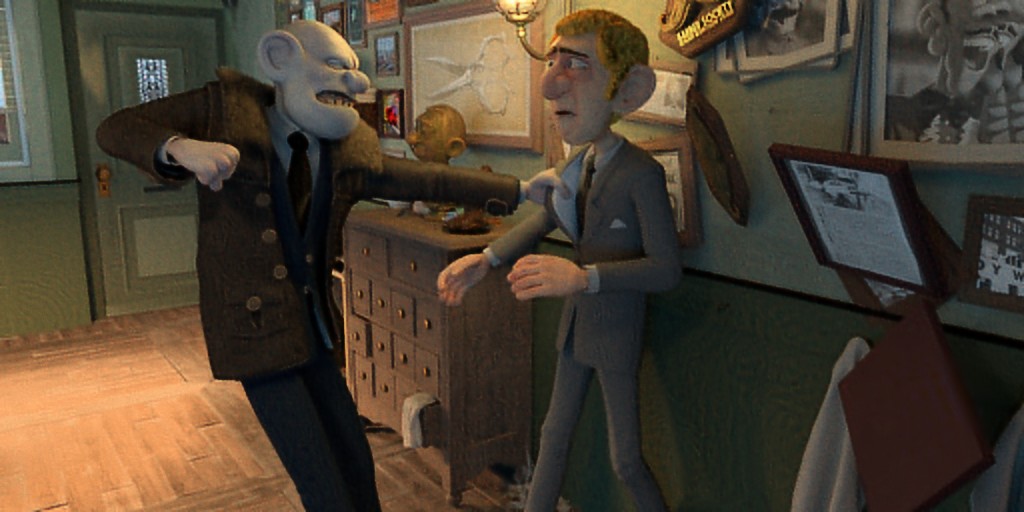} \\ 
    \includegraphics[width=0.25\linewidth]{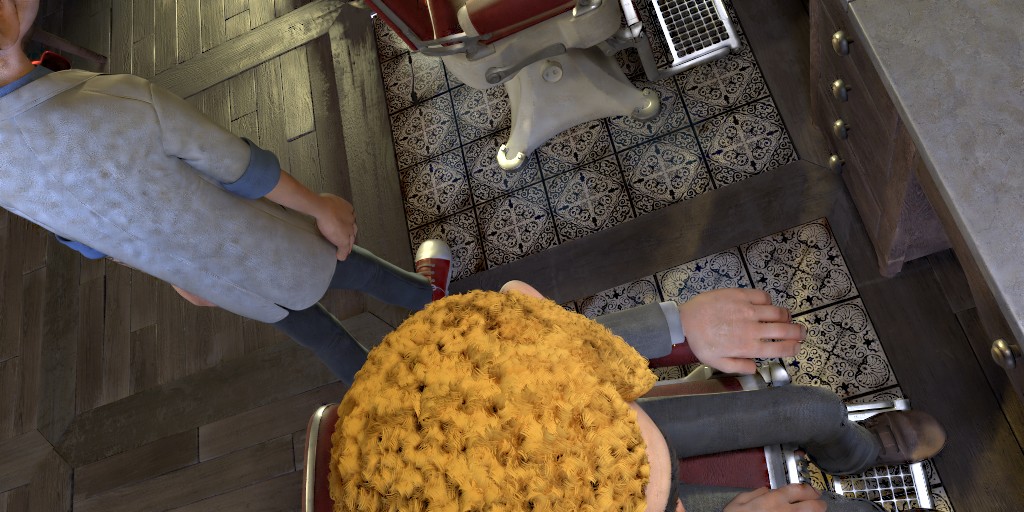} & 
    \includegraphics[width=0.25\linewidth]{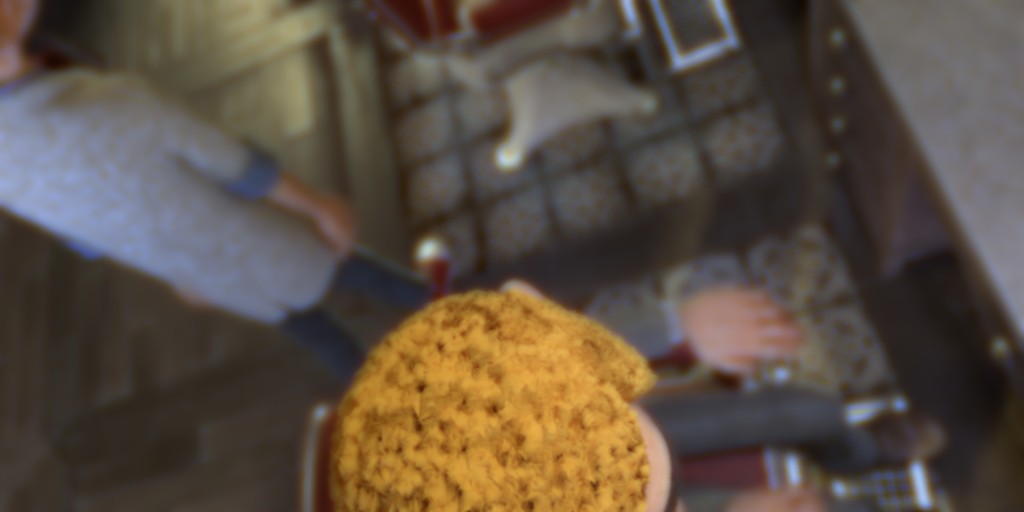} & 
    \includegraphics[width=0.25\linewidth]{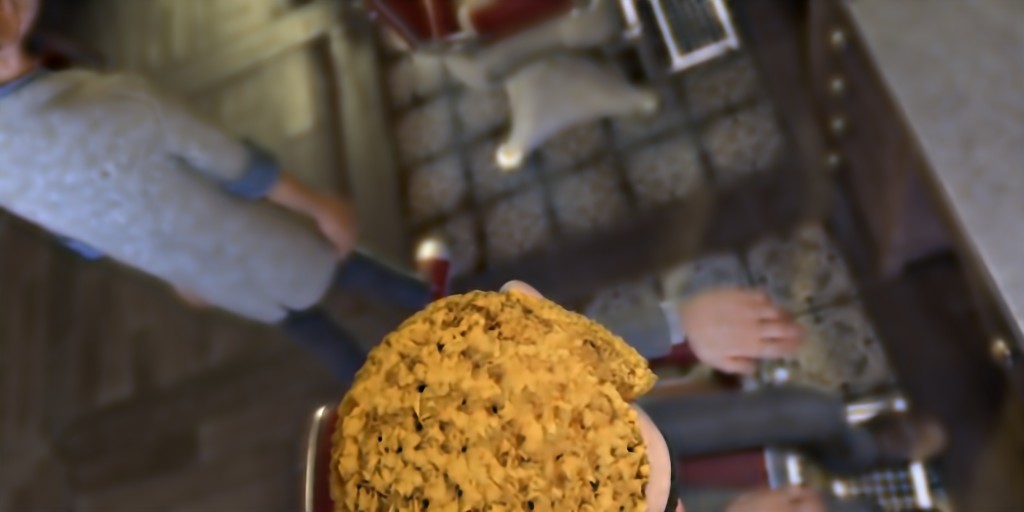} & 
    \includegraphics[width=0.25\linewidth]{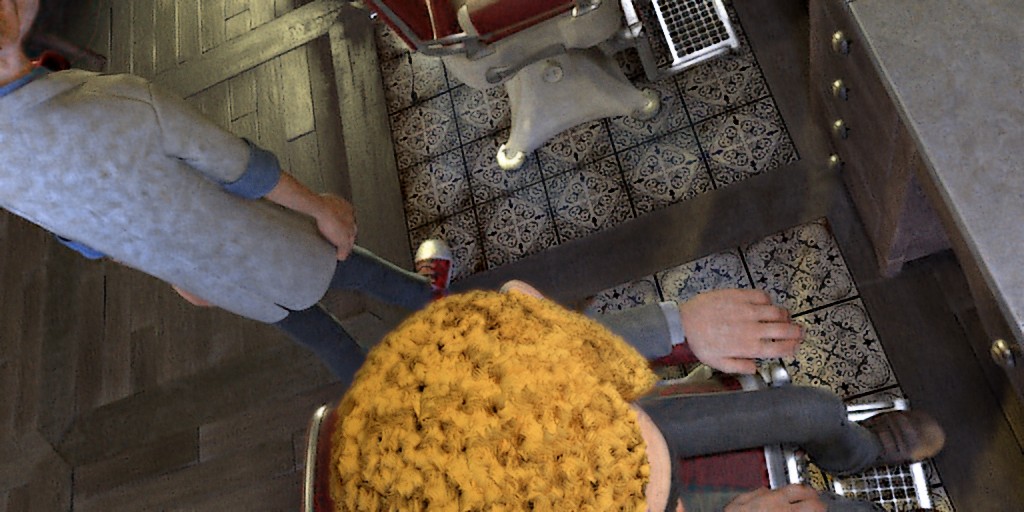} \\ 
    \end{tabular}
    }
    \caption{\textbf{A qualitative comparison of image reconstruction.} Our method produced a much more accurate all-in-focus images. The existing EDOF baseline \cite{elmalem2018learned} produced over-smooth all-in-focus images compared to our method.}
    \label{fig:edof}
\end{figure*}

To assess our method performance in the task of extended depth of field (EDOF) we compared ourselves to \cite{elmalem2018learned} which built upon the same phase-coded imaging system. We also follow the same optical parameters to produce the same acquired images. In this case, the original method was trained on a different synthetic dataset, so we could test all the images in our subset. Table \ref{tab:all_in_focus} shows a quantitative comparison in terms of PSNR and SSIM. Visual results can be found in Figure \ref{fig:edof}.
Our method outperformed the baseline by $5 dB$ in PSNR and 0.13 in SSIM on average and showed an improvement in terms of image reconstruction metrics. To better show our method's advantages we did additional effort optimizing EDOF architecture and hyperparameters. We were able to improve from the original work by $\sim 2dB$ on average. Yet, we were still $\sim 2dB$ on average below our zero-shot method.
Overall our method outperformed the original EDOF baseline \cite{elmalem2018learned} and our improved EDOF baseline both in terms of PSNR [dB] and SSIM. Remarkably, relying only on a single given phase-coded image. Table \ref{tab:all_in_focus} shows a quantitative comparison using the agent subset. Qualitative results can be found in Figure \ref{fig:edof}. More examples can be found in the supplementary materials.

\begin{table}[htbp]
    \caption{\textbf{A quantitative comparison of image reconstruction (PSNR [dB] $\uparrow$/ SSIM $\uparrow$)}. Our method outperforms the existing supervised baseline significantly. The PSNR average improvement of our method is $\sim 5dB$. We re-trained EDOF with an improved architecture and optimization process. DPCIP performs better in $\sim2dB$ over the improved supervised baseline.}    
    \vspace{-0.3cm}
    \label{tab:all_in_focus}
    \centering
    \begin{tabular}{lccc}
        \toprule
        & Supervised & Supervised & 0-shot \\
        Test Scene & Original EDOF & Improved EDOF & DCPIP \\
        \midrule
        City & 24.01/0.65 & 28.75/0.82 & \textbf{29.73/0.91} \\
        Headbutt & 28.54/0.92 & 27.673/0.8346 & \textbf{31.28/0.94} \\
        Sitting & 26.12/0.81 & 29.12/0.84 & \textbf{30.04/0.91} \\
        WallSlam & 26.94/0.87 & \textbf{32.54/0.91} & 31.45/0.93 \\
        WuManchu & 23.67/0.71 & 26.546/0.71 & \textbf{30.86/0.93} \\
        \bottomrule
    \end{tabular}
\end{table}

\subsubsection{Deblurring}
We extend our experiments to solve blind image deblurring. For image deblurring, we aim to reconstruct a sharp image from a blurred one assuming we have a single blur kernel.
Image deblurring can be thought of as a reduction of the defocus problem as in defocus we have a depth-dependent blur kernel. Since our imaging system has an effective range, everything outside the effective range will be clipped to the range's limits.
Thus we can simulate a deblurring setup from our defocus problem and the system's physical limitations. To simulate a deblurring experiments, we chose images where ground-truth depth values are all outside the imaging system's effective range, values outside the effective range will result a clip of the $\psi$ values, hence we will have acquired images with a single PSF blur kernel- which is a exactly a deblurring scenario.

We compare our method with Neural Deblurring \cite{ren2020neural}, a single image deblurring approach that shares several key components with our method, such as using DIP as a generator and a learned module (fully-connected network) to recover the sharp image and the blurring kernel respectively. 
As mentioned, to mimic a deblurring setup we took from the `agent' dataset 13 examples from 3 scenes. Each example has a constant depth map ($\psi=-4$).
We ran DPCIP and Neural Deblurring on all images and compared PSNR and SSIM for image reconstruction.
However, Neural Deblurring fatally fails in scenarios where the deblurring kernel is large (the PSF kernels in the DCM are 71x71). In addition, our blur kernels are channel-dependent. Since comparing to Neural deblurring `out of the box' is not practical, we modify Neural Deblurring to work with our setup (e.g. phase mask kernel sizes). First, we let it predict each channel separately to match the models' capacity to our problem. In addition, to overcome the kernel size limitation, we effectively reduced the predicted kernel from the original size of 71x71 to 19x19. The size reduction can be done since the pre-defined PSF kernels are centered (See more details in the supplementary material). 

\begin{figure*} []   
    \resizebox{1.0\linewidth}{!}{
    \begin{tabular}{cccc}
    & & 0-shot & 0-shot\\
    GT & Input & Neural-Deblurring\cite{ren2020neural} & DPCIP \\
    \includegraphics[width=0.25\linewidth]{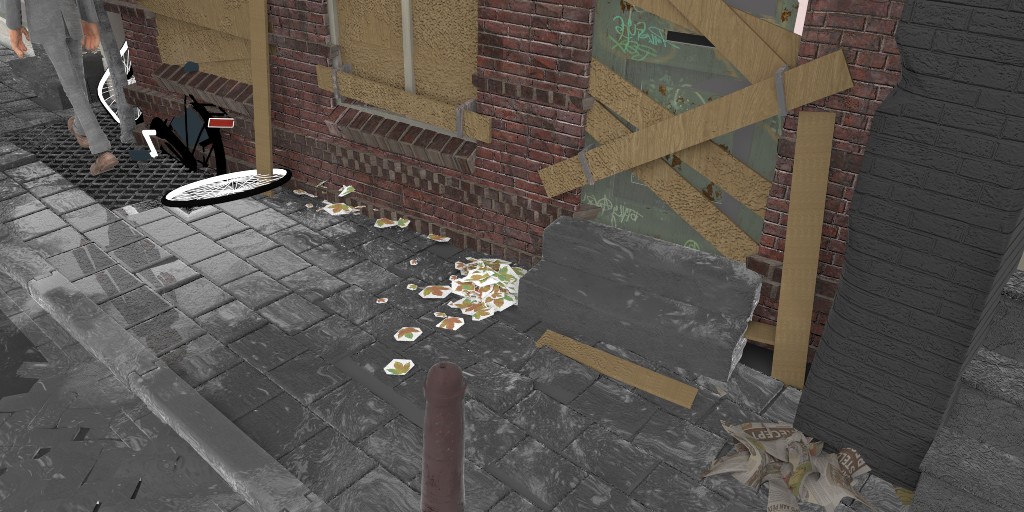} & 
    \includegraphics[width=0.25\linewidth]{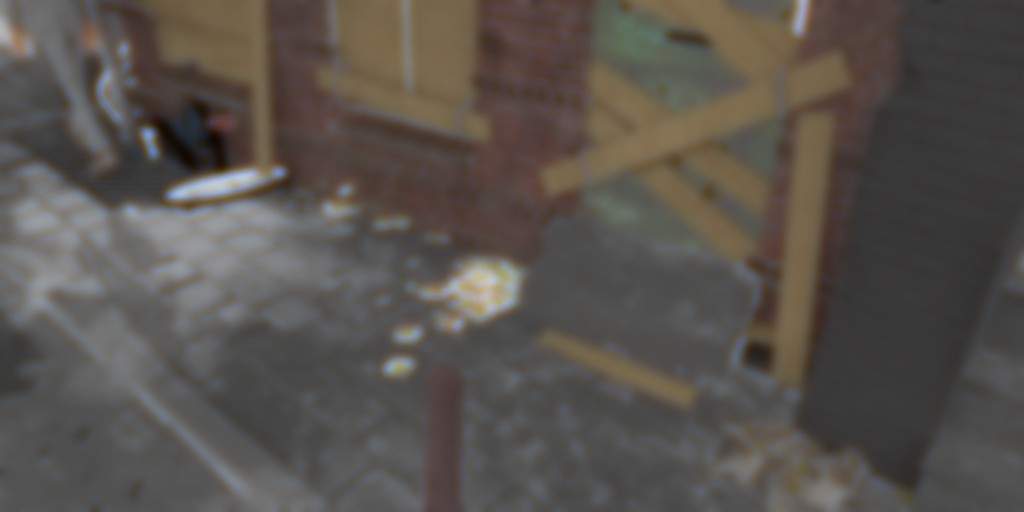} & 
    \includegraphics[width=0.25\linewidth]{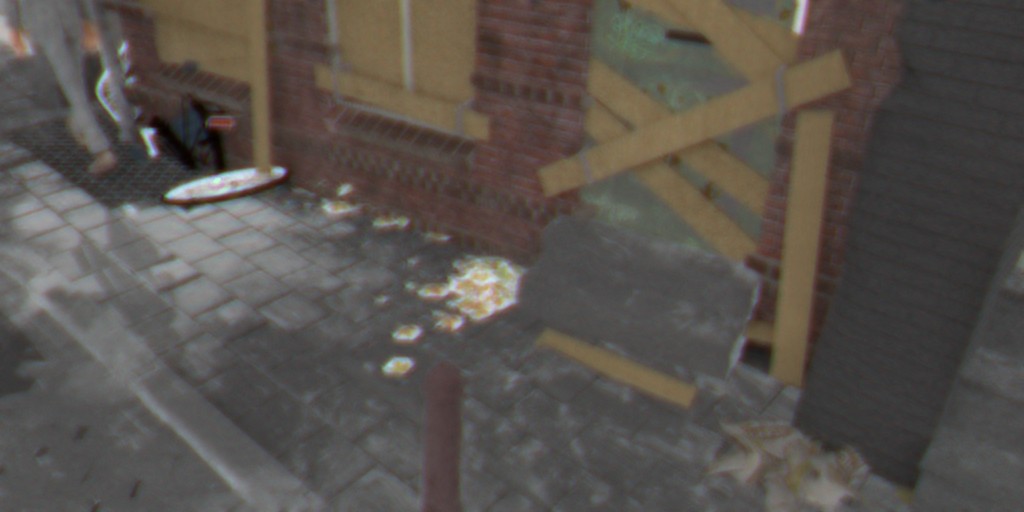} & 
    \includegraphics[width=0.25\linewidth]{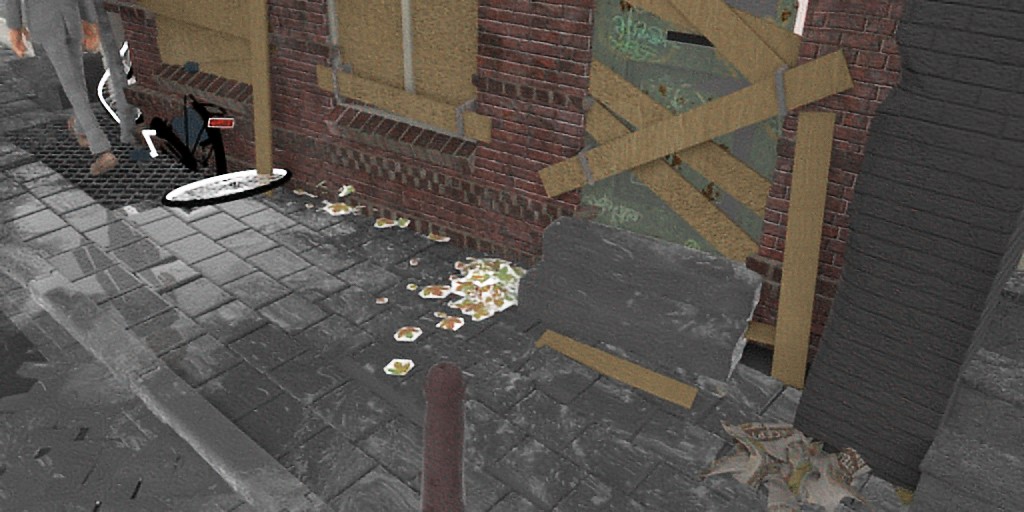} \\ 
    \includegraphics[width=0.25\linewidth]{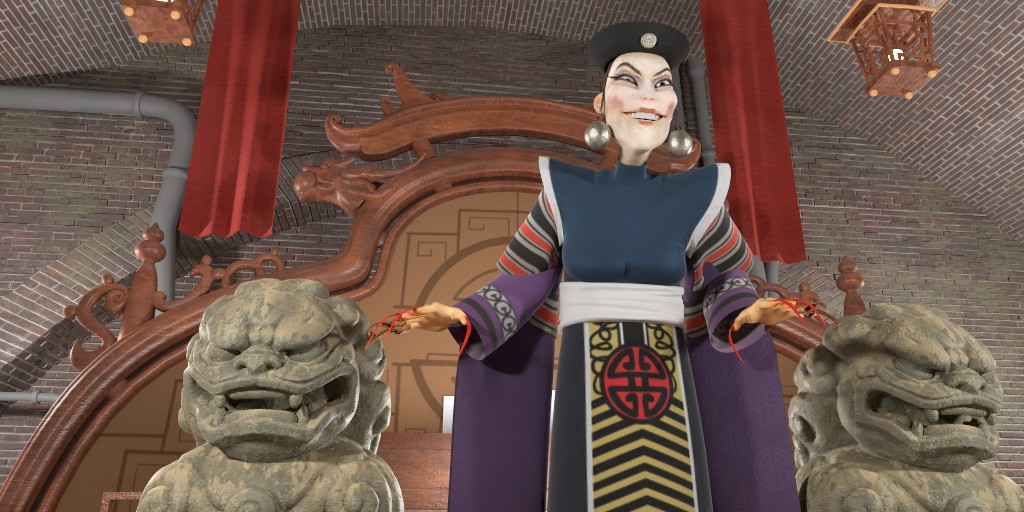} & 
    \includegraphics[width=0.25\linewidth]{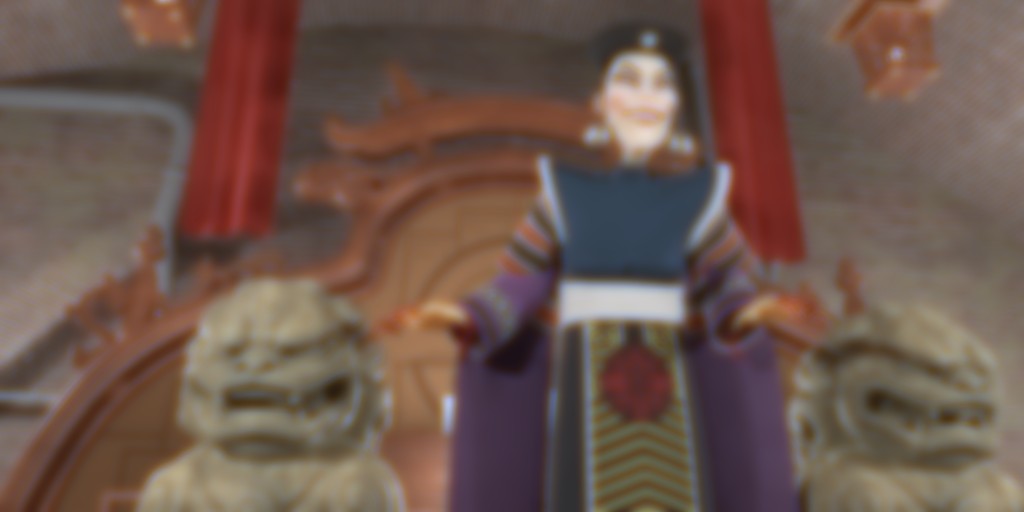} & 
    \includegraphics[width=0.25\linewidth]{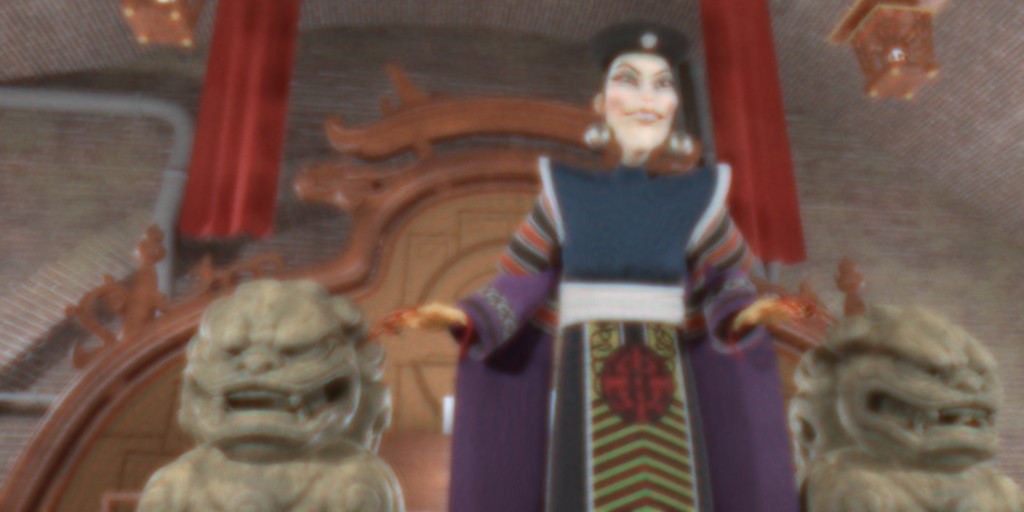} & 
    \includegraphics[width=0.25\linewidth]{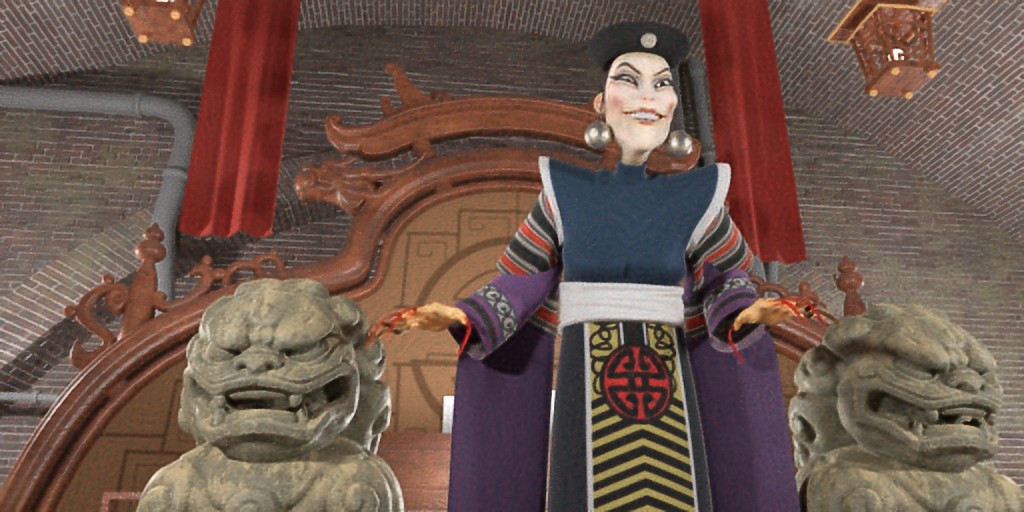} \\     
    \end{tabular}
    }
    \caption{\textbf{Image deblurring}. Our method outperforms Neural Deblurring \cite{ren2020neural} Even though we adapt Neural Deblurring to our deblurring subset, our method still outperforms it by a significant margin, making our approach more usable for real-world applications where the PSF kernels can be large.}
    \label{fig:deblurring}
\end{figure*}

To conclude our blind image deblurring experiments, Table \ref{tab:deblurring} summarizes our results quantitatively and Figure \ref{fig:deblurring} demonstrates the results qualitatively for the deblurring case (more examples can be found in the supplementary materials).
Overall we can see that our method manages to outperform in blind image deblurring by effectively recovering a single-valued depth map.

\begin{table}[!ht]
\caption{\textbf{Image deblurring}. Our method outperforms previous blind image deblurring works (even after adaptation) both in PSNR [dB] and SSIM.}
    \centering    
    \resizebox{0.7\linewidth}{!}{
    \begin{tabular}{l|cc}
    \toprule
    & Neural Deblurring \cite{ren2020neural} & DPCIP \\
    \midrule
    Deblurring & 19.6/0.67 & \textbf{26.38/0.834} \\    
    \bottomrule
\end{tabular}
}
\label{tab:deblurring}
\end{table}

\subsection{Ablation studies}
\label{sec:ablation}
\subsubsection{Implicit generator effect}
One of the key ideas of our method is the idea of considering DIP as an implicit model \cite{ulyanov2018deep}. Since there has been a lot of work in the area of implicit neural representations (INR), we compared our method to several implicit models to see the effect of the generator type along with its positional input encoding.
We conducted a study on the effect of the implicit generator. 15 images collected from 3 scenes were used as our test set. We tested three implicit generators: DIP \cite{ulyanov2018deep}, SIREN \cite{sitzmann2020implicit}, and PIP \cite{shabtay2022pip} on both passive depth estimation and all-in-focus image reconstruction. Table \ref{tab:ablation_generator} summarizes the implicit generator effect. Overall DIP outperformed both PIP and SIREN in terms of image reconstruction. In depth estimation, PIP and DIP are comparable and they are both better than SIREN.

\begin{table}[]
\caption{\textbf{Generator effect}. We evaluated three types of generators in our method. DIP outperformed other generators in image reconstruction and was comparable to PIP for depth estimation. Both DIP and PIP outperformed SIREN in both tasks.}
    \centering 
     \resizebox{1.0\linewidth}{!}{
    \begin{tabular}{l|ccc}
    \toprule
    & DIP & PIP & SIREN \\
    \midrule
    Depth Estimation (RMSE [m] $\downarrow$) & \textbf{0.005} & 0.006 & 0.19 \\    
    Image Reconstruction (PSNR / SSIM $\uparrow$) & \textbf{30.15 / 0.916} & 26.34 / 0.82 & 25.85 / 0.81\\    
    \bottomrule
\end{tabular}
}
\label{tab:ablation_generator}
\end{table}

\subsubsection{SSIM Loss effect}
We justify the switch from L2 loss to SSIM during the optimization using the following ablation study. We collected 15 images from 3 scenes to be used as our test set. We ran the same optimizations on all images, with and without the switch to SSIM after 500 iterations. Table \ref{tab:loss_effect} concludes the effect. We can see that both for image reconstruction and depth estimation switching from L2 to SSIM greatly improved our overall results.

\begin{table}[htbp]
    \centering
    \caption{\textbf{SSIM Loss effect}. The SSIM loss improved the results for all image reconstruction metrics (PSNR and SSIM) and also for the depth estimation.}
    \vspace{-0.3cm}
    \label{tab:loss_effect}
    \begin{tabular}{lccc}
        \toprule       
         & PSNR [dB]& SSIM & Depth error [m] \\
        \midrule
        L2 & 25.97 & 0.813 & 0.008 \\
        L2+SSIM & \textbf{30.20} & \textbf{0.920} & \textbf{0.003} \\        
        \bottomrule
    \end{tabular}
\end{table}

\subsubsection{Mismatch in the optical system}
We analyze the robustness of our method with respect to a mismatch in the optical parameters. We took the same 15-image test set used throughout the ablation study.  We ran our method several times on the same test set of images, where each time our method had slightly different PSF kernels. The mismatch was done by changing the phase mask, each time by a different factor ($1\% - 10\%$). We can see in Table \ref{tab:mismatch_optics} that with small mismatches in the PSF, the generations of our method are still fairly robust. 

\begin{table}[]
\vspace{-0.5cm}
    \caption{\textbf{Analysis of mismatch in the optical system (mismatch in the phase of the mask)} Our model is fairly robust to small errors in the optical model.}
    \vspace{-0.3cm}
    \centering
    \resizebox{1.\linewidth}{!}{
    \label{tab:mismatch_optics}
    \begin{tabular}{lccccc}       
        \toprule
        Phase change & $0\%$ & $1\%$ & $2\%$ & $5\%$ & $10\%$ \\        
        \midrule  
        PSNR[dB] / SSIM & 30.15/0.916 & 29.57/0.911 & 27.90/0.88 & 25.90/0.83 & 24.41/0.77 \\
        \bottomrule        
    \end{tabular}
    }
\end{table}

\subsection{Real-World Results}
\begin{figure*} []
\centering
\resizebox{1.0\linewidth}{!}{
    \begin{tabular}{cccc}
    & Supervised & 0-shot & 0-shot \\
    Blur input image & EDOF\cite{elmalem2018learned} / Mono from \cite{gil2019monster} & DPCIP (DIP) & DPCIP (PIP) \\
    \multirow{2}{*}[40pt]{\includegraphics[width=0.25\linewidth]{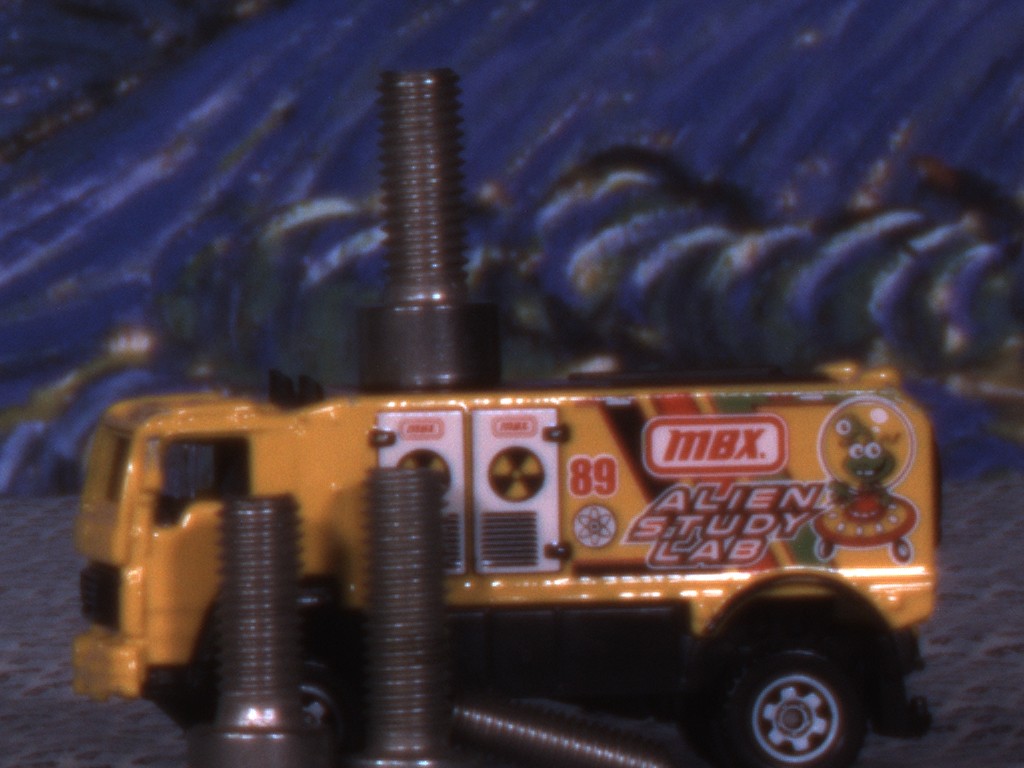} }& 
    \includegraphics[width=0.25\linewidth]{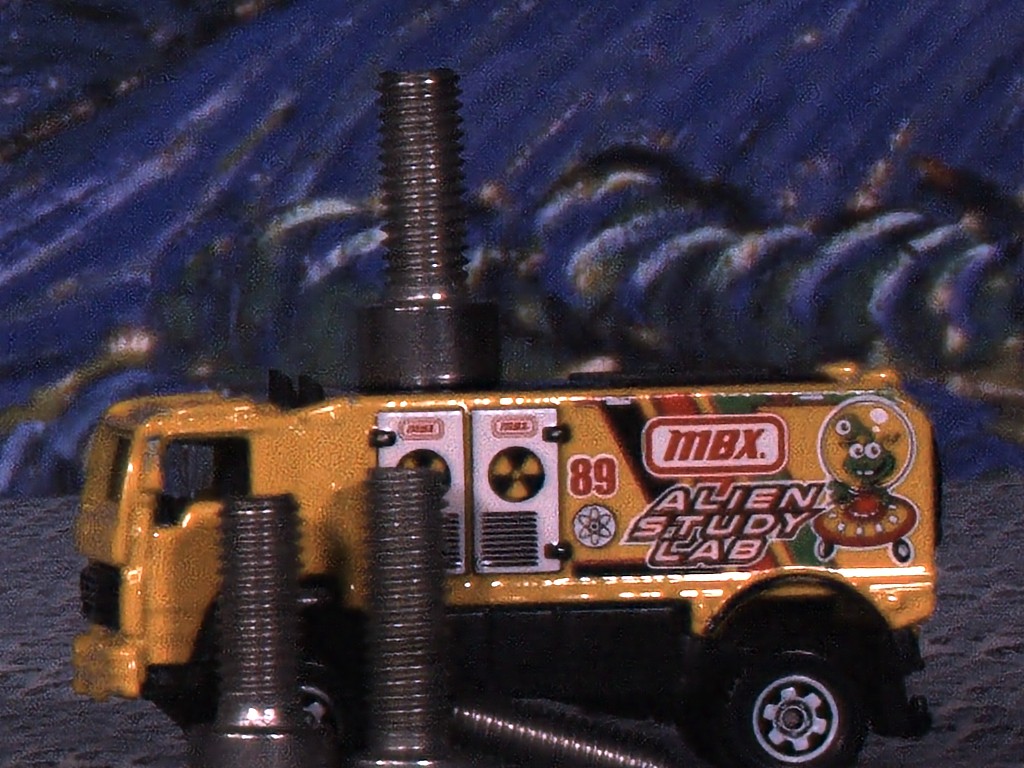} & 
    \includegraphics[width=0.25\linewidth]{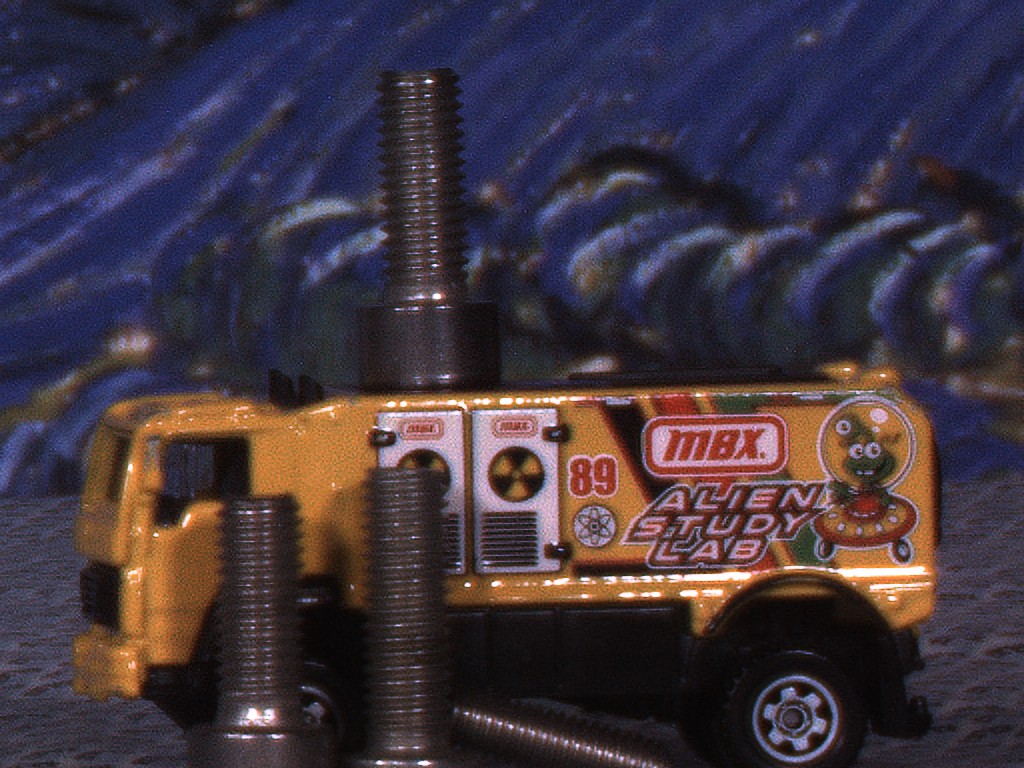} &
    \includegraphics[width=0.25\linewidth]{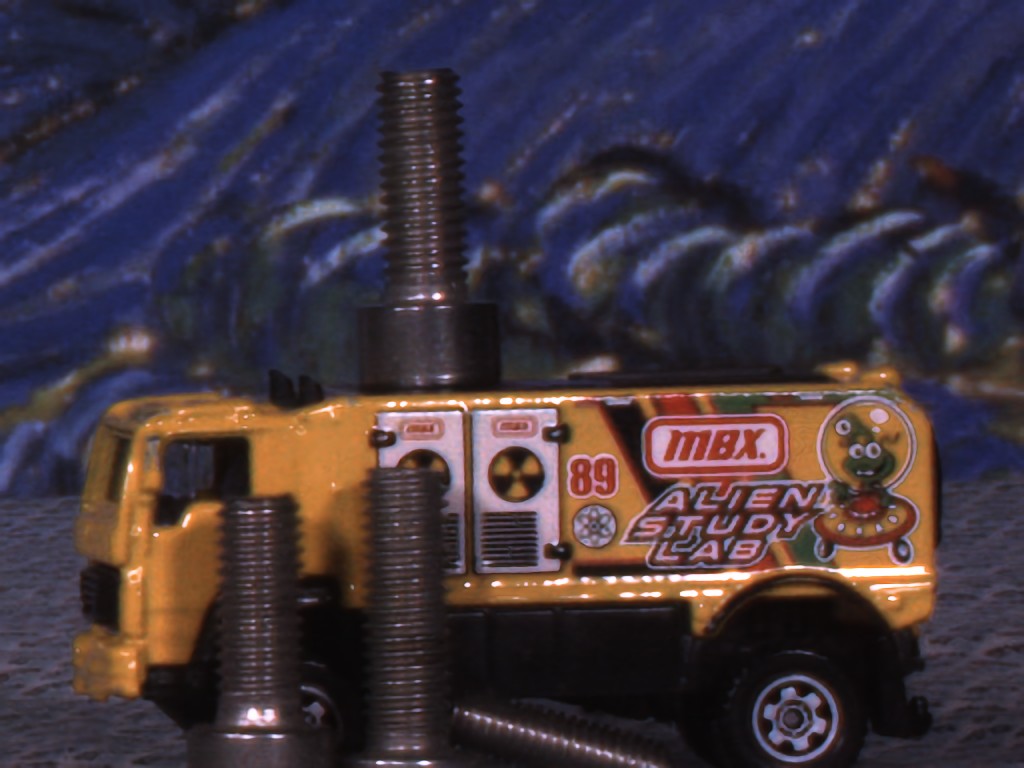} \\
     & 
    \includegraphics[width=0.25\linewidth]{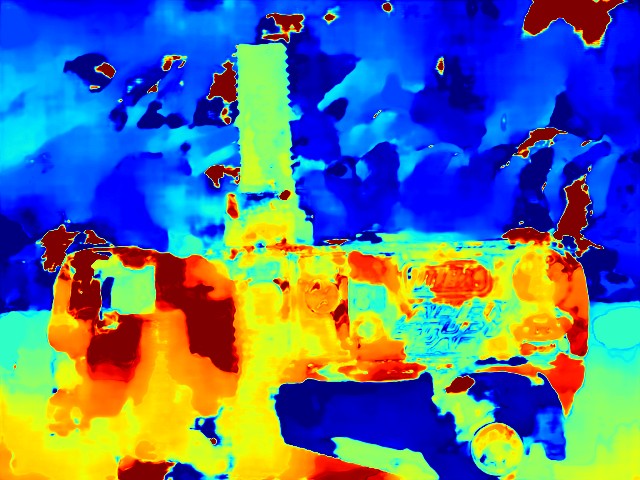} & 
    \includegraphics[width=0.25\linewidth]{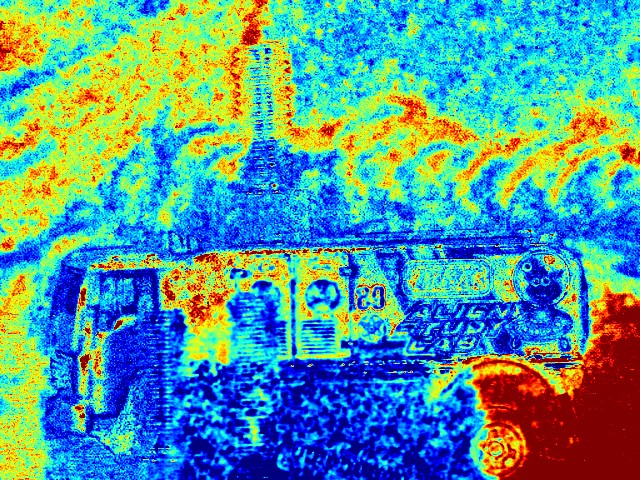} &
    \includegraphics[width=0.25\linewidth]{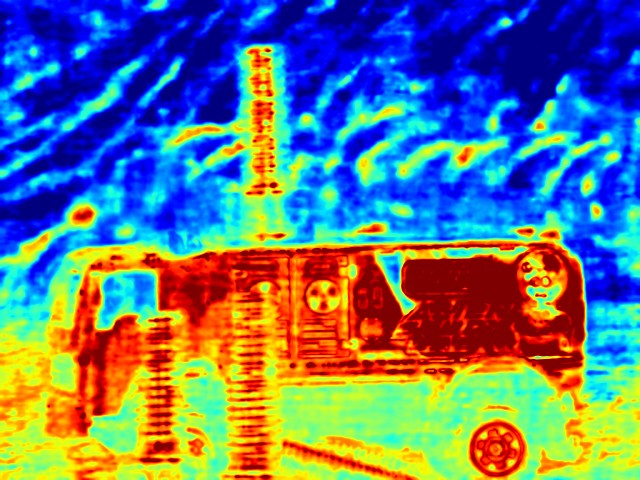} 
    \vspace{10pt}
    \\
    & Supervised & 0-shot & 0-shot \\
    Blur input image & EDOF\cite{elmalem2018learned} / Mono from \cite{gil2019monster} & DPCIP (DIP) & DPCIP (PIP) \\
    \multirow{2}{*}[40pt]{\includegraphics[width=0.25\linewidth]{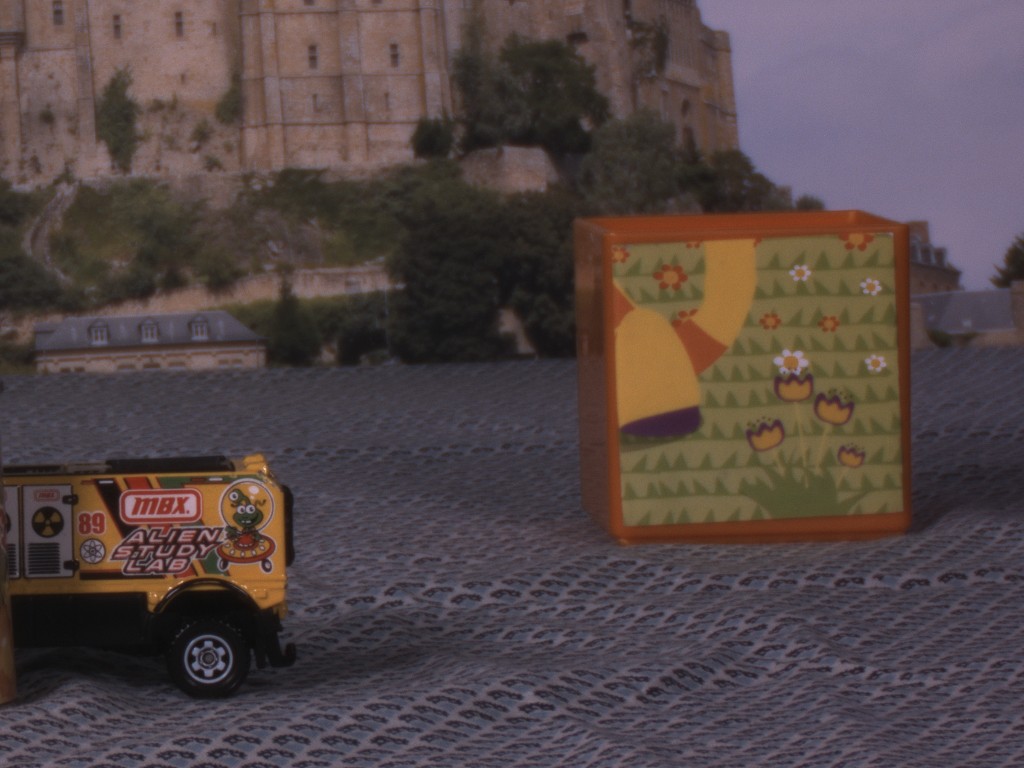}} & 
    \includegraphics[width=0.25\linewidth]{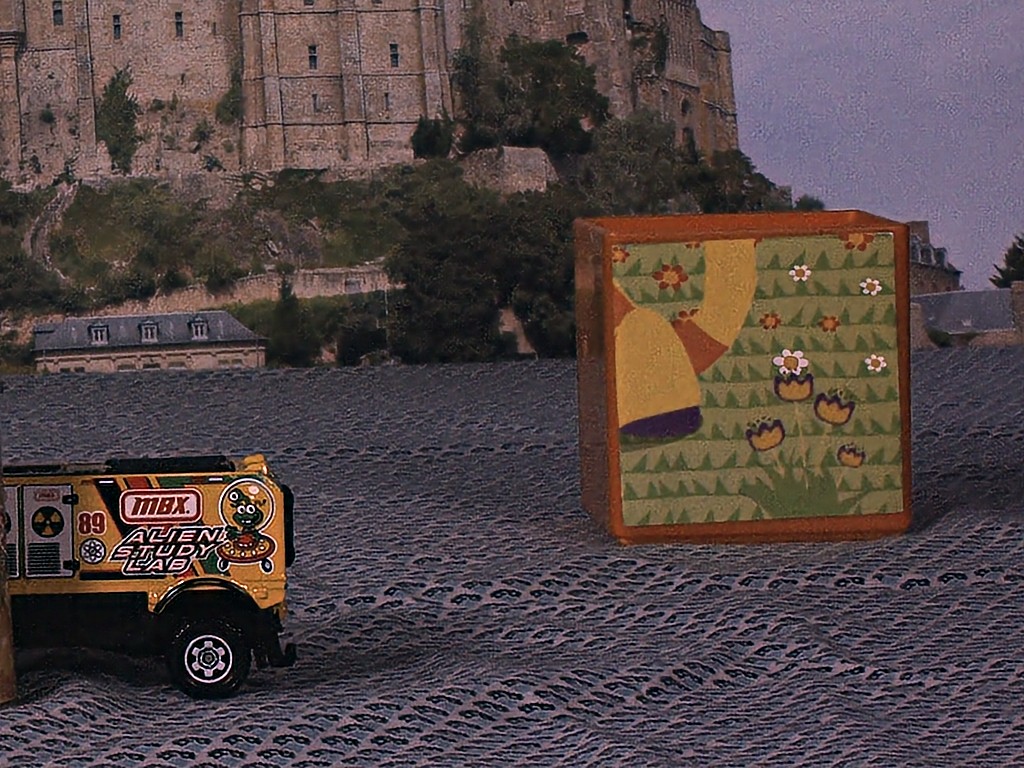} &    
    \includegraphics[width=0.25\linewidth]{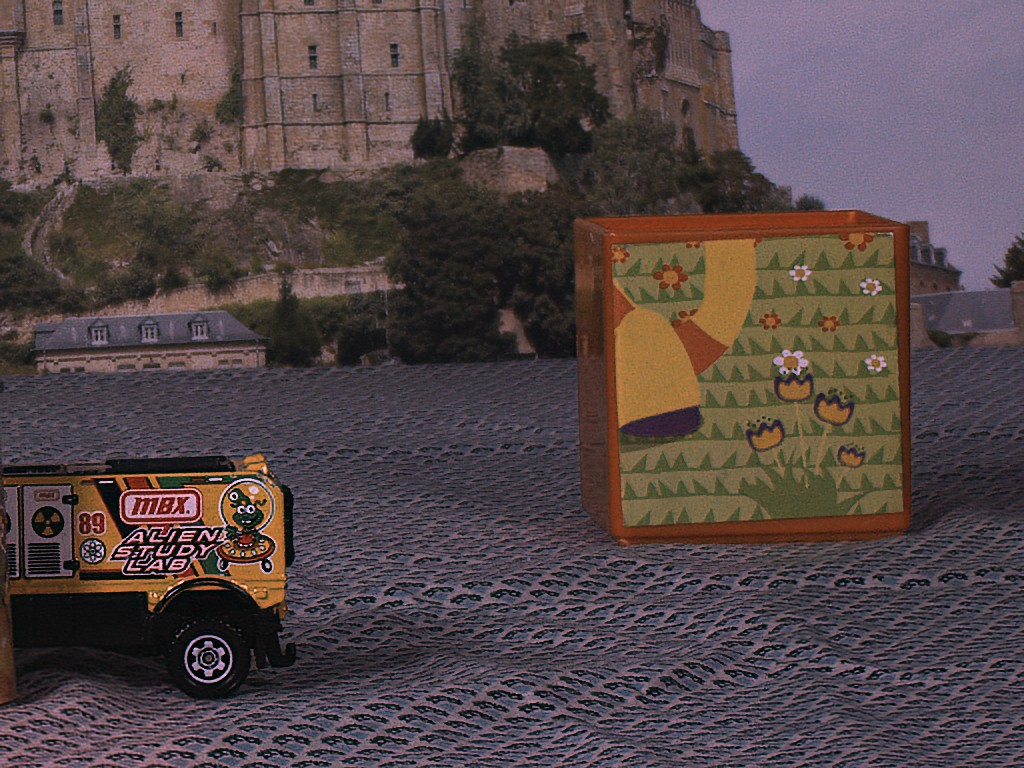} & 
    \includegraphics[width=0.25\linewidth]{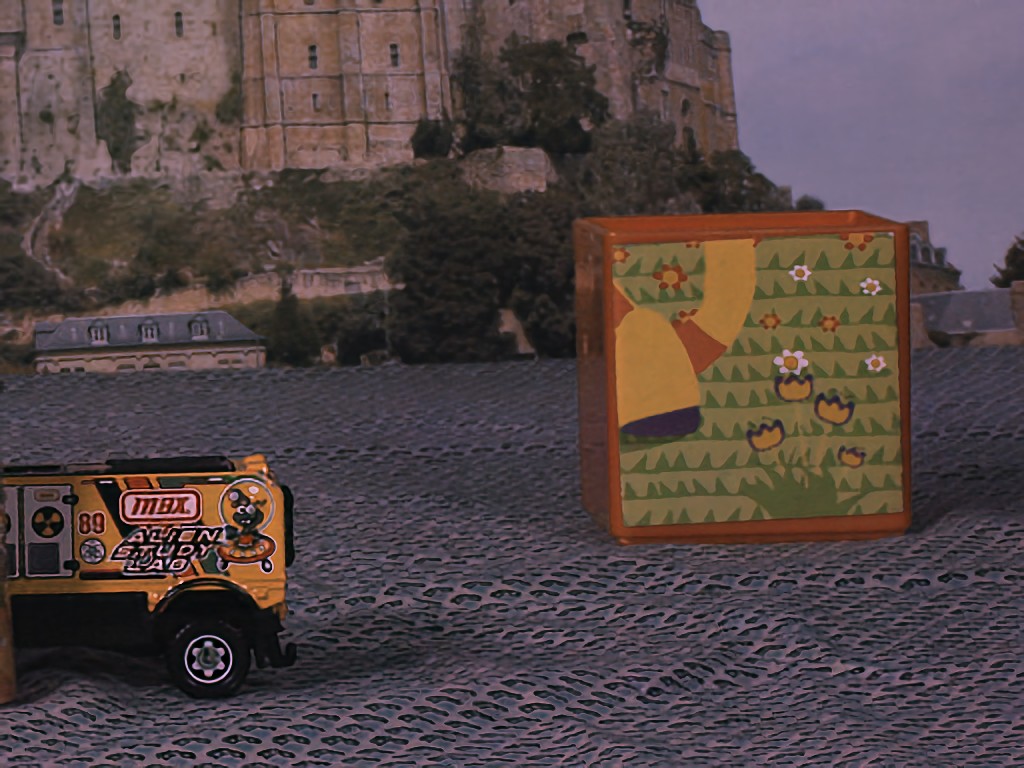}
    \\  
     & 
    \includegraphics[width=0.25\linewidth]{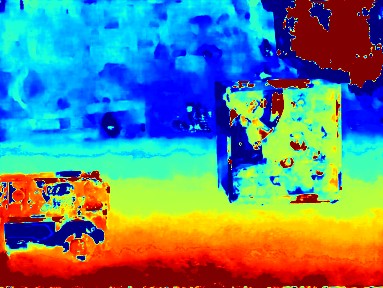} &    
    \includegraphics[width=0.25\linewidth]{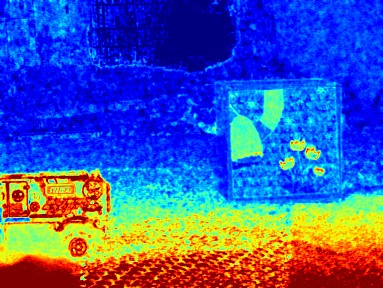} & 
    \includegraphics[width=0.25\linewidth]{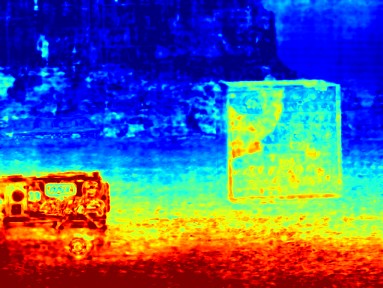} \\  
    \end{tabular}
    }
    \caption{\textbf{Real world examples.} We can see that using PIP as the generator generates the most accurate depth maps and also the best all-in-focus images. Using DIP as a generator tends to produce too detailed depth maps, but with good all-in-focus images. This means that PIP as a generator disentangles better between depth map estimation and all-in-focus restoration. In all cases, our zero-shot method is comparable to or better than the supervised baselines. For depth estimation, the mono network from \cite{gil2019monster} seems to produce large areas with incorrect depths while our method manages to be more consistent. In the all-in-focus task, the EDOF network \cite{elmalem2018learned} produced a bit noisier results than ours; see Figure \ref{fig:edof_noisy} for zoom-in visualizations.}
    \label{fig:realworld1}
\end{figure*}

We further extend our simulation results to real-world examples using a dedicated camera as done in \cite{haim2018depth, elmalem2018learned, gil2019monster}. For each image crop, we compare ourselves to the relevant baseline in both image reconstruction and passive depth estimation. Figure \ref{fig:realworld1} provide a qualitative comparison of the results.

Since real-world images suffer from much higher noise levels we add TV regularization to help DPCIP to better estimate the depth maps.
For each image we ran DPCIP twice, with DIP and PIP \cite{shabtay2022pip} as generators. PIP has the ability to control over the initialization scheme of the input code. From the visual results in Figure \ref{fig:realworld1} we can see that PIP manages to better disentangle between the depth maps and the image content, thus leading to better results both in terms of image reconstruction and depth maps. 

\begin{figure*} []
\centering
\resizebox{1.0\linewidth}{!}{
    \begin{tabular}{cccc}
    & Supervised & 0-shot & 0-shot \\
    Blur img & EDOF\cite{elmalem2018learned} / Mono from \cite{gil2019monster} & DPCIP (DIP) & DPCIP (PIP) \\
    \includegraphics[width=0.25\linewidth]{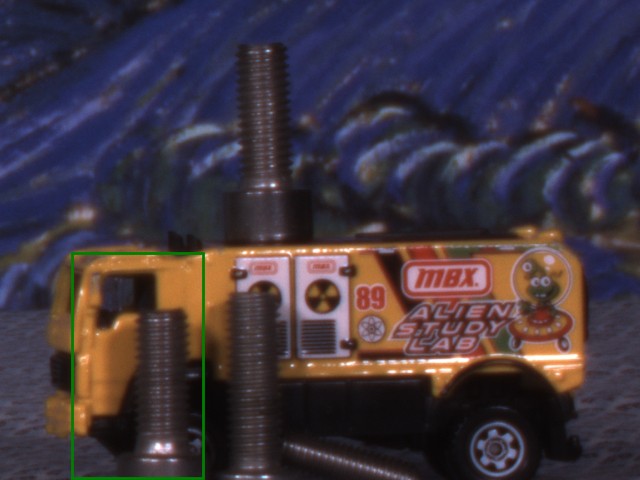} & 
    \includegraphics[width=0.25\linewidth]{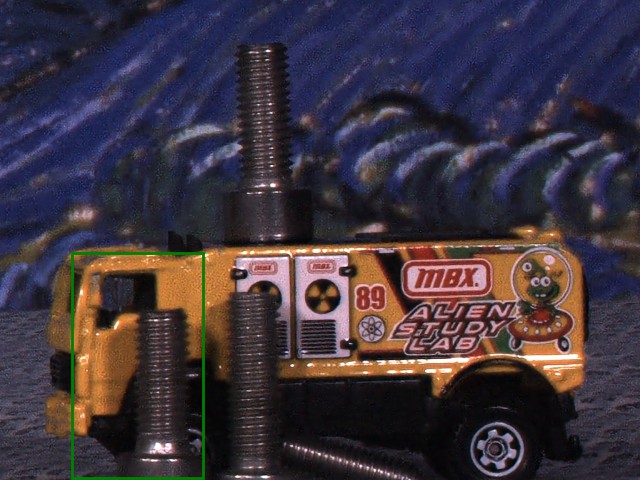} & 
    \includegraphics[width=0.25\linewidth]{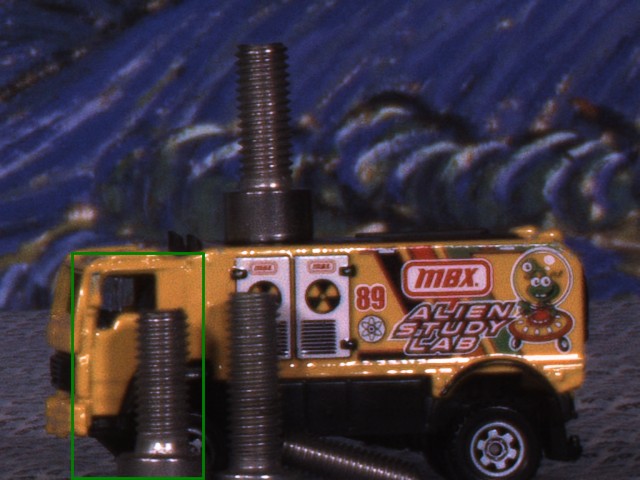} &
    \includegraphics[width=0.25\linewidth]{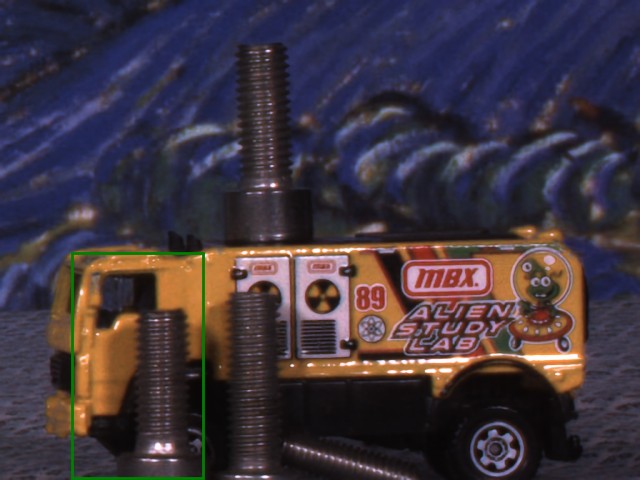} \\
    \includegraphics[width=0.25\linewidth]{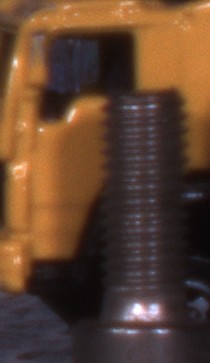} & 
    \includegraphics[width=0.25\linewidth]{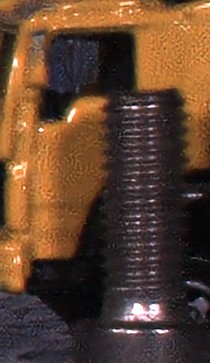} & 
    \includegraphics[width=0.25\linewidth]{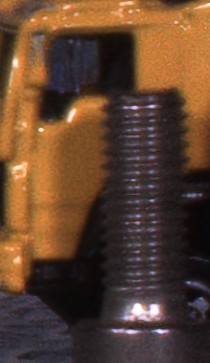} &
    \includegraphics[width=0.25\linewidth]{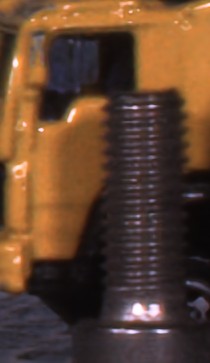} \\
    \end{tabular}
    }
    \caption{\textbf{Zoom-in reconstruction results} We can see that DPCIP produced smoother results, but kept the details sharp as opposed to EDOF competitor that generated noisier reconstruction results.}
    \label{fig:edof_noisy}
\end{figure*}

We further extend our experiments with our dataset to \cite{ikoma2021depth}, which train a supervised network to extract depth from defocus cues. The results on our dataset are worse than our method or \cite{gil2019monster}. An explanation for the degradation in performance is needed for a large dataset. In the original paper, the training dataset was $\times 10$ more than our dataset, which strengthens DPCIP's advantages as a zero-shot method that does not rely on a training dataset at all. The supplementary material contains visual examples.
In the visual results we can see that regarding depth estimation, the supervised baseline \cite{gil2019monster} produced smoother depth maps, but can also produce regions with incorrect depths, as opposed to our zero-shot method which has noisier depths but significantly smaller areas with incorrect depths.
Compared to the image reconstruction supervised baseline, our zero-shot method produces sharp images that are less noisy. See Figure \ref{fig:edof_noisy} for zoom-in visualization.

Overall, even though supervised methods are considered as upper-bound, our zero-shot method produced comparable or even better results without the need for any dataset for training which can be considered a significant advantage when it comes to real-world scenarios where the limited amount of existing datasets might not suffice for training a suitable supervised method. More viusal examples can be found at the supplementary materials.




\section{Conclusion}
In this work, we have proposed a self-supervised method for recovering an absolute depth map and an all-in-focus image using a phase-coded imaging system. As opposed to previous methods that rely on datasets for recovering the depth map or the all-in-focus image, we do it solely on the given captured image and the knowledge of the optical setup by leveraging implicit models as the image prior.

Our method reduces the need for using training datasets since collecting such datasets can be hard and requires external graphics rendering software. In addition, the current datasets can suffer from generalization and domain gap issues. Providing a solution for real-world setups and users who seek to use an existing imaging system without the need for any additional information or data other than optical parameters.
In addition, due to the high quality of restoration, our method can provide ``pseudo GT" to be used as supervision for training a neural network to restore an all-in-focus image from the blurred acquired image. This way we can enjoy both high quality of reconstruction and fast inference time.

\ifpeerreview \else
\section*{Acknowledgments}
This work was supported by the European research council under Grant ERC-StG 757497. This work was supported by Tel Aviv University Center for AI and Data Science (TAD).\\
We would like to thank Shay Elmalem for his useful comments and advice in addition to providing us the simulation code we based our work on.
\fi
\bibliographystyle{IEEEtran}
\bibliography{main}

\begin{thebibliography}{10}
\providecommand{\url}[1]{#1}
\csname url@samestyle\endcsname
\providecommand{\newblock}{\relax}
\providecommand{\bibinfo}[2]{#2}
\providecommand{\BIBentrySTDinterwordspacing}{\spaceskip=0pt\relax}
\providecommand{\BIBentryALTinterwordstretchfactor}{4}
\providecommand{\BIBentryALTinterwordspacing}{\spaceskip=\fontdimen2\font plus
\BIBentryALTinterwordstretchfactor\fontdimen3\font minus
  \fontdimen4\font\relax}
\providecommand{\BIBforeignlanguage}[2]{{%
\expandafter\ifx\csname l@#1\endcsname\relax
\typeout{** WARNING: IEEEtran.bst: No hyphenation pattern has been}%
\typeout{** loaded for the language `#1'. Using the pattern for}%
\typeout{** the default language instead.}%
\else
\language=\csname l@#1\endcsname
\fi
#2}}
\providecommand{\BIBdecl}{\relax}
\BIBdecl

\bibitem{ulyanov2018deep}
D.~Ulyanov, A.~Vedaldi, and V.~Lempitsky, ``Deep image prior,'' in
  \emph{Proceedings of the IEEE conference on computer vision and pattern
  recognition}, 2018, pp. 9446--9454.

\bibitem{levin2007image}
A.~Levin, R.~Fergus, F.~Durand, and W.~T. Freeman, ``Image and depth from a
  conventional camera with a coded aperture,'' \emph{ACM transactions on
  graphics (TOG)}, vol.~26, no.~3, pp. 70--es, 2007.

\bibitem{haim2018depth}
H.~Haim, S.~Elmalem, R.~Giryes, A.~M. Bronstein, and E.~Marom, ``Depth
  estimation from a single image using deep learned phase coded mask,''
  \emph{IEEE Transactions on Computational Imaging}, vol.~4, no.~3, pp.
  298--310, 2018.

\bibitem{gil2019monster}
Y.~Gil, S.~Elmalem, H.~Haim, E.~Marom, and R.~Giryes, ``Online training of
  stereo self-calibration using monocular depth estimation,'' \emph{IEEE
  Transactions on Computational Imaging}, vol.~7, pp. 812--823, 2021.

\bibitem{chang2019deep}
J.~Chang and G.~Wetzstein, ``Deep optics for monocular depth estimation and 3d
  object detection,'' in \emph{Proceedings of the IEEE/CVF International
  Conference on Computer Vision}, 2019, pp. 10\,193--10\,202.

\bibitem{wu2019phasecam3d}
Y.~Wu, V.~Boominathan, H.~Chen, A.~Sankaranarayanan, and A.~Veeraraghavan,
  ``Phasecam3d — learning phase masks for passive single view depth
  estimation,'' in \emph{2019 IEEE International Conference on Computational
  Photography (ICCP)}, 2019, pp. 1--12.

\bibitem{elmalem2018learned}
S.~Elmalem, R.~Giryes, and E.~Marom, ``Learned phase coded aperture for the
  benefit of depth of field extension,'' \emph{Optics express}, vol.~26,
  no.~12, pp. 15\,316--15\,331, 2018.

\bibitem{akpinar2021wavefront}
U.~Akpinar, E.~Sahin, M.~Meem, R.~Menon, and A.~Gotchev, ``Learning wavefront
  coding for extended depth of field imaging,'' \emph{IEEE Transactions on
  Image Processing}, vol.~30, pp. 3307--3320, 2021.

\bibitem{Gkioulekas2021defocus}
I.~Gkioulekas, J.~Chen, J.~T. Barron, N.~Wadhwa, P.~Srinivasan, R.~Garg,
  S.~Xin, and T.~Xue, ``Defocus map estimation and blur removal from a single
  dual-pixel image,'' 2021.

\bibitem{sitzmann2018end}
V.~Sitzmann, S.~Diamond, Y.~Peng, X.~Dun, S.~Boyd, W.~Heidrich, F.~Heide, and
  G.~Wetzstein, ``End-to-end optimization of optics and image processing for
  achromatic extended depth of field and super-resolution imaging,'' \emph{ACM
  Transactions on Graphics (TOG)}, vol.~37, no.~4, pp. 1--13, 2018.

\bibitem{yosef2021video}
E.~Yosef, S.~Elmalem, and R.~Giryes, ``Video reconstruction from a single
  motion blurred image using learned dynamic phase coding,'' \emph{Scientific
  Reports}, vol.~13, no.~1, p. 13625, 2023.

\bibitem{gandelsman2019double}
Y.~Gandelsman, A.~Shocher, and M.~Irani, ``" double-dip": unsupervised image
  decomposition via coupled deep-image-priors,'' in \emph{Proceedings of the
  IEEE/CVF Conference on Computer Vision and Pattern Recognition}, 2019, pp.
  11\,026--11\,035.

\bibitem{ren2020neural}
D.~Ren, K.~Zhang, Q.~Wang, Q.~Hu, and W.~Zuo, ``Neural blind deconvolution
  using deep priors,'' in \emph{Proceedings of the IEEE/CVF Conference on
  Computer Vision and Pattern Recognition}, 2020, pp. 3341--3350.

\bibitem{Yokota_2019_ICCV}
T.~Yokota, K.~Kawai, M.~Sakata, Y.~Kimura, and H.~Hontani, ``Dynamic pet image
  reconstruction using nonnegative matrix factorization incorporated with deep
  image prior,'' in \emph{Proceedings of the IEEE/CVF International Conference
  on Computer Vision (ICCV)}, October 2019.

\bibitem{Hashimoto_2022}
\BIBentryALTinterwordspacing
F.~Hashimoto, K.~Ote, and Y.~Onishi, ``{PET} image reconstruction incorporating
  deep image prior and a forward projection model,'' \emph{{IEEE} Transactions
  on Radiation and Plasma Medical Sciences}, vol.~6, no.~8, pp. 841--846, nov
  2022. [Online]. Available: \url{https://doi.org/10.1109%2Ftrpms.2022.3161569}
\BIBentrySTDinterwordspacing

\bibitem{8581448}
K.~Gong, C.~Catana, J.~Qi, and Q.~Li, ``Pet image reconstruction using deep
  image prior,'' \emph{IEEE Transactions on Medical Imaging}, vol.~38, no.~7,
  pp. 1655--1665, 2019.

\bibitem{9576711}
------, ``Direct reconstruction of linear parametric images from dynamic pet
  using nonlocal deep image prior,'' \emph{IEEE Transactions on Medical
  Imaging}, vol.~41, no.~3, pp. 680--689, 2022.

\bibitem{Mataev2019DeepREDDI}
G.~Mataev, M.~Elad, and P.~Milanfar, ``Deepred: Deep image prior powered by
  red,'' \emph{ArXiv}, vol. abs/1903.10176, 2019.

\bibitem{fermanian:hal-03310533}
\BIBentryALTinterwordspacing
R.~Fermanian, M.~Le~Pendu, and C.~Guillemot, ``{Regularizing the Deep Image
  Prior with a Learned Denoiser for Linear Inverse Problems},'' in \emph{{MMSP
  2021 - IEEE 23rd International Workshop on Multimedia Siganl
  Processing}}.\hskip 1em plus 0.5em minus 0.4em\relax Tampere, Finland:
  {IEEE}, Oct. 2021, pp. 1--6. [Online]. Available:
  \url{https://hal.archives-ouvertes.fr/hal-03310533}
\BIBentrySTDinterwordspacing

\bibitem{kurniawan2022noise}
E.~Kurniawan, Y.~Park, and S.~Lee, ``Noise-resistant demosaicing with deep
  image prior network and random rgbw color filter array,'' \emph{Sensors},
  vol.~22, no.~5, p. 1767, 2022.

\bibitem{chen2020dip}
Y.-C. Chen, C.~Gao, E.~Robb, and J.-B. Huang, ``Nas-dip: Learning deep image
  prior with neural architecture search,'' in \emph{European Conference on
  Computer Vision}.\hskip 1em plus 0.5em minus 0.4em\relax Springer, 2020, pp.
  442--459.

\bibitem{shabtay2022pip}
N.~Shabtay, E.~Schwartz, and R.~Giryes, ``Pip: Positional-encoding image
  prior,'' \emph{arXiv preprint arXiv:2211.14298}, 2022.

\bibitem{tancik2020fourier}
M.~Tancik, P.~Srinivasan, B.~Mildenhall, S.~Fridovich-Keil, N.~Raghavan,
  U.~Singhal, R.~Ramamoorthi, J.~Barron, and R.~Ng, ``Fourier features let
  networks learn high frequency functions in low dimensional domains,''
  \emph{Advances in Neural Information Processing Systems}, vol.~33, pp.
  7537--7547, 2020.

\bibitem{mildenhall2020nerf}
B.~Mildenhall, P.~P. Srinivasan, M.~Tancik, J.~T. Barron, R.~Ramamoorthi, and
  R.~Ng, ``Nerf: Representing scenes as neural radiance fields for view
  synthesis,'' in \emph{Computer Vision--ECCV 2020: 16th European Conference,
  Glasgow, UK, August 23--28, 2020, Proceedings, Part I}, 2020, pp. 405--421.

\bibitem{sitzmann2019siren}
V.~Sitzmann, J.~N. Martel, A.~W. Bergman, D.~B. Lindell, and G.~Wetzstein,
  ``Implicit neural representations with periodic activation functions,'' in
  \emph{arXiv}, 2020.

\bibitem{Yu_2021_ICCV}
A.~Yu, R.~Li, M.~Tancik, H.~Li, R.~Ng, and A.~Kanazawa, ``Plenoctrees for
  real-time rendering of neural radiance fields,'' in \emph{Proceedings of the
  IEEE/CVF International Conference on Computer Vision (ICCV)}, October 2021,
  pp. 5752--5761.

\bibitem{barron2023zip}
J.~T. Barron, B.~Mildenhall, D.~Verbin, P.~P. Srinivasan, and P.~Hedman,
  ``Zip-nerf: Anti-aliased grid-based neural radiance fields,'' \emph{arXiv
  preprint arXiv:2304.06706}, 2023.

\bibitem{muller2022instant}
T.~M{\"u}ller, A.~Evans, C.~Schied, and A.~Keller, ``Instant neural graphics
  primitives with a multiresolution hash encoding,'' \emph{ACM Transactions on
  Graphics (ToG)}, vol.~41, no.~4, pp. 1--15, 2022.

\bibitem{Barron_2021_ICCV}
J.~T. Barron, B.~Mildenhall, M.~Tancik, P.~Hedman, R.~Martin-Brualla, and P.~P.
  Srinivasan, ``Mip-nerf: A multiscale representation for anti-aliasing neural
  radiance fields,'' in \emph{Proceedings of the IEEE/CVF International
  Conference on Computer Vision (ICCV)}, October 2021, pp. 5855--5864.

\bibitem{Barron_2022_CVPR}
J.~T. Barron, B.~Mildenhall, D.~Verbin, P.~P. Srinivasan, and P.~Hedman,
  ``Mip-nerf 360: Unbounded anti-aliased neural radiance fields,'' in
  \emph{Proceedings of the IEEE/CVF Conference on Computer Vision and Pattern
  Recognition (CVPR)}, June 2022, pp. 5470--5479.

\bibitem{Tancik_2022_CVPR}
M.~Tancik, V.~Casser, X.~Yan, S.~Pradhan, B.~Mildenhall, P.~P. Srinivasan,
  J.~T. Barron, and H.~Kretzschmar, ``Block-nerf: Scalable large scene neural
  view synthesis,'' in \emph{Proceedings of the IEEE/CVF Conference on Computer
  Vision and Pattern Recognition (CVPR)}, June 2022, pp. 8248--8258.

\bibitem{yu2021pixelnerf}
A.~Yu, V.~Ye, M.~Tancik, and A.~Kanazawa, ``pixelnerf: Neural radiance fields
  from one or few images,'' in \emph{Proceedings of the IEEE/CVF Conference on
  Computer Vision and Pattern Recognition}, 2021, pp. 4578--4587.

\bibitem{Mildenhall_2022_CVPR}
B.~Mildenhall, P.~Hedman, R.~Martin-Brualla, P.~P. Srinivasan, and J.~T.
  Barron, ``Nerf in the dark: High dynamic range view synthesis from noisy raw
  images,'' in \emph{Proceedings of the IEEE/CVF Conference on Computer Vision
  and Pattern Recognition (CVPR)}, June 2022, pp. 16\,190--16\,199.

\bibitem{ramasinghe2022beyond}
S.~Ramasinghe and S.~Lucey, ``Beyond periodicity: towards a unifying framework
  for activations in coordinate-mlps,'' in \emph{Computer Vision--ECCV 2022:
  17th European Conference, Tel Aviv, Israel, October 23--27, 2022,
  Proceedings, Part XXXIII}.\hskip 1em plus 0.5em minus 0.4em\relax Springer,
  2022, pp. 142--158.

\bibitem{hertz2021sape}
A.~Hertz, O.~Perel, R.~Giryes, O.~Sorkine-Hornung, and D.~Cohen-Or, ``Sape:
  Spatially-adaptive progressive encoding for neural optimization,''
  \emph{Advances in Neural Information Processing Systems}, vol.~34, pp.
  8820--8832, 2021.

\bibitem{lindell2021bacon}
D.~B. Lindell, D.~Van~Veen, J.~J. Park, and G.~Wetzstein, ``Bacon: Band-limited
  coordinate networks for multiscale scene representation,'' in \emph{CVPR},
  2022.

\bibitem{Butler:ECCV:2012}
D.~J. Butler, J.~Wulff, G.~B. Stanley, and M.~J. Black, ``A naturalistic open
  source movie for optical flow evaluation,'' in \emph{European Conf. on
  Computer Vision (ECCV)}, ser. Part IV, LNCS 7577, {A. Fitzgibbon et al.
  (Eds.)}, Ed.\hskip 1em plus 0.5em minus 0.4em\relax Springer-Verlag, Oct.
  2012, pp. 611--625.

\bibitem{silberman2012indoor}
N.~Silberman, D.~Hoiem, P.~Kohli, and R.~Fergus, ``Indoor segmentation and
  support inference from rgbd images,'' in \emph{Computer Vision--ECCV 2012:
  12th European Conference on Computer Vision, Florence, Italy, October 7-13,
  2012, Proceedings, Part V 12}.\hskip 1em plus 0.5em minus 0.4em\relax
  Springer, 2012, pp. 746--760.

\bibitem{mayer2016large}
N.~Mayer, E.~Ilg, P.~Hausser, P.~Fischer, D.~Cremers, A.~Dosovitskiy, and
  T.~Brox, ``A large dataset to train convolutional networks for disparity,
  optical flow, and scene flow estimation,'' in \emph{Proceedings of the IEEE
  conference on computer vision and pattern recognition}, 2016, pp. 4040--4048.

\bibitem{goodman2005introduction}
J.~W. Goodman, \emph{Introduction to Fourier optics}.\hskip 1em plus 0.5em
  minus 0.4em\relax Roberts and Company publishers, 2005.

\bibitem{wang2004image}
Z.~Wang, A.~C. Bovik, H.~R. Sheikh, and E.~P. Simoncelli, ``Image quality
  assessment: from error visibility to structural similarity,'' \emph{IEEE
  transactions on image processing}, vol.~13, no.~4, pp. 600--612, 2004.

\bibitem{kingma2014adam}
D.~P. Kingma and J.~Ba, ``Adam: A method for stochastic optimization,''
  \emph{arXiv preprint arXiv:1412.6980}, 2014.

\bibitem{sitzmann2020implicit}
V.~Sitzmann, J.~Martel, A.~Bergman, D.~Lindell, and G.~Wetzstein, ``Implicit
  neural representations with periodic activation functions,'' \emph{Advances
  in Neural Information Processing Systems}, vol.~33, pp. 7462--7473, 2020.

\bibitem{ikoma2021depth}
H.~Ikoma, C.~M. Nguyen, C.~A. Metzler, Y.~Peng, and G.~Wetzstein, ``Depth from
  defocus with learned optics for imaging and occlusion-aware depth
  estimation,'' in \emph{2021 IEEE International Conference on Computational
  Photography (ICCP)}.\hskip 1em plus 0.5em minus 0.4em\relax IEEE, 2021, pp.
  1--12.

\end{thebibliography}

\ifpeerreview \else





\fi







\ifpeerreview \else

\begin{IEEEbiography}[{\includegraphics[width=1in,height=1.25in,clip,keepaspectratio]{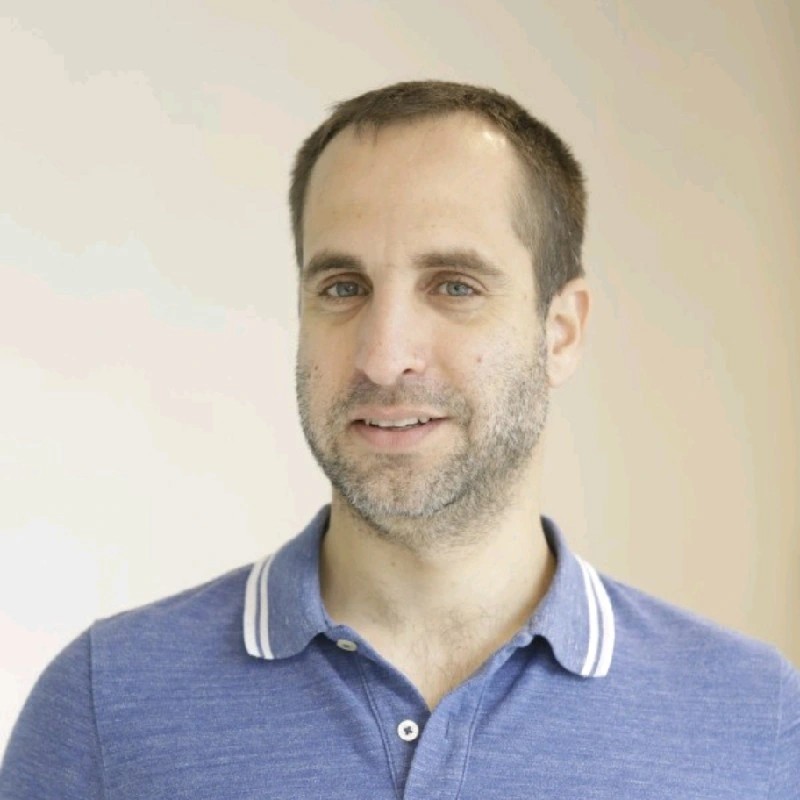}}]{Nimrod Shabtay}
received the B.Sc. degree in electrical engineering from Tel-Aviv university in 2016. From 2022 Mr. Shabtay is working towards a PhD degree (direct track) in electrical engineering in Tel-Aviv university under the supervision of Prof. Raja Giryes. His intrests are leveraging zero-shot methods for the benefits of computer vision and multi-modal tasks. He was an AI technology leader at Nanit, a computer vision startup for the benefit of parents of new born babies.
\end{IEEEbiography}

\begin{IEEEbiography}[{\includegraphics[width=1in,height=1.25in,clip,keepaspectratio]{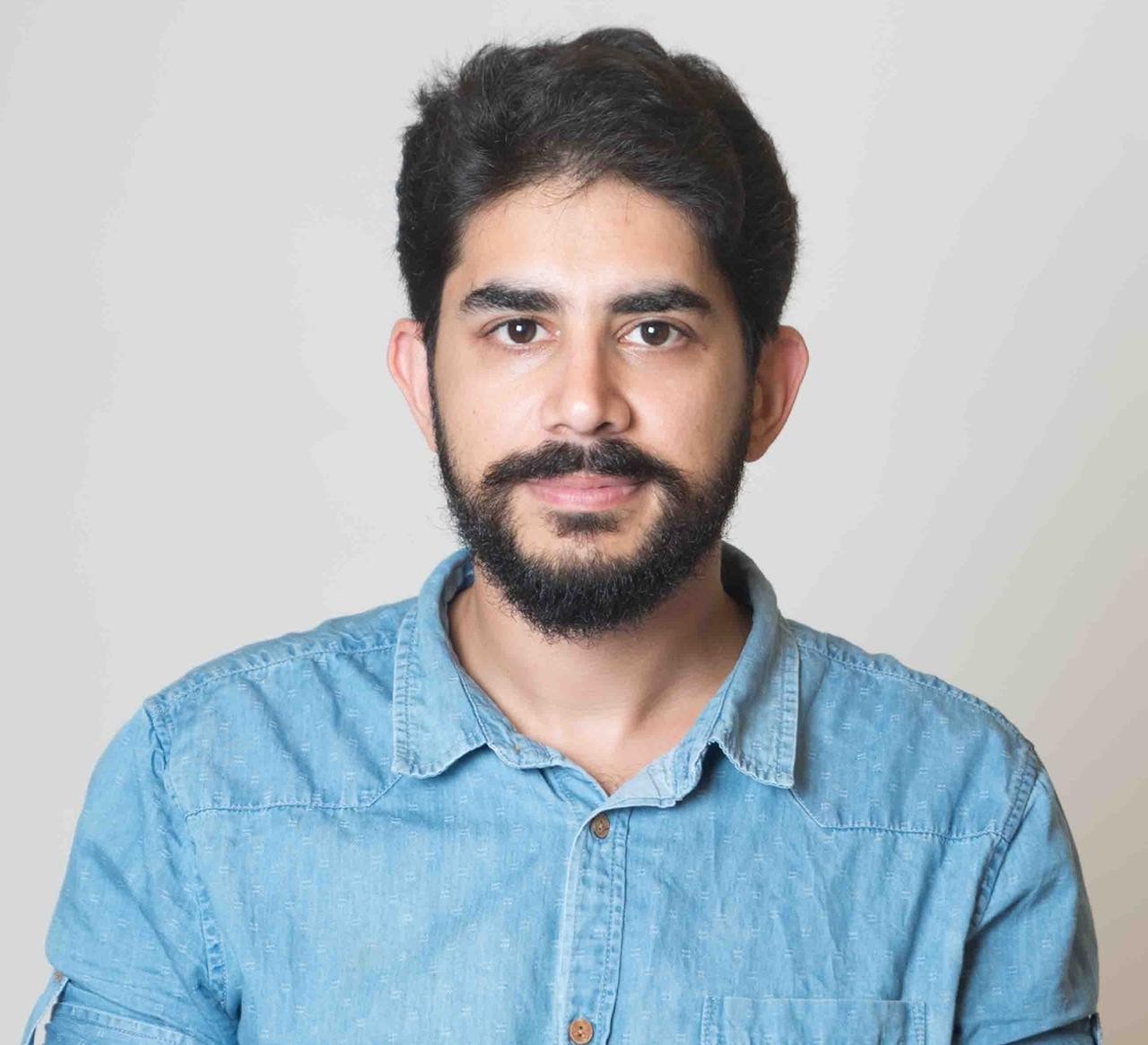}}]{Eli Scwartz}
is a research scientist at IBM Research. His research interests range from computational photography to object recognition and vision-language foundation models. He received his Ph.D. (2023) and M.Sc. (2018) from Tel Aviv University, Tel Aviv, Israel. He was the recipient of the IBM Ph.D. Fellowship in 2020-2021. Dr. Schwartz has authored a couple of dozen academic papers published in top journals and conferences and regularly organizes workshops at major conferences. He co-founded and served as the CTO of Inka Robotics, a startup that developed the world’s first tattooing robot, from 2015 to 2017. Before that, he worked at Microsoft, developing algorithms for the HoloLens AR headset from 2013 to 2016.
\end{IEEEbiography}

\begin{IEEEbiography}[{\includegraphics[width=1in,height=1.25in,clip,keepaspectratio]
{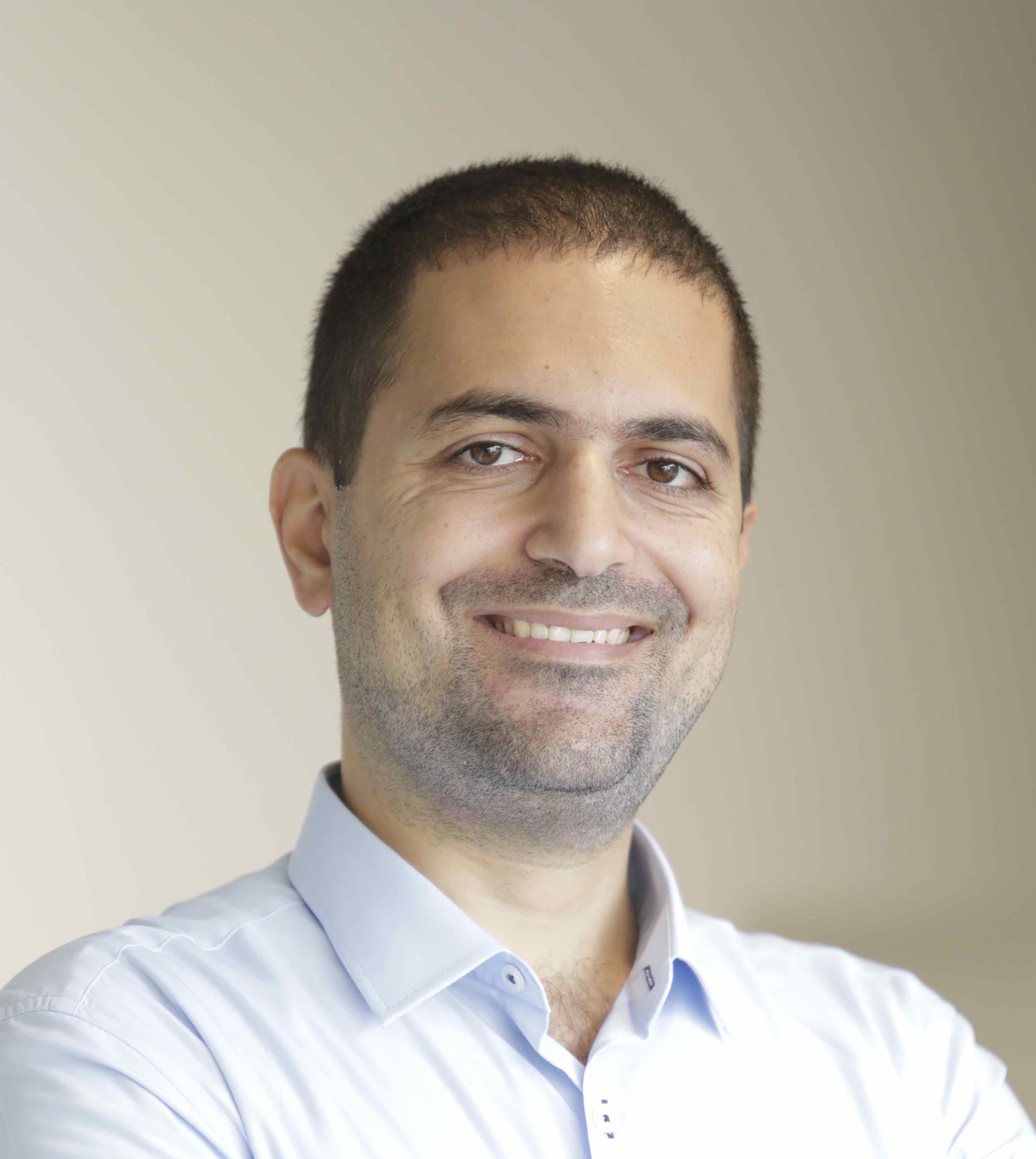}}]{Raja Giryes} (Senior Member, IEEE) is an Associate Professor in the school of electrical engineering at Tel Aviv University. His research interests lie at the intersection between signal and image processing and machine learning, and in particular, in deep learning, inverse problems, sparse representations, computational photography, and signal and image modeling. Raja received the EURASIP best P.hD. award, the ERC-StG grant, Maof prize for excellent young faculty (2016-2019), VATAT scholarship for excellent postdoctoral fellows (2014-2015), Intel Research and Excellence Award (2005, 2013), the Excellence in Signal Processing Award (ESPA) from Texas Instruments (2008) and was part of the Azrieli Fellows program (2010-2013). He is an associate editor in IEEE Transactions on Pattern Recognition and Machine Intelligence, IEEE Transactions on Image Processing and Elsevier Pattern Recognition, served as area chair in NeurIPS 2023-2024 and in ICASSP 2019-2022, and has organized workshops and tutorials on deep learning theory in various conferences including ICML, CVPR, ECCV and ICCV. He is also a co-organizer of the Israel computer vision day and the AI on chip workshop. He is a IEEE Senior Member and a member of the Israeli Young Academy since 2022.
\end{IEEEbiography}


\fi

\end{document}


\maketitle

\section{Depth Estimation}
\begin{figure}[H]    
    \resizebox{1\linewidth}{!}{
    \begin{tabular}{cccccc}    
    & & Supervised & & 0-shot &\\
    GT & Input & Mono from \cite{gil2019monster} & Error map (Mono-GT) & DPCIP & Error map (DPCIP-GT)\\
    \includegraphics[width=0.25\linewidth]{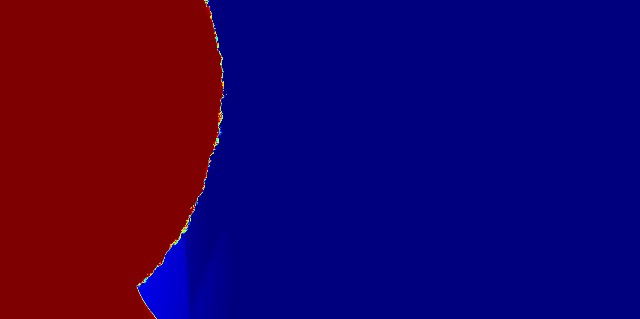} & 
    \includegraphics[width=0.25\linewidth]{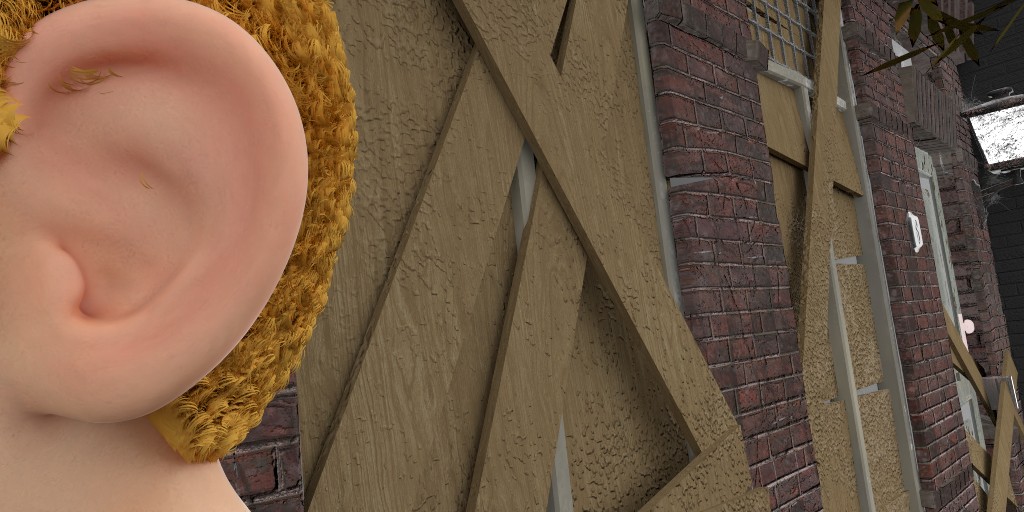} & 
    \includegraphics[width=0.25\linewidth]{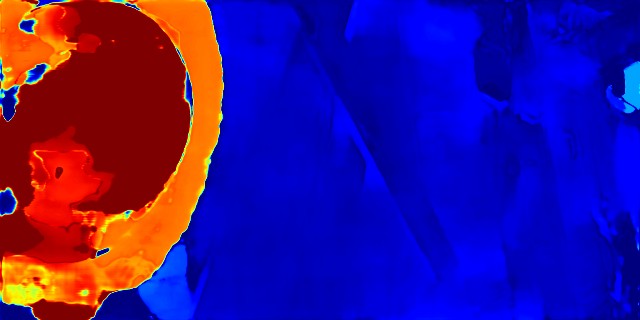} & 
    \includegraphics[width=0.285\linewidth]{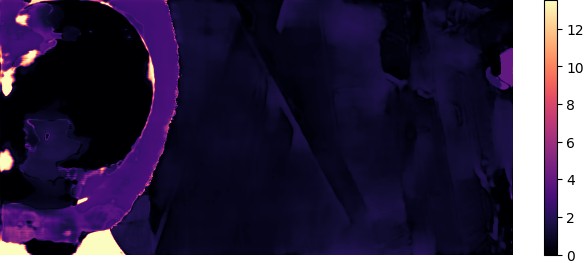} & 
    \includegraphics[width=0.25\linewidth]{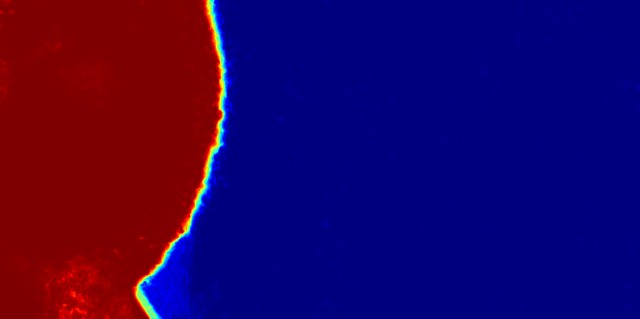} &
    \includegraphics[width=0.285\linewidth]{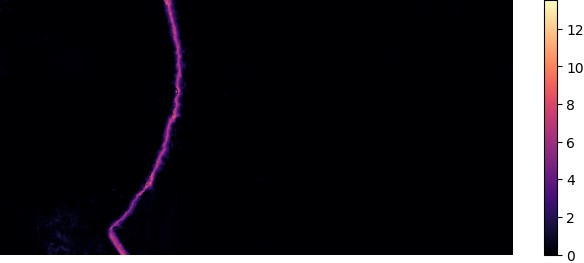} \\
    \includegraphics[width=0.25\linewidth]{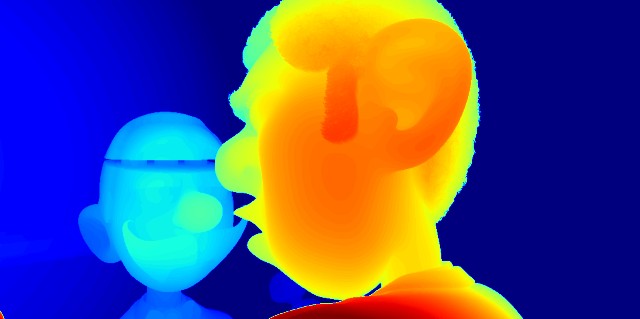} & 
    \includegraphics[width=0.25\linewidth]{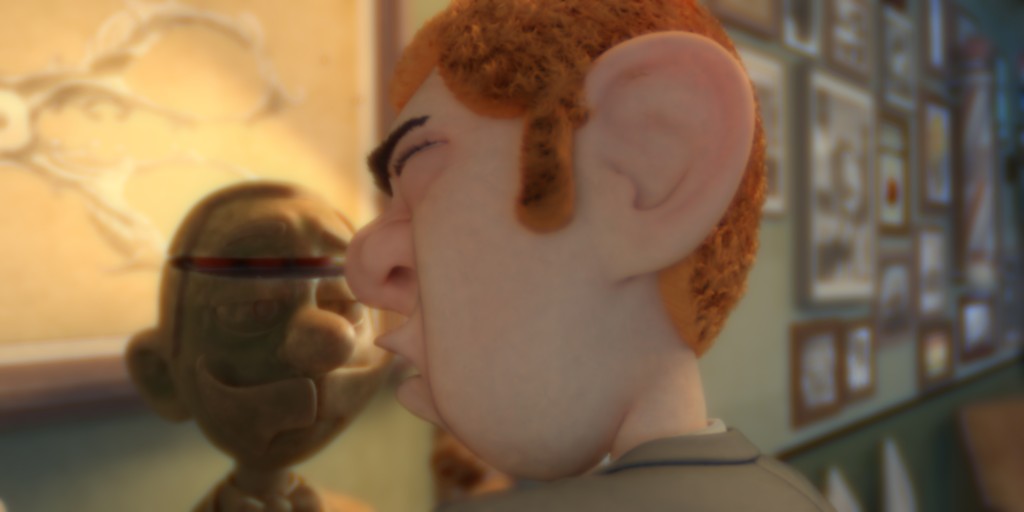} & 
    \includegraphics[width=0.25\linewidth]{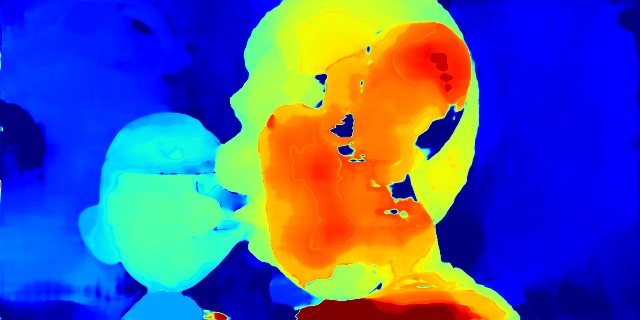} & 
    \includegraphics[width=0.285\linewidth]{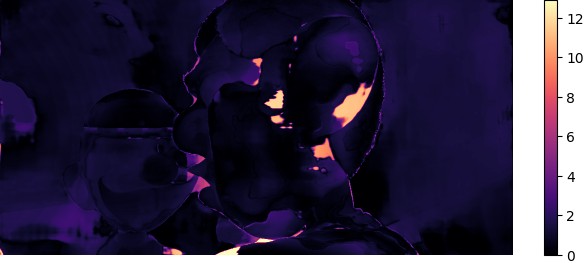} & 
    \includegraphics[width=0.25\linewidth]{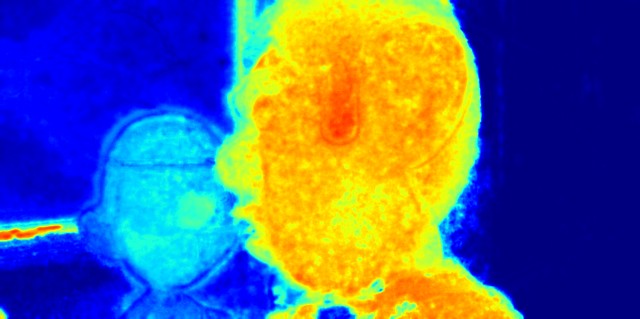} &
    \includegraphics[width=0.285\linewidth]{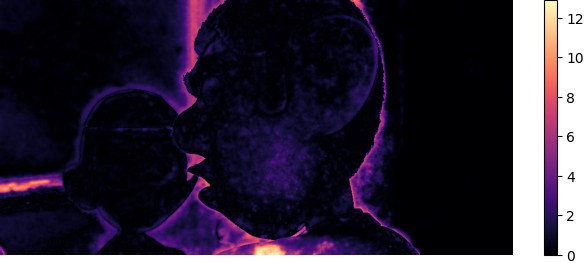} \\ 
    \includegraphics[width=0.25\linewidth]{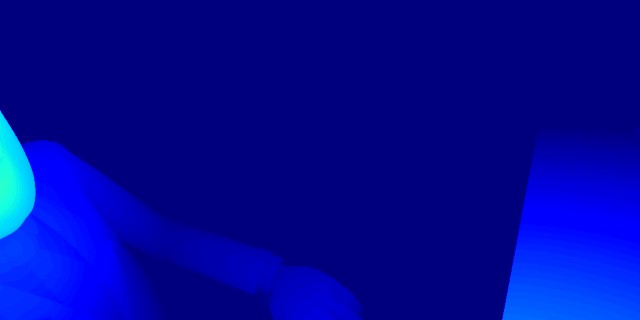} & 
    \includegraphics[width=0.25\linewidth]{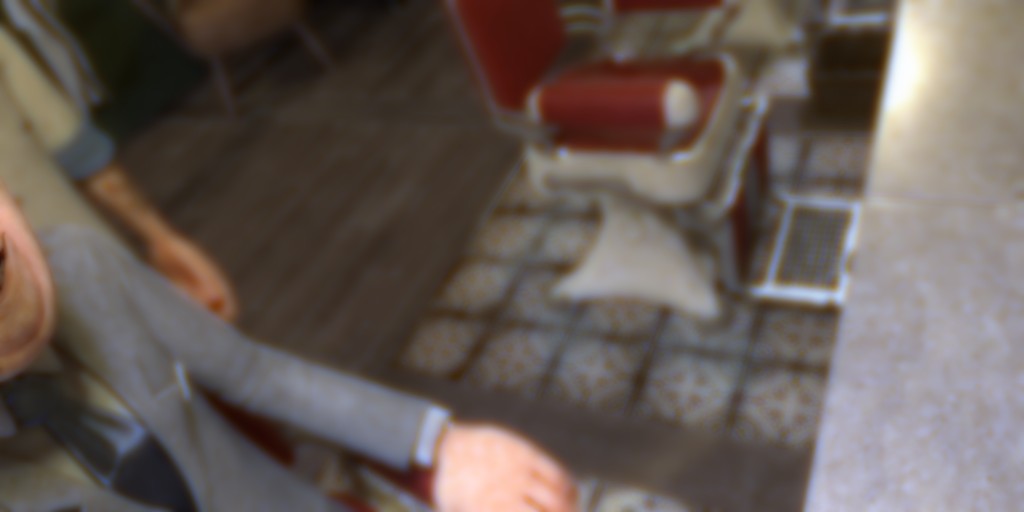} & 
    \includegraphics[width=0.25\linewidth]{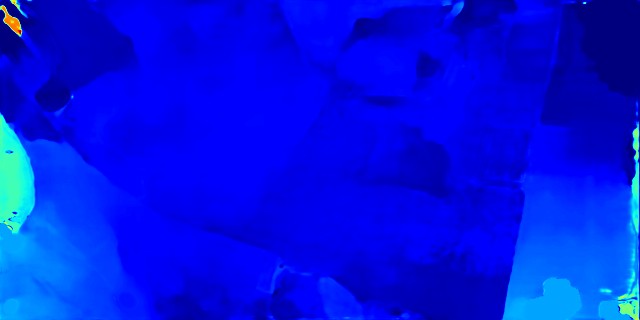} & 
    \includegraphics[width=0.285\linewidth]{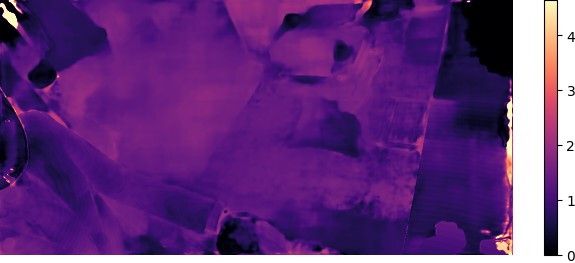} & 
    \includegraphics[width=0.25\linewidth]{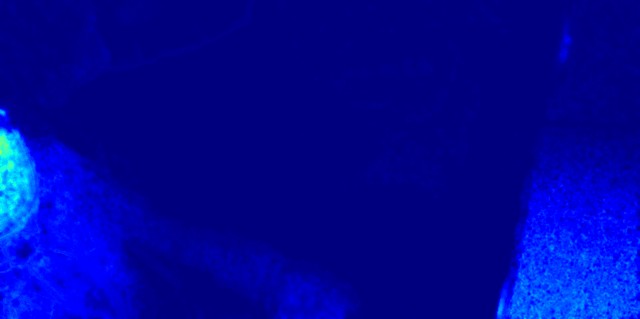} &
    \includegraphics[width=0.285\linewidth]{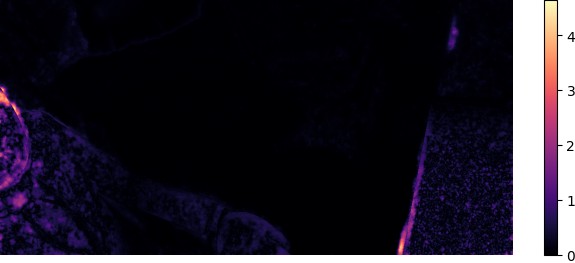} \\ 
    \includegraphics[width=0.25\linewidth]{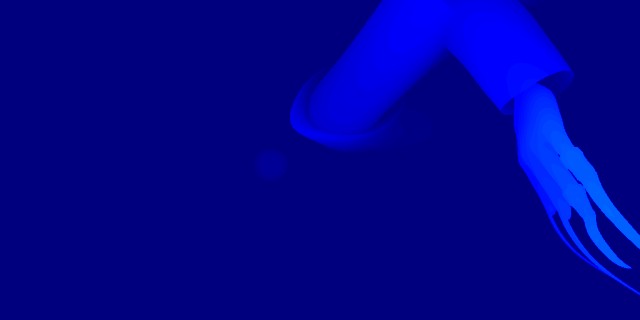} & 
    \includegraphics[width=0.25\linewidth]{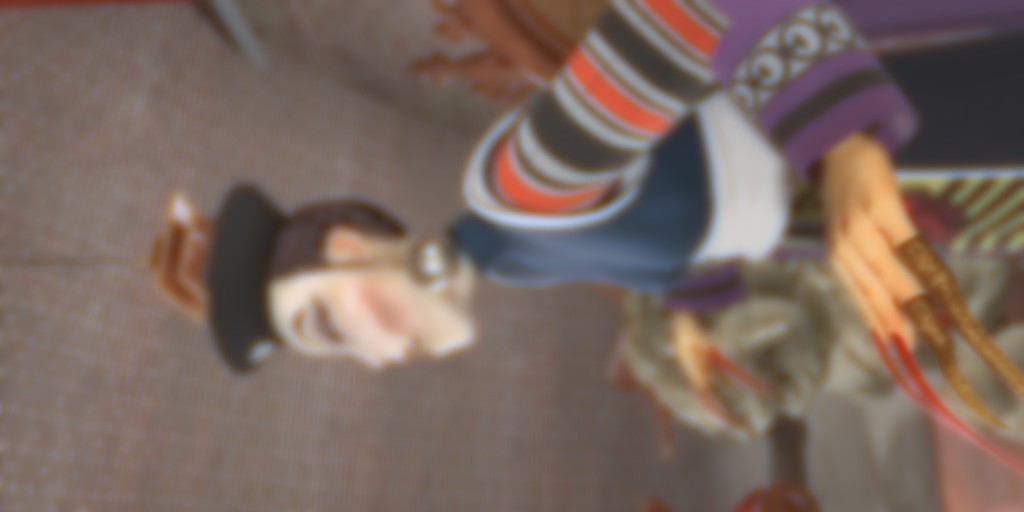} & 
    \includegraphics[width=0.25\linewidth]{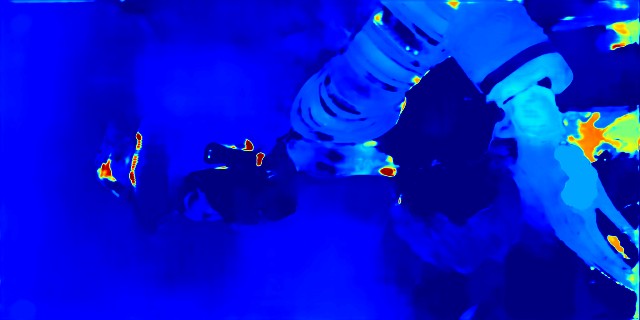} & 
    \includegraphics[width=0.285\linewidth]{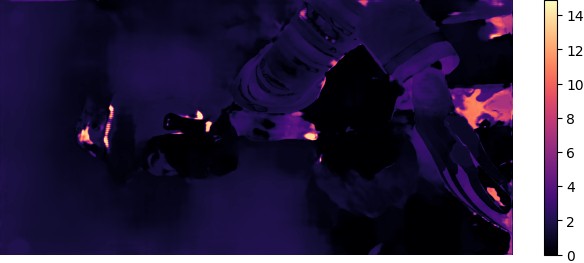} & 
    \includegraphics[width=0.25\linewidth]{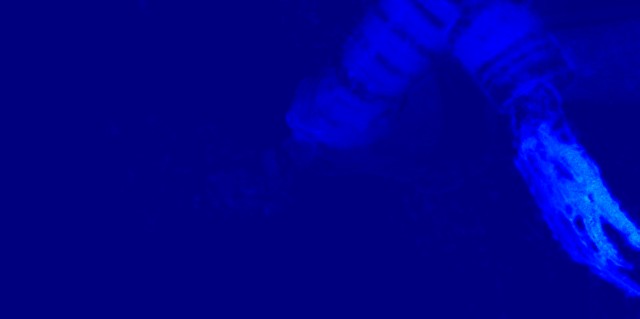} &
    \includegraphics[width=0.285\linewidth]{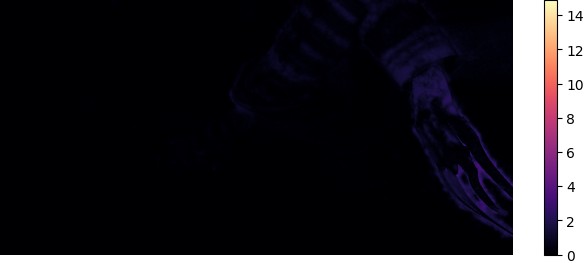} \\ 
    \end{tabular}
    }
    \caption{\textbf{Depth Estimation visual examples}. DPCIP produced more accurate depth maps, even on scenes the mono network was trained on. Note that the depths are clipped to the physical range of the imaging system.}
    \label{fig:depth_sup}
\end{figure}

We further extend our depth estimation visualizations to all the scenes in our dataset, in Figure \ref{fig:depth_sup} we show a single image from each scene including the error maps between the ground-truth and each model predictions. DPCIP produced more accurate and consistent depth maps across all scenes. 

\clearpage
\section{All in focus}
\begin{figure}[H]     
    \resizebox{1\linewidth}{!}{
    \begin{tabular}{cccc}
    & & Supervised & 0-shot\\
    GT & Input & EDOF\cite{elmalem2018learned} & DPCIP \\
    \includegraphics[width=0.25\linewidth]{SupMatFigures/AllInFocus/City_0140.png} & 
    \includegraphics[width=0.25\linewidth]{SupMatFigures/AllInFocus/City_0140.png} & 
    \includegraphics[width=0.25\linewidth]{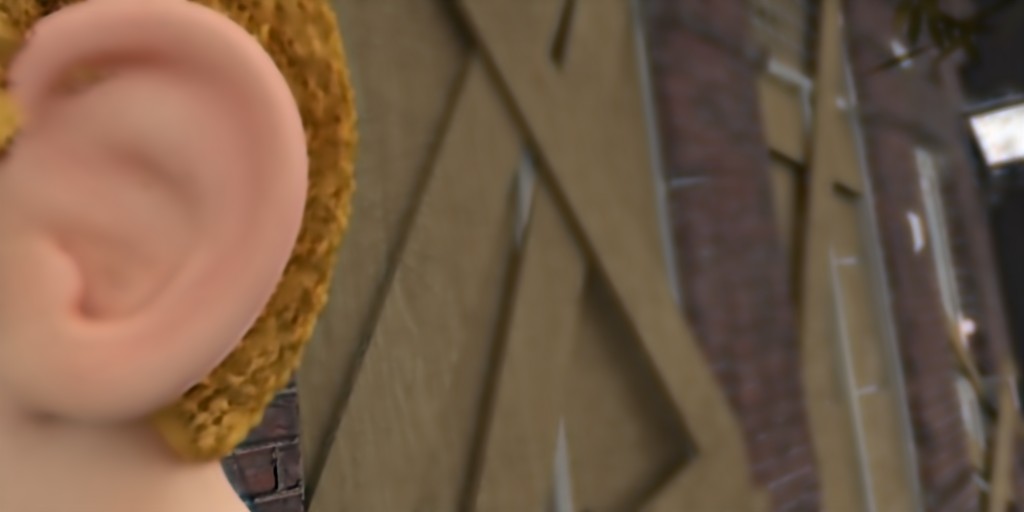} & 
    \includegraphics[width=0.25\linewidth]{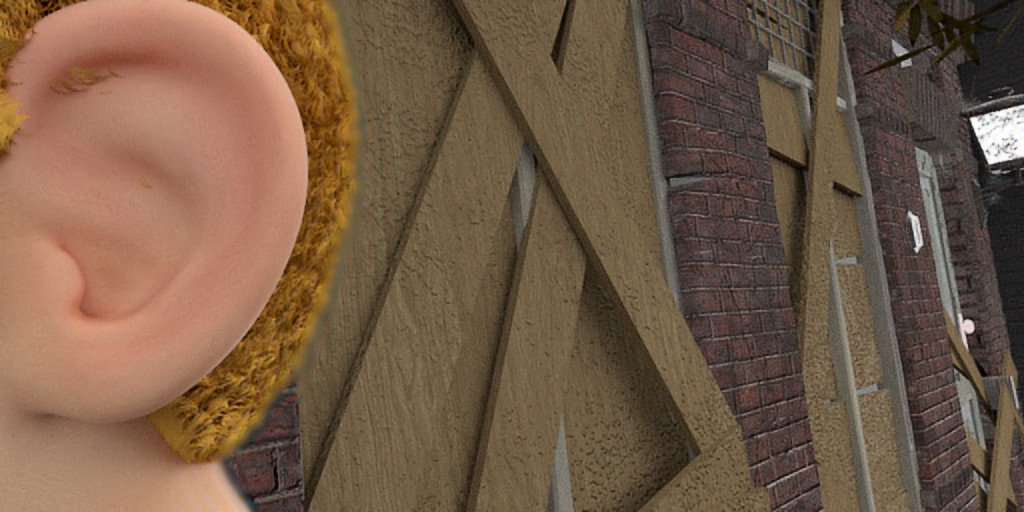} \\ 
    \includegraphics[width=0.25\linewidth]{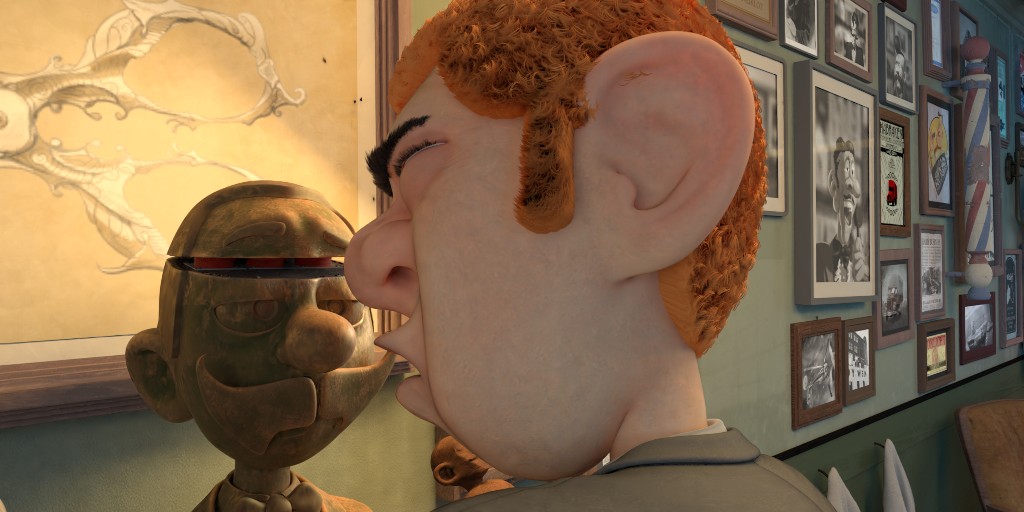} & 
    \includegraphics[width=0.25\linewidth]{SupMatFigures/AllInFocus/Headbutt_0081_maskImg.png} & 
    \includegraphics[width=0.25\linewidth]{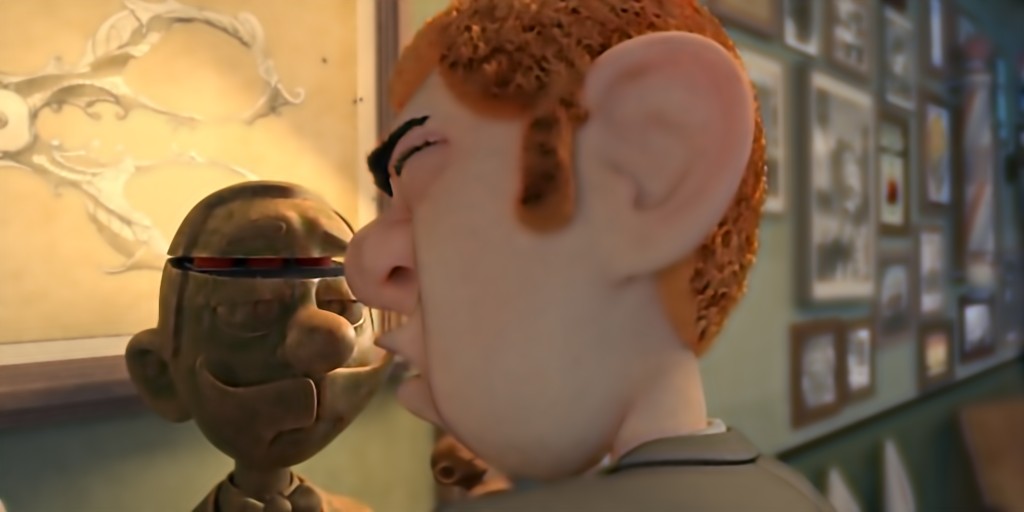} & 
    \includegraphics[width=0.25\linewidth]{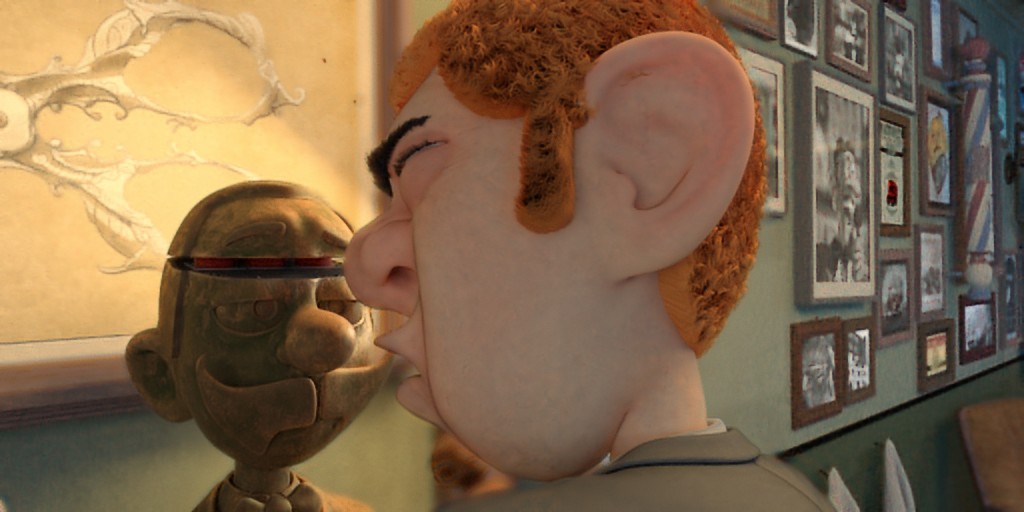} \\ 
    \includegraphics[width=0.25\linewidth]{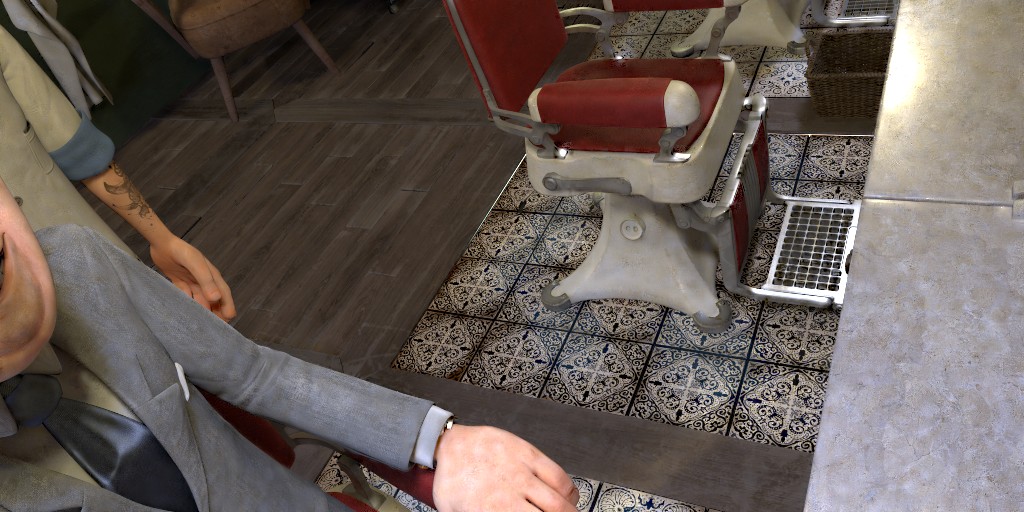} & 
    \includegraphics[width=0.25\linewidth]{SupMatFigures/Depth/Sitting_0498_maskImg.png} & 
    \includegraphics[width=0.25\linewidth]{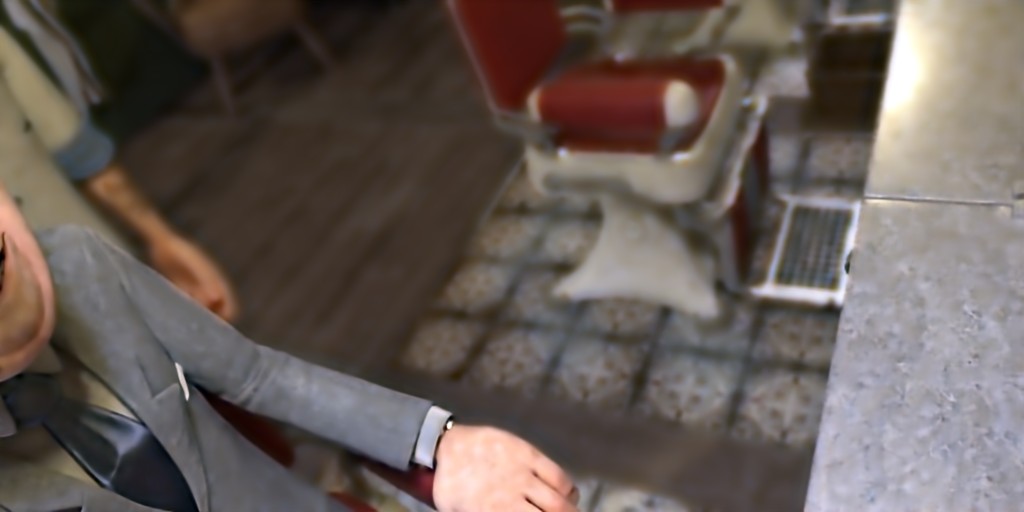} & 
    \includegraphics[width=0.25\linewidth]{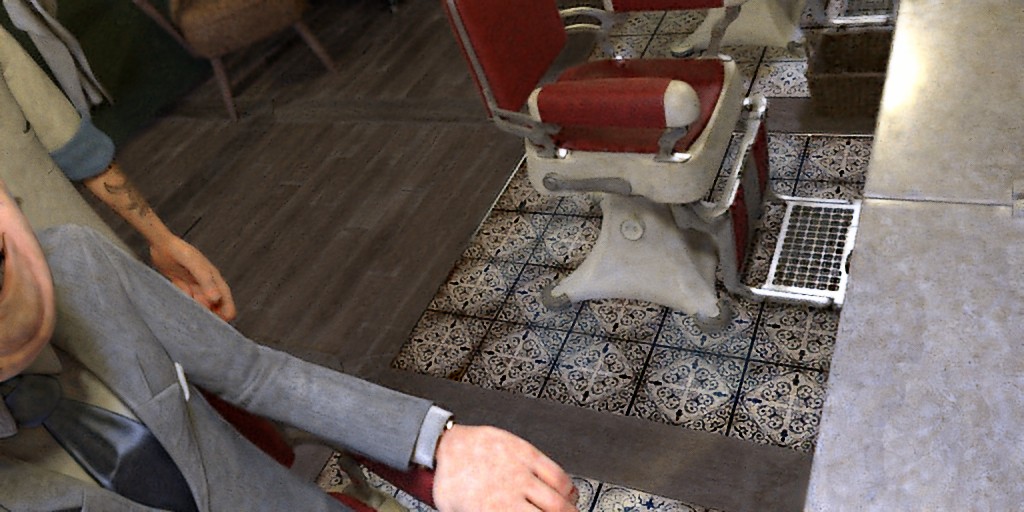} \\ 
    \includegraphics[width=0.25\linewidth]{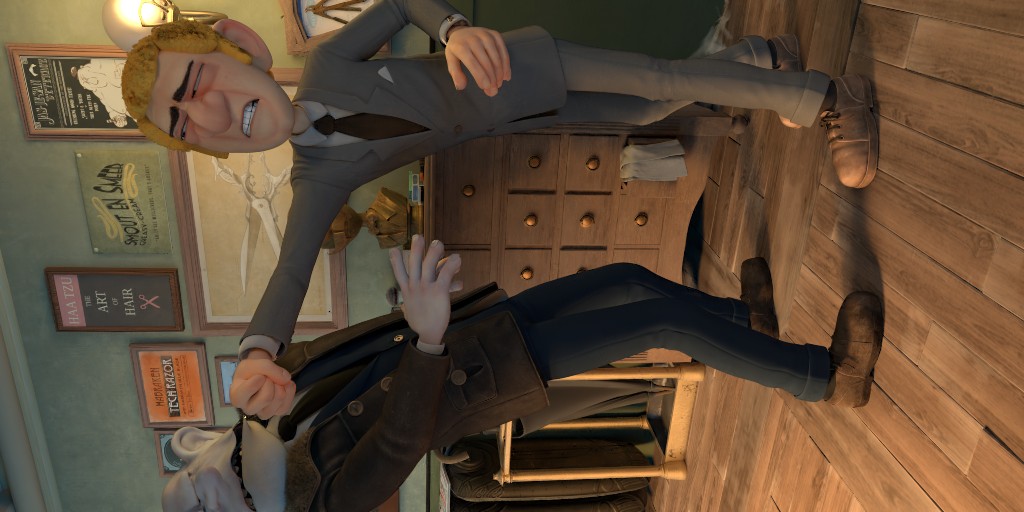} & 
    \includegraphics[width=0.25\linewidth]{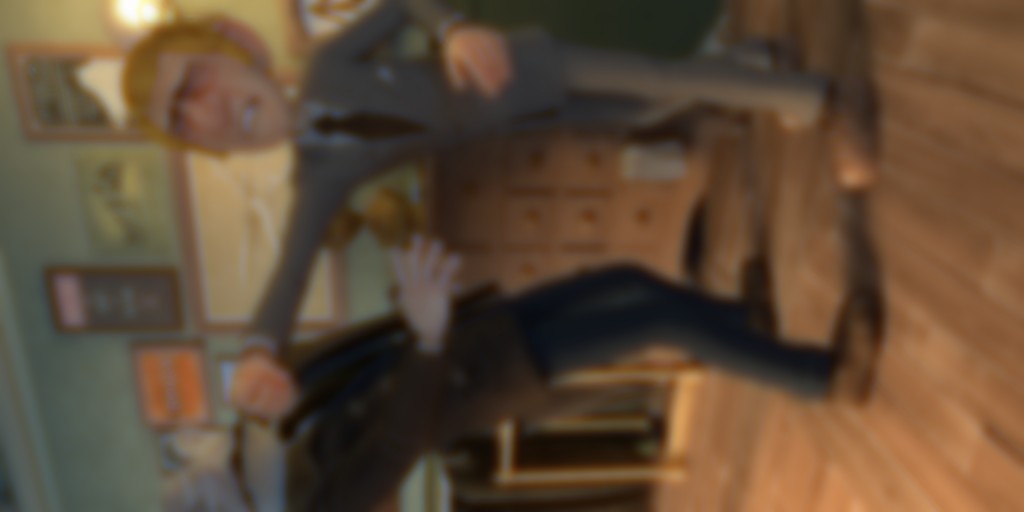} & 
    \includegraphics[width=0.25\linewidth]{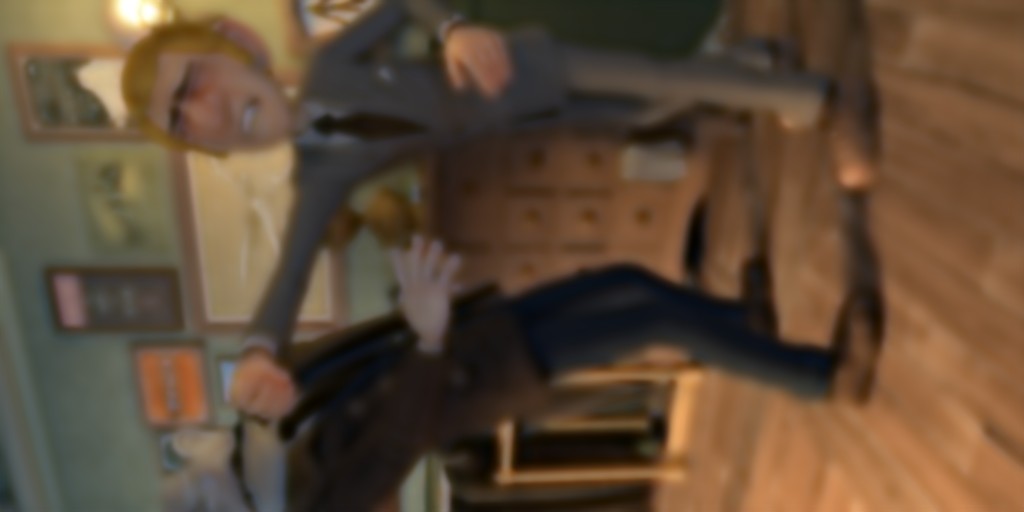} & 
    \includegraphics[width=0.25\linewidth]{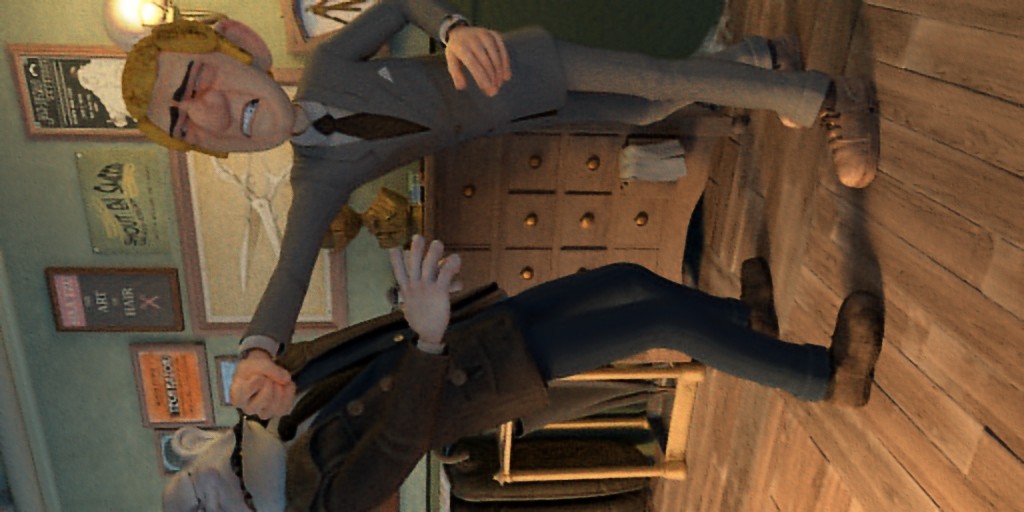} \\ 
    \includegraphics[width=0.25\linewidth]{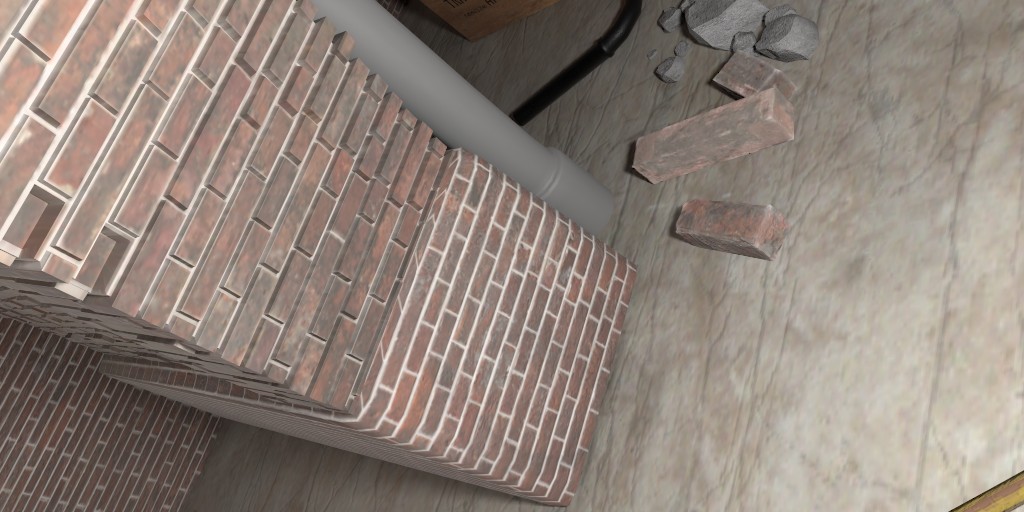} & 
    \includegraphics[width=0.25\linewidth]{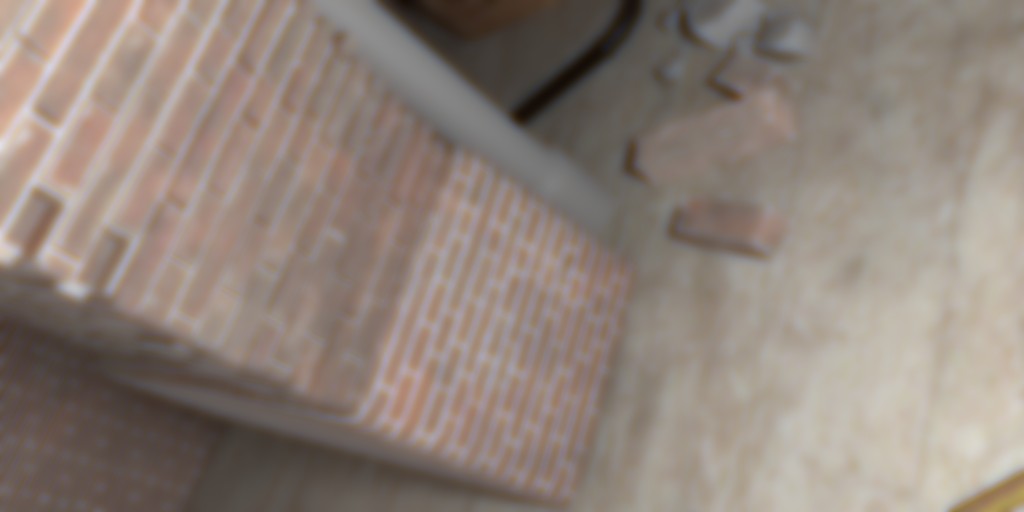} & 
    \includegraphics[width=0.25\linewidth]{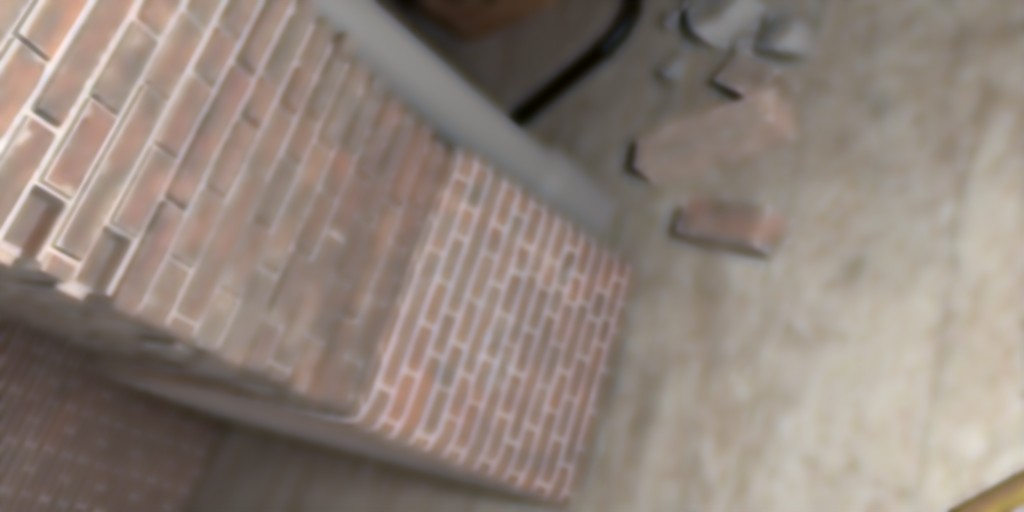} & 
    \includegraphics[width=0.25\linewidth]{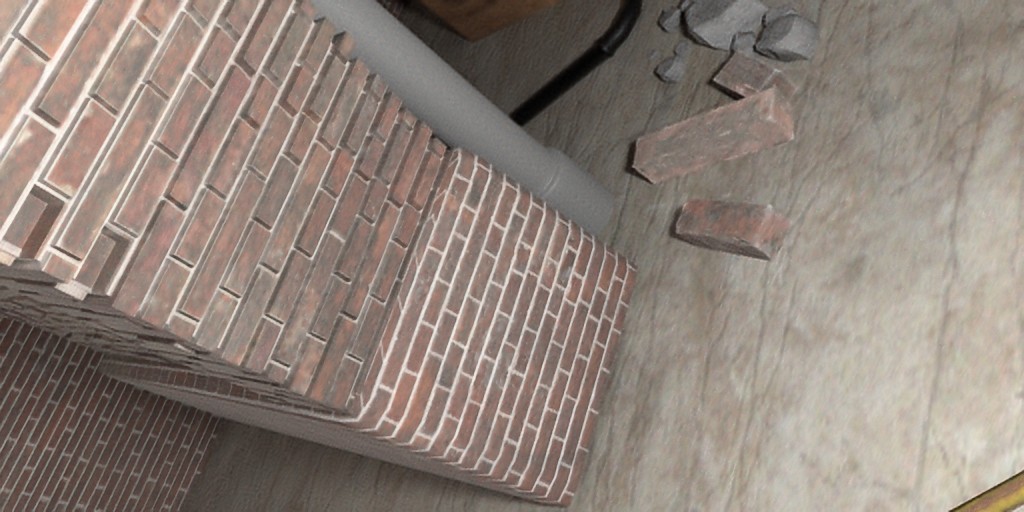} \\ 
    \end{tabular}
    }
    \caption{\textbf{All-in-Focus Image Reconstruction visual examples.} An example from each scene. Our method outperforms the baseline and produced significantly higher quality reconstruction.}
    \label{fig:all_in_focus_sup}    
\end{figure}

We further extend our image reconstruction visualizations to all the scenes in our dataset, in Figure \ref{fig:all_in_focus_sup} we show a single image from each scene. DPCIP produced better reconstructions across all images and scenes. 

\clearpage
\section{Image Deblurring}
\begin{figure}[H]    
    \resizebox{1.\linewidth}{!}{
    \begin{tabular}{cccc}
    & & 0-shot & 0-shot \\
    GT & Input & Neural Deblurring \cite{ren2020neural} & DPCIP \\
    \includegraphics[width=0.25\linewidth]{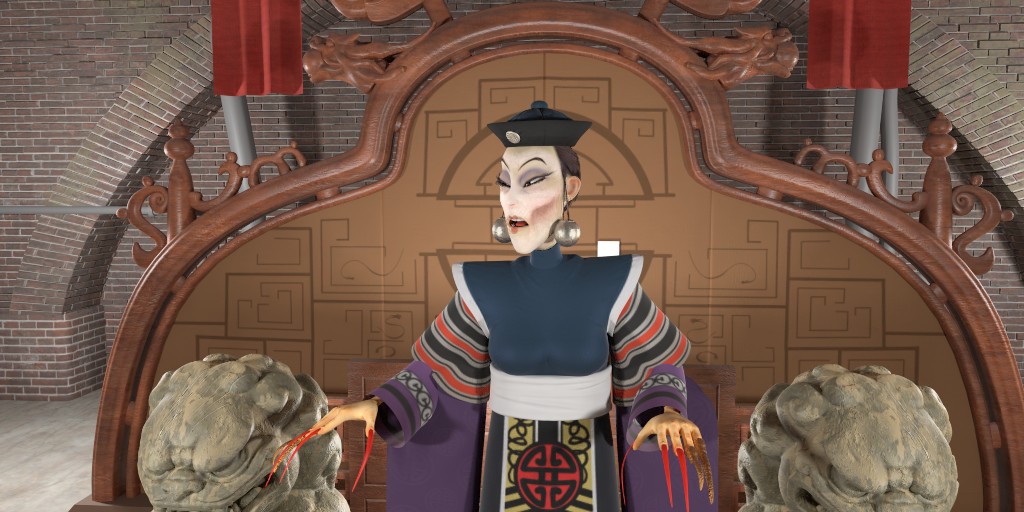} & 
    \includegraphics[width=0.25\linewidth]{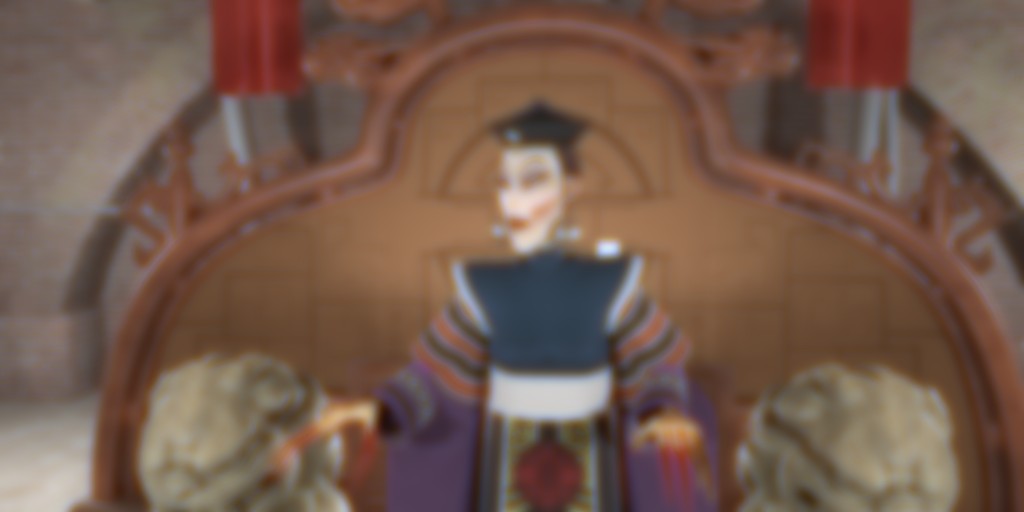} & 
    \includegraphics[width=0.25\linewidth]{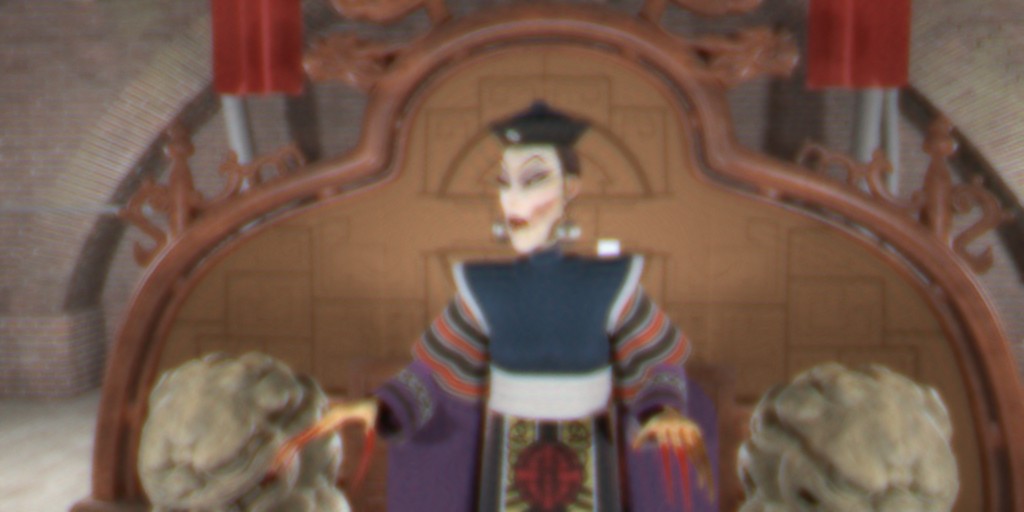} & 
    \includegraphics[width=0.25\linewidth]{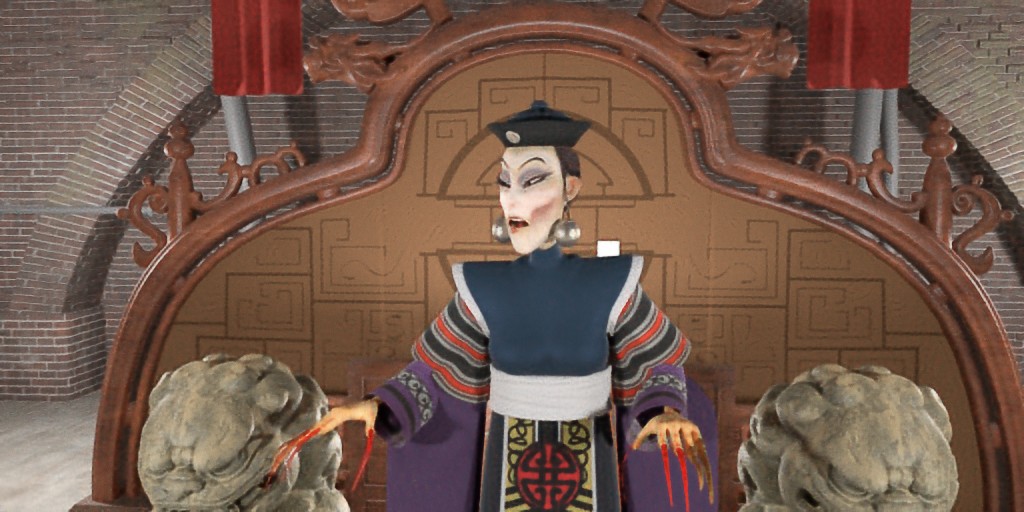} \\ 
    \end{tabular}
    }
    \caption{\textbf{Image Deblurring visual examples.} Our method produces an accurate sharp images, and is also capabale of handling large blurring kernels while SelfDeblur struggles.}
    \label{fig:deblurring_sup}    
\end{figure}

\begin{figure}[H]   
    \centering
    \resizebox{0.5\linewidth}{!}{
    \includegraphics[width=0.3\linewidth]{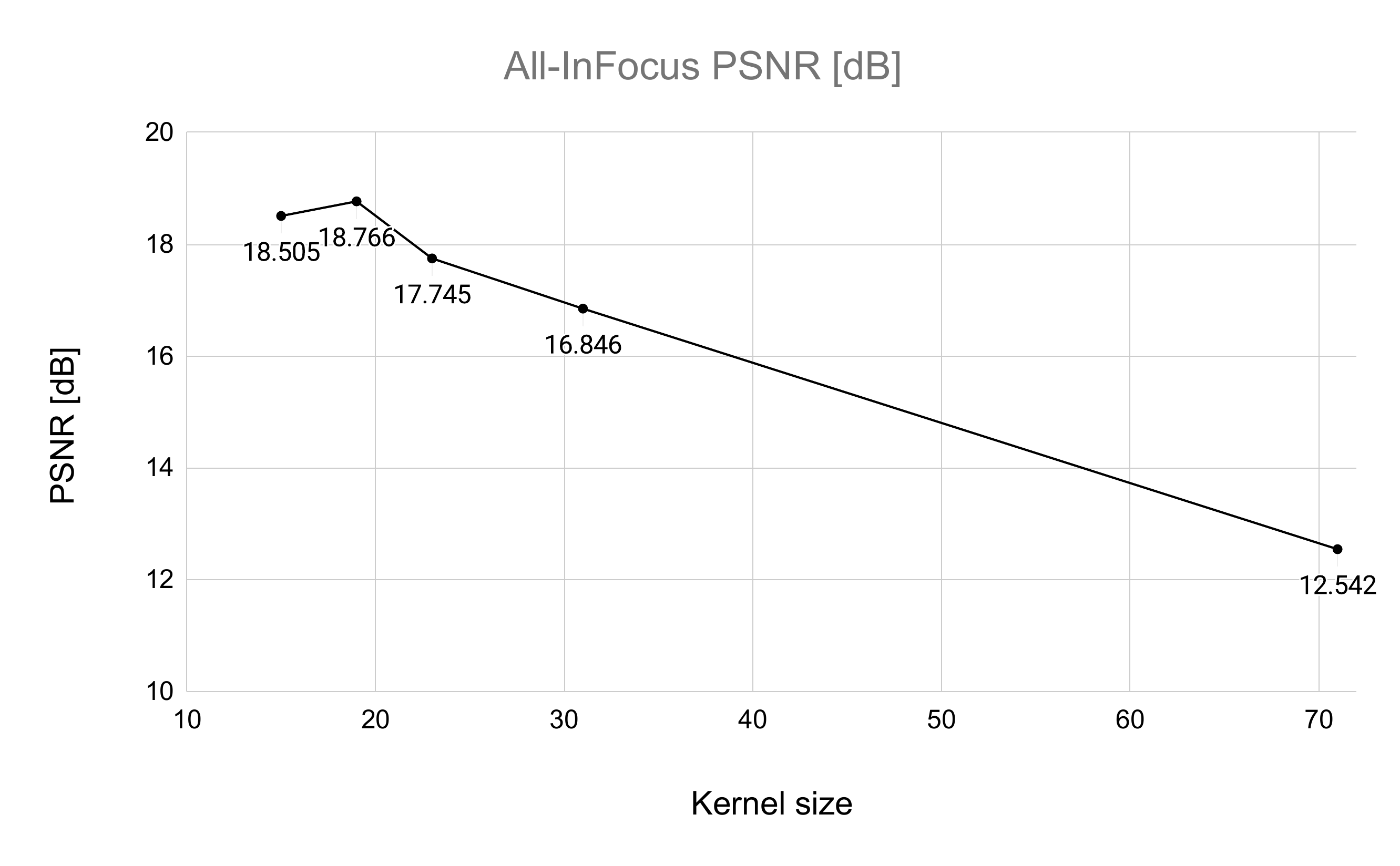}
    }
    \caption{\textbf{Blind image deblurring as a function of the kernel size}. The graph shows the average performance on a subset of samples to choose the best kernel size forNeural deblurring on the dataset. We can see that as the kernel size drops the performance improves, up to a certain level where the loss of deblurring signal is too large.}
    \label{fig:deblurring_kernel_size}
\end{figure}

We justify our modifications for the deblurring comparison by analyzing Nueral deblurring \cite{ren2020neural} performance based on the kernel size that it needs to estimate. Figure \ref{fig:deblurring_kernel_size} show the effect of kernel size on the reconstruction performance over a small subset of images. In addition Table \ref{tab:psf_kernel_size_reduction} analyze the energy lost for each psf kernel based on the kernel decimation. From both \ref{tab:psf_kernel_size_reduction, fig:deblurring_kernel_size} we can see that the can see that the chosen kernel size of $19\times19$ has limited energy loss compared to the performance boost it gives, thus justify our modification for chosing this size for our comparisons. In addition in Figure \ref{fig:deblurring_sup} we add more visuals showing DPCIP outperforming Nerual deblurring.

\begin{table}[H]
    \centering  
    \resizebox{0.3\linewidth}{!}{
    \begin{tabular}{l|ccc}
    \toprule
    Kernel size & R & G & B \\
    \midrule
    71x71 & 100\% & 100\% & 100\% \\    
    31x31 & 94.72\% & 96.69\% & 98.49\% \\    
    23x23 & 87.31\% & 94.30\% & 97.32\% \\    
    19x19 & 77.20\% & 89.10\% & 96.25\% \\    
    15x15 & 56.24\% & 78.81\% & 94.28\% \\    
    \bottomrule
\end{tabular}
}
\vspace{3pt}
\caption{\textbf{PSF Kernel energy spread analysis}. Compared to the full signal, the Blue kernel is the most centered, while the Red kernel is the most spread out. We can see that reducing the kernel by half nearly effect the energy, but as we keep reducing the kernel size The loss of energy becomes severe.}
\label{tab:psf_kernel_size_reduction}
\end{table}

\section{Real World}
We add more samples of real-world images and comparing DPCIP both for depth map estimation vs the mono network from \cite{gil2019monster} and all-in-focus reconstruction vs \cite{elmalem2018learned}.
In addition we add visuals of DPCIP compared to our trained version of \cite{ikoma2021depth}. We can see that \cite{ikoma2021depth} is struggling vs. DPCIP probably due to a limited dataset for training, which emphasize the advantage of DPCIP as a zero-shot method.
\begin{figure}[H] 
\centering
\resizebox{1.\linewidth}{!}{
    \begin{tabular}{cccc}
    & Supervised & 0-shot & 0-shot \\
    Blur img & EDOF\cite{elmalem2018learned} / Mono from \cite{gil2019monster} & DPCIP (DIP) & DPCIP (PIP) \\
    \multirow{2}{*}[40pt]{\includegraphics[width=0.25\linewidth]{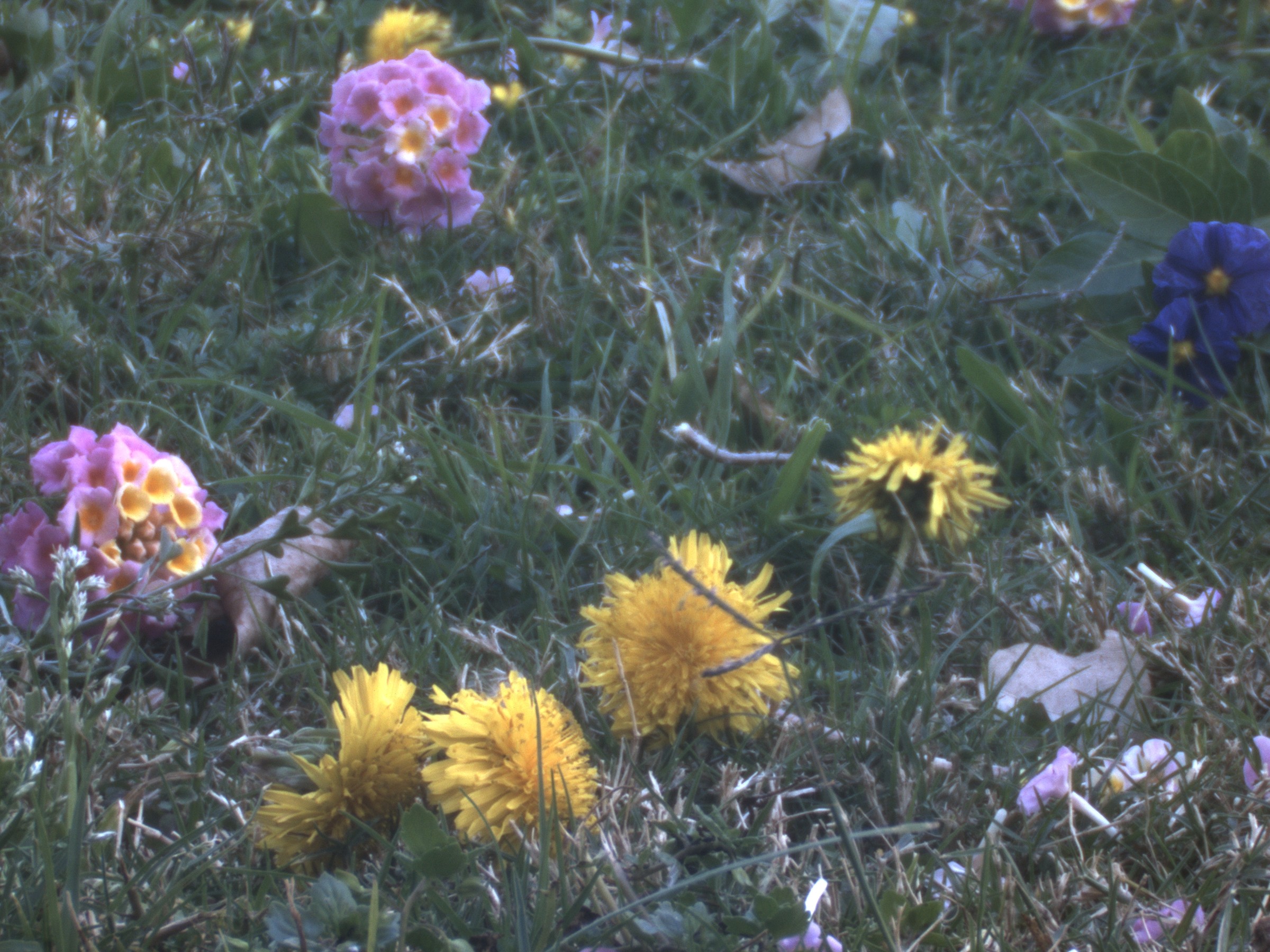}} & 
    \includegraphics[width=0.25\linewidth]{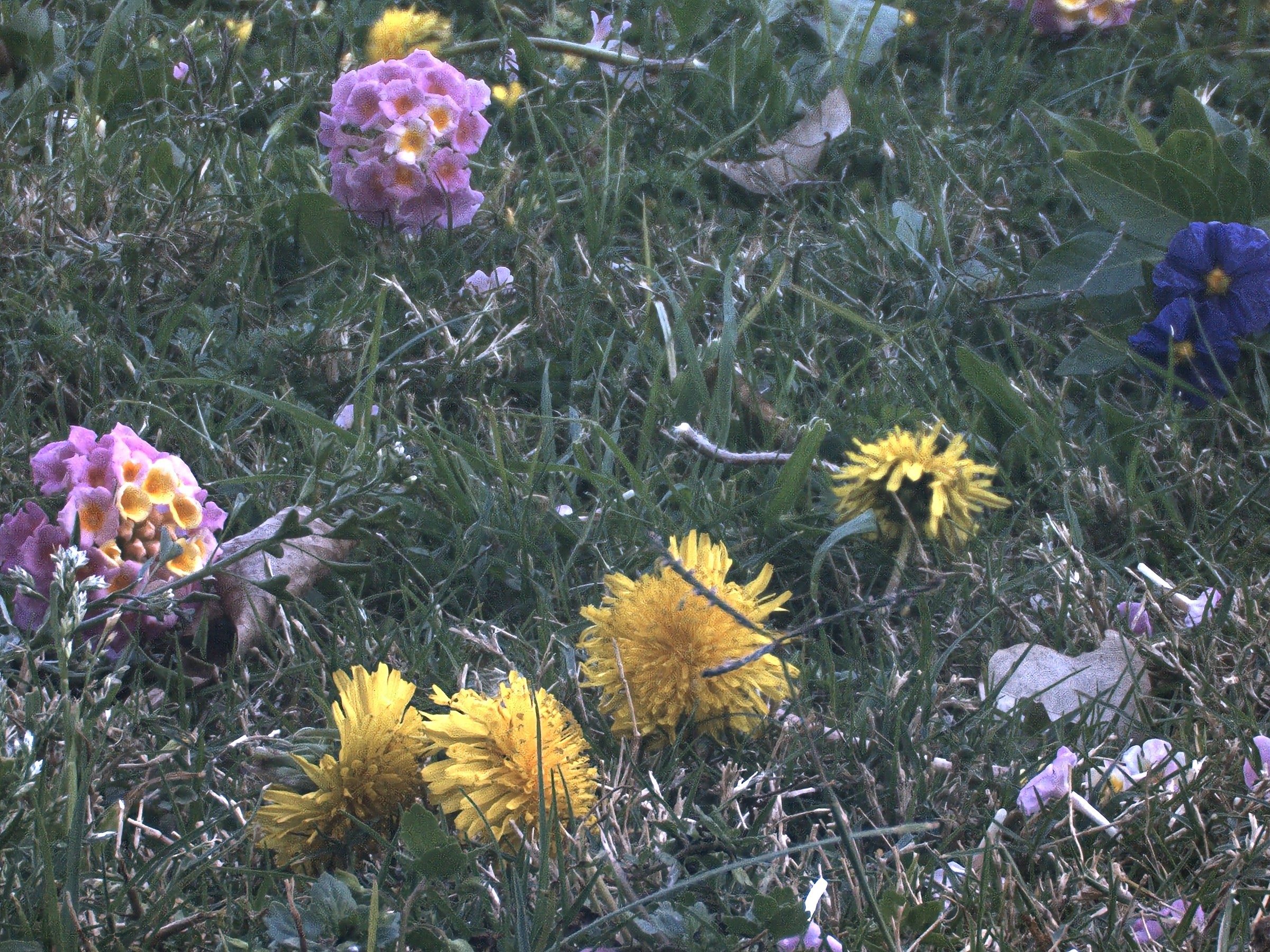} & 
    \includegraphics[width=0.25\linewidth]{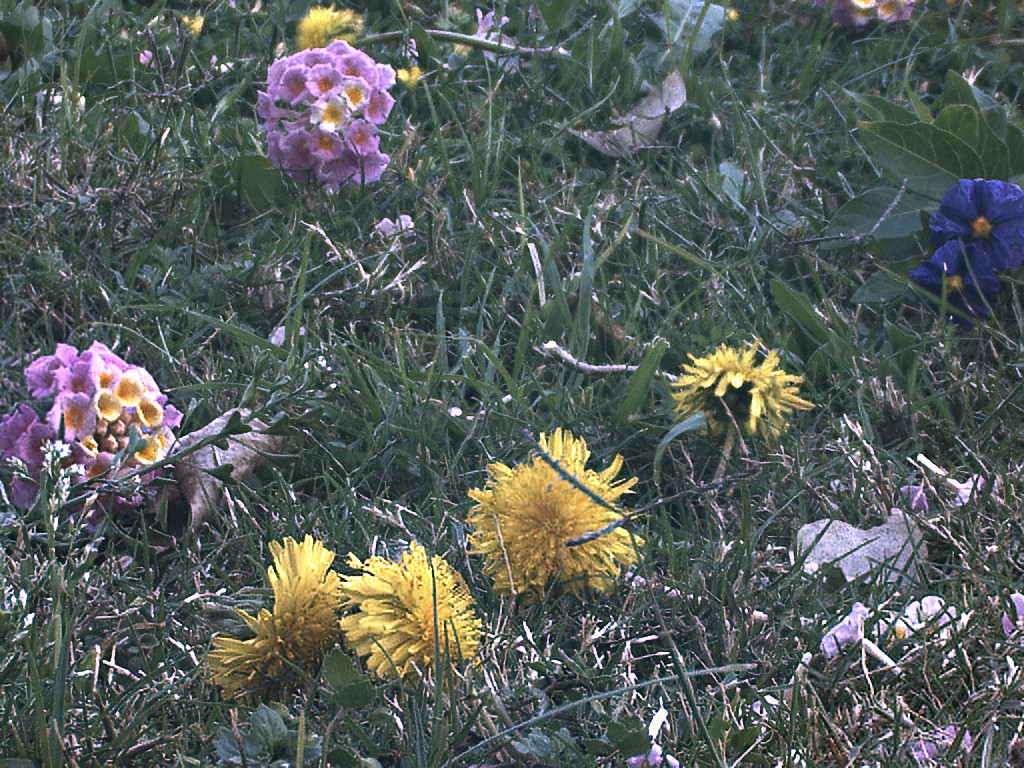} &
    \includegraphics[width=0.25\linewidth]{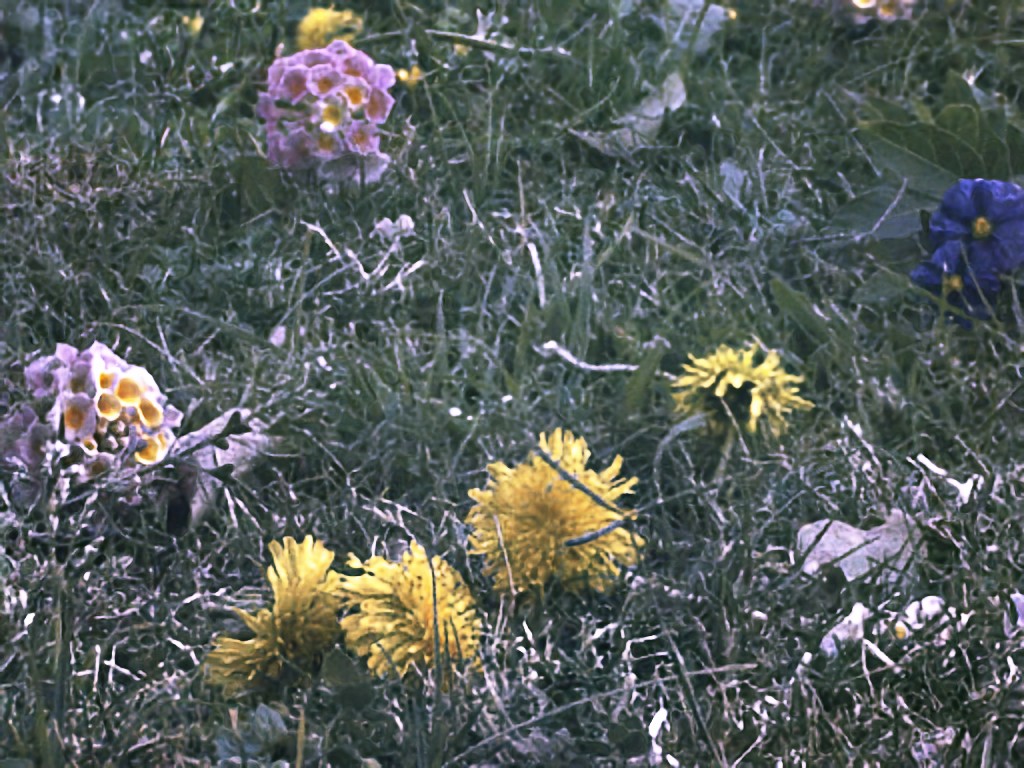} \\
    &
    \includegraphics[width=0.25\linewidth]{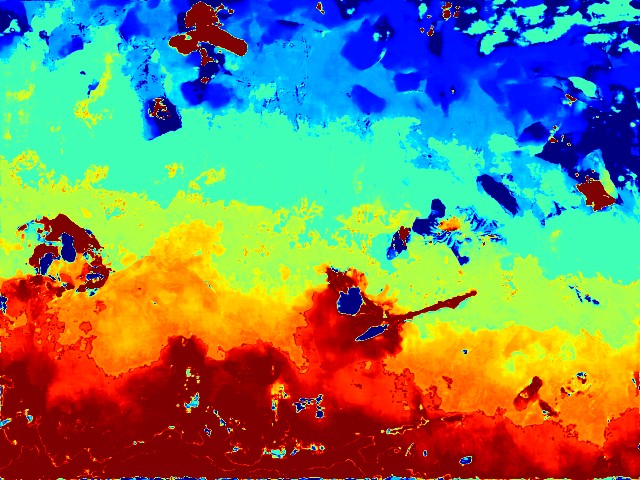} & 
    \includegraphics[width=0.25\linewidth]{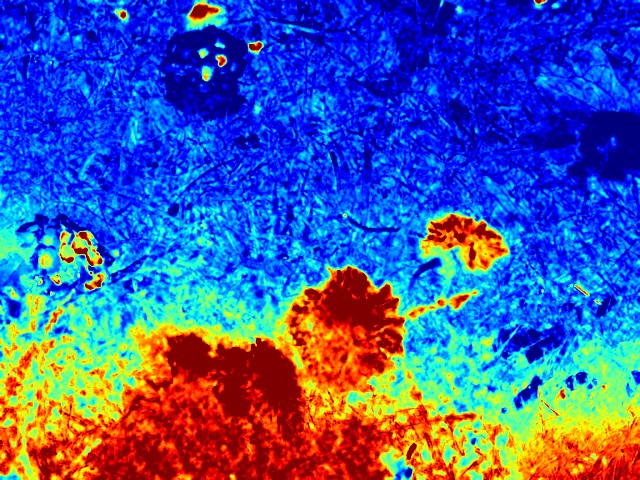} &
    \includegraphics[width=0.25\linewidth]{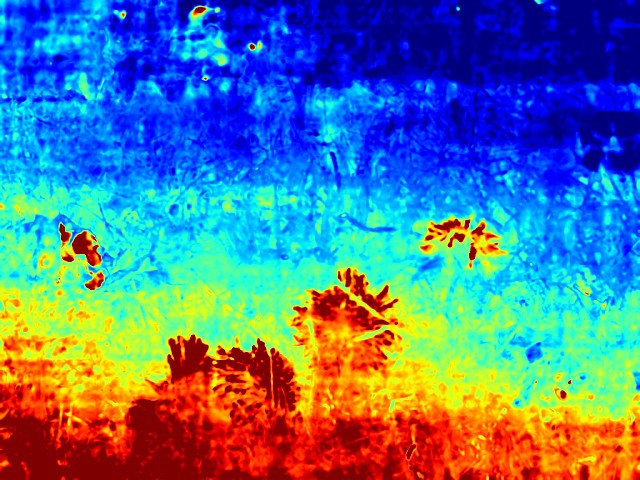} \\
    \vspace{10pt}
    \multirow{2}{*}[40pt]{\includegraphics[width=0.25\linewidth]{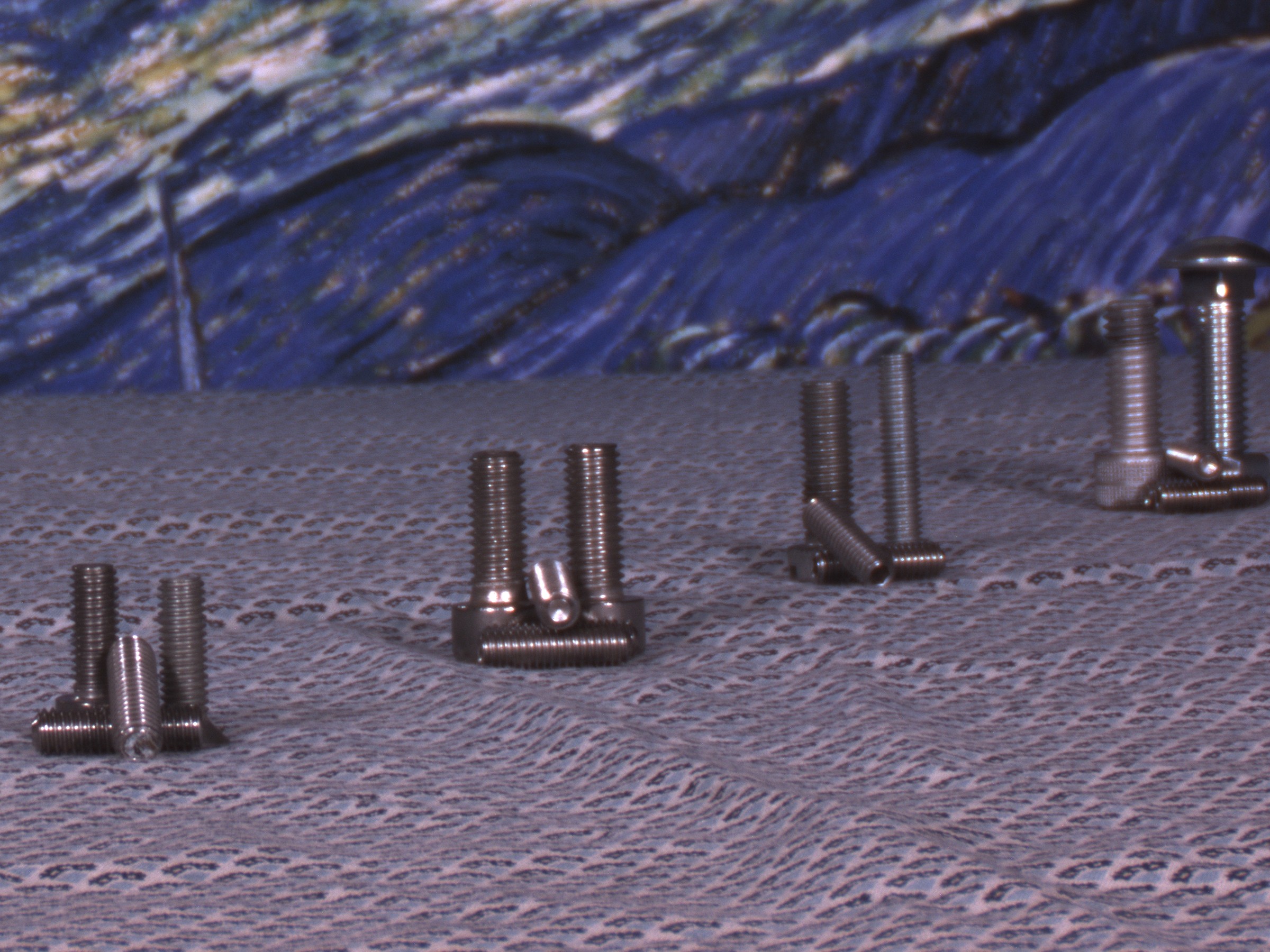}} & 
    \includegraphics[width=0.25\linewidth]{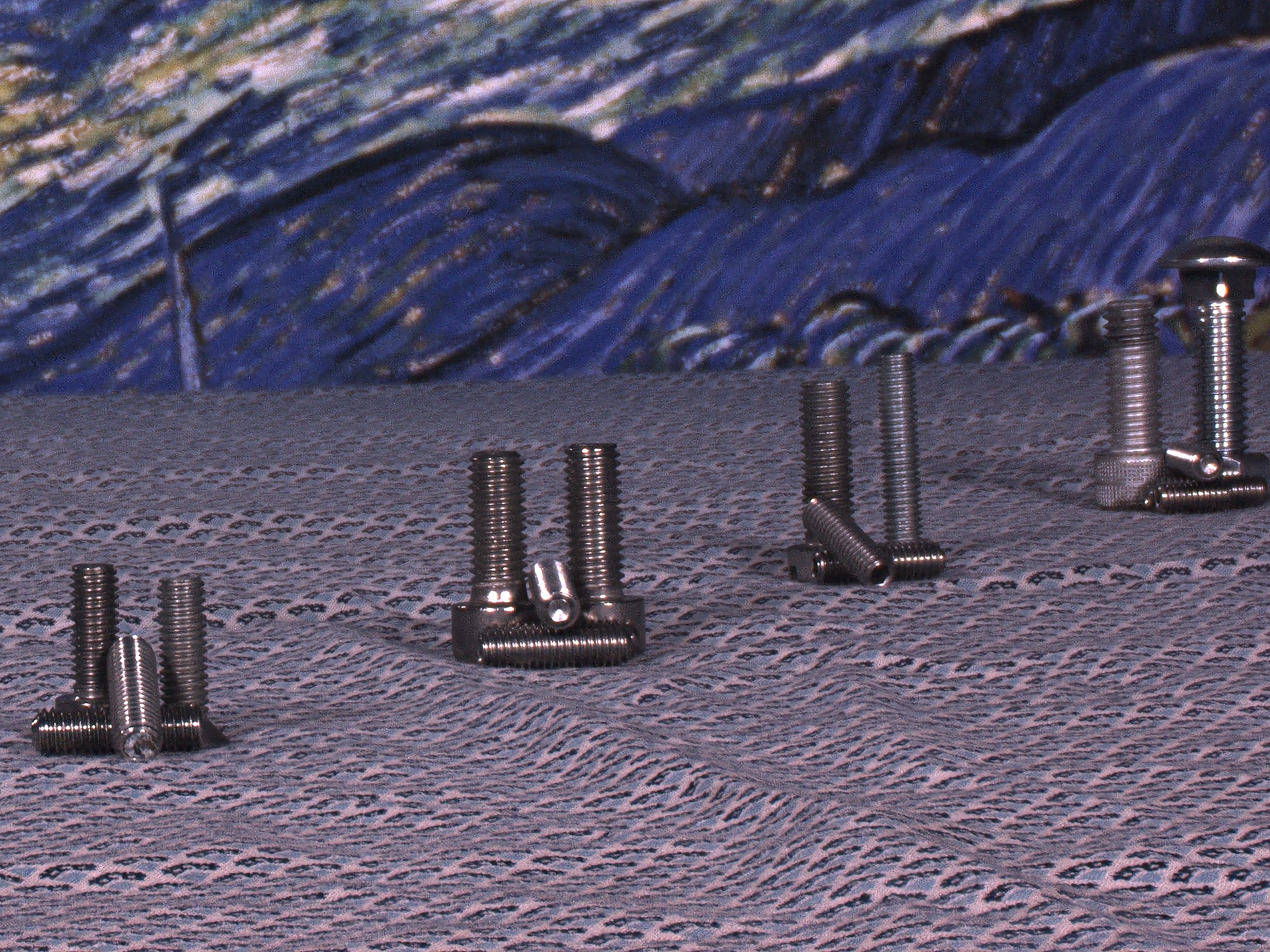} & 
    \includegraphics[width=0.25\linewidth]{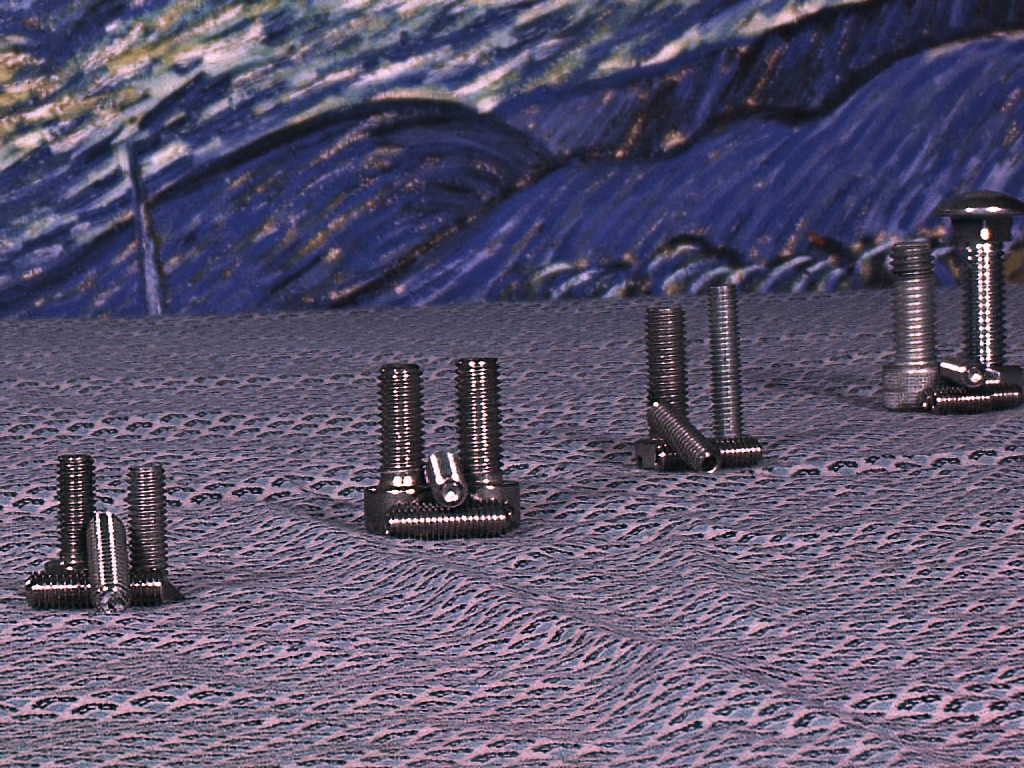} &
    \includegraphics[width=0.25\linewidth]{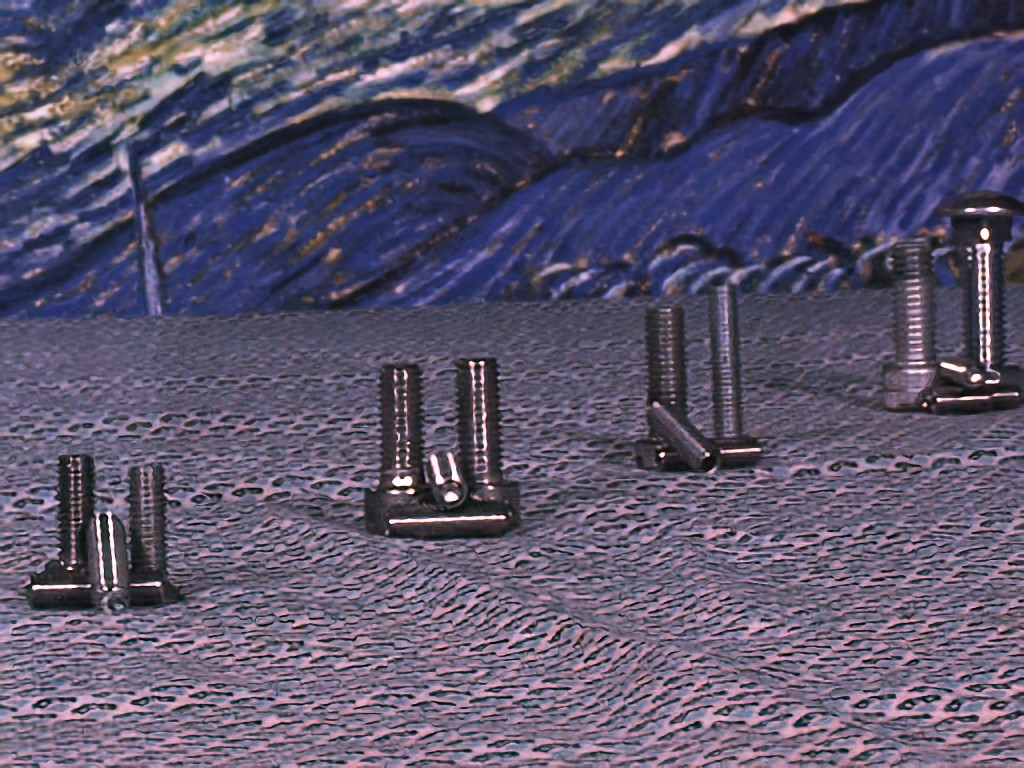} \\
    & 
    \includegraphics[width=0.25\linewidth]{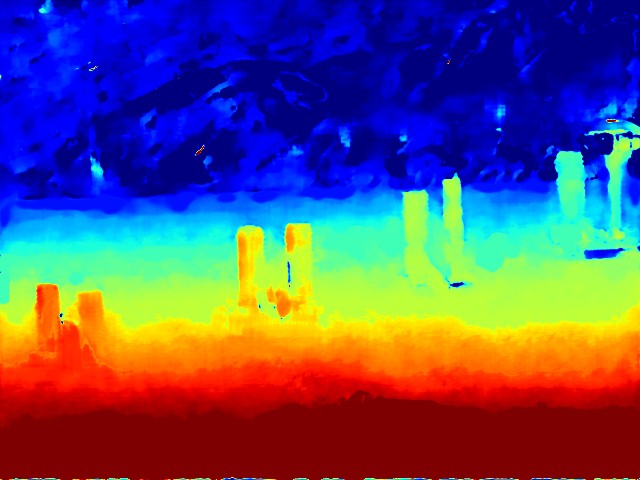} & 
    \includegraphics[width=0.25\linewidth]{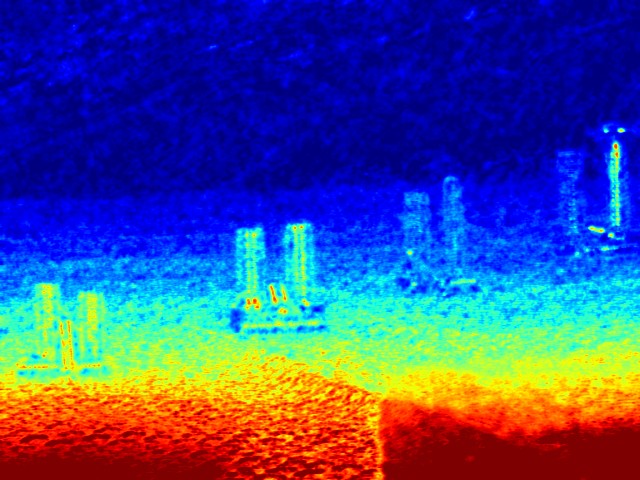} &
    \includegraphics[width=0.25\linewidth]{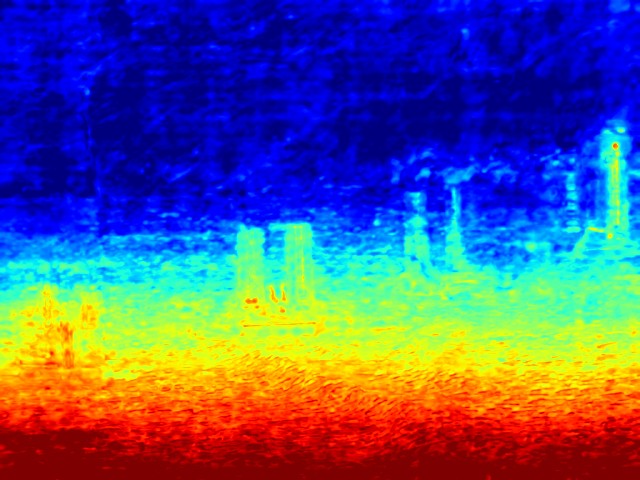} \\
    \end{tabular}
    }
    \caption{\textbf{Real world qualitative examples.} DPCIP produced higher-quality reconstruction images and depth maps compared to the baselines.}
    \label{fig:realworld_sup}
\end{figure}

\begin{figure}[H] 
\centering
\resizebox{1.\linewidth}{!}{
    \begin{tabular}{cccc}
    & Supervised & 0-shot & 0-shot\\
    Blur img & snapshot \cite{ikoma2021depth} & DPCIP (DIP) &  DPCIP (PIP)\\
    \includegraphics[width=0.25\linewidth]{Figures/Blur/out_6_RGB_inp2.png} & 
    \includegraphics[width=0.25\linewidth]{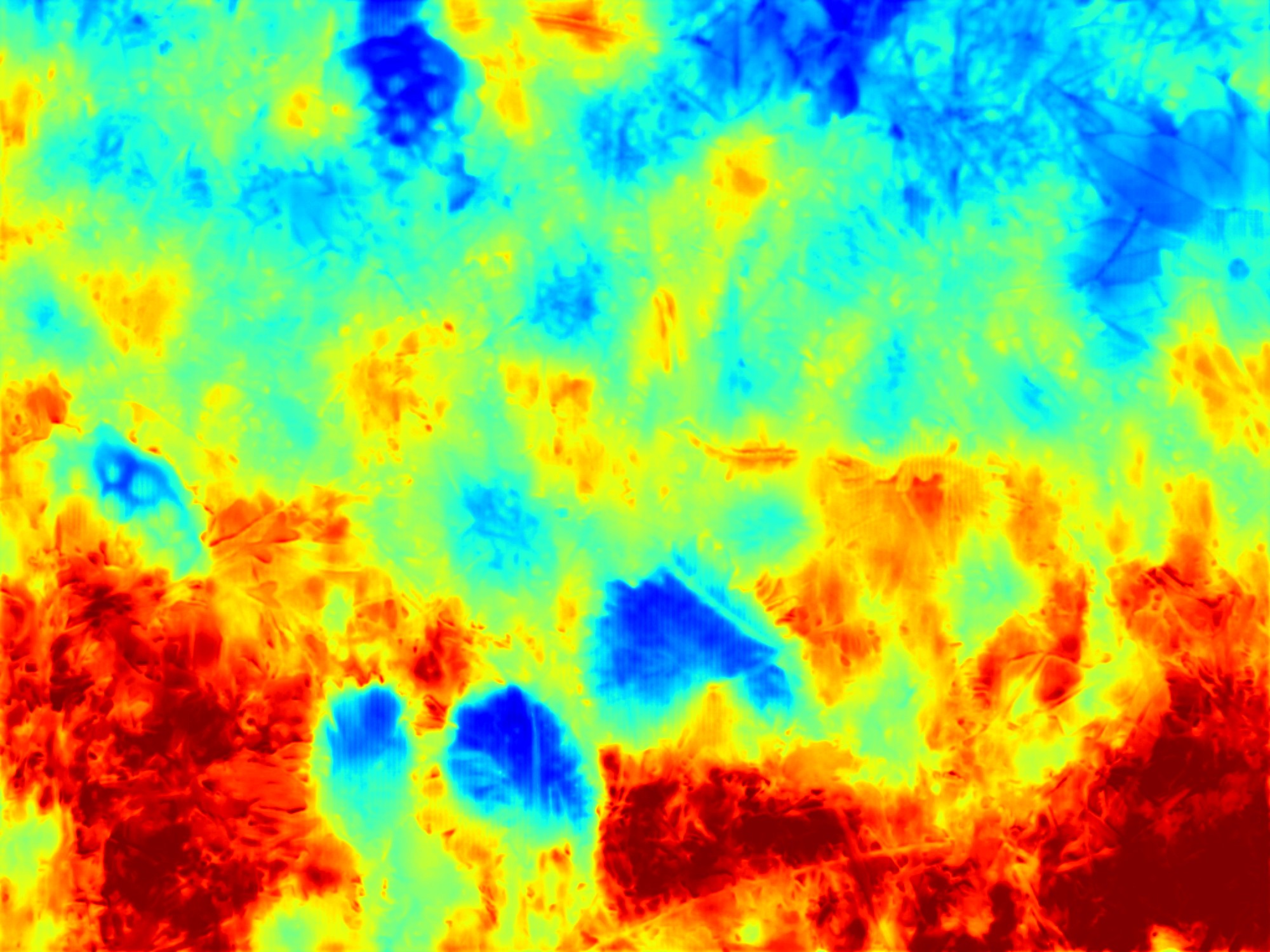} & 
    \includegraphics[width=0.25\linewidth]{Figures/Ours/RealWorld/out_6_DPCIP_dip_depth_jet.png} &
    \includegraphics[width=0.25\linewidth]{Figures/Ours/RealWorld/out_6_DPCIP_pip_depth_jet.png} \\
    \includegraphics[width=0.25\linewidth]{Figures/Blur/in1_RGB_inp2.png} & 
    \includegraphics[width=0.25\linewidth]{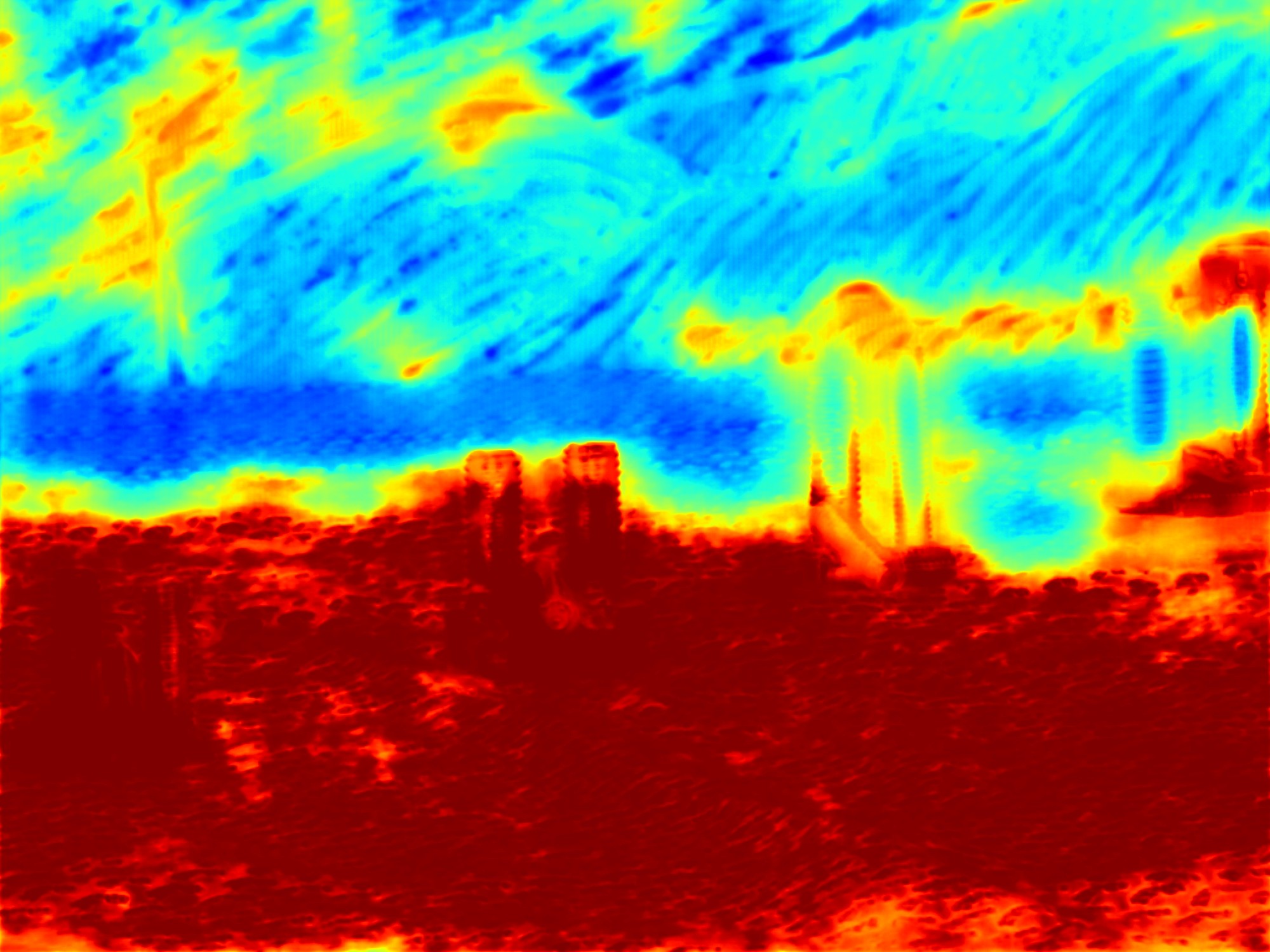} & 
    \includegraphics[width=0.25\linewidth]{Figures/Ours/RealWorld/in_1_DPCIP_dip_depth_jet.png} &
    \includegraphics[width=0.25\linewidth]{Figures/Ours/RealWorld/in_1_DPCIP_pip_depth_jet.png} \\    
    \end{tabular}
    }
    \caption{\textbf{Visual comparison to \cite{ikoma2021depth}} - We can see that our 0-shot method produced better depth maps than the supervised method trained on our small dataset, this strenghten our advantage as 0-shot method that do not need datasets to produce depth maps.}
    \label{fig:realworld_snapshot}
\end{figure}

\bibliographystyle{IEEEtran}
\bibliography{main}